\documentclass{aa}
\usepackage[utf8]{inputenc}
\usepackage[varg]{txfonts}
\usepackage{graphicx}
\usepackage{color}
\usepackage{caption}
\usepackage{natbib}
\usepackage{tabularx}
\usepackage{textgreek}
\usepackage{amsmath} 
\usepackage{amsbsy}
\usepackage{geometry}
\usepackage{makecell}
\usepackage{xspace}
\usepackage{booktabs,caption}
\usepackage[para, online]{threeparttable}

\newcommand{\lya}{\textup{Ly$\alpha$}\xspace}

\newcommand{\flya}{\textup{$f_{\mathrm{esc}}^{\lya}$}\xspace}
\newcommand{\flyc}{\textup{$f_{\mathrm{esc}}^{\mathrm{LyC}}$}\xspace}
\newcommand{\Msun}{\textup{$M_{\odot}$}\xspace}
\newcommand{\Llyaout}{\textup{$L_{\mathrm{esc}}^{\lya}$}\xspace}
\newcommand{\Llyain}{\textup{$L_{\mathrm{int}}^{\lya}$}\xspace}
\newcommand{\Llycout}{\textup{$L_{\mathrm{esc}}^{\mathrm{LyC}}$}\xspace}
\newcommand{\Llycin}{\textup{$L_{\mathrm{int}}^{\mathrm{LyC}}$}\xspace}
\newcommand{\Llycnineout}{\textup{$L_{\mathrm{esc}}^{\mathrm{900}}$}\xspace}
\newcommand{\Llycninein}{\textup{$L_{\mathrm{int}}^{\mathrm{900}}$}\xspace}
\newcommand{\flycnine}{\textup{$f_{\mathrm{esc}}^{\mathrm{900}}$}\xspace}
\newcommand{\radj}{\textup{$R_{\mathrm{adj}}^2$}\xspace}
\newcommand{\lten}{\textup{$\mathrm{log}_{10}$}\xspace}
\newcommand{\specialcell}[2][c]{%
  \begin{tabular}[#1]{@{}c@{}}#2\end{tabular}}

\begin{document}

\title{Predicting LyC emission of galaxies using their physical and \lya emission properties}
\author{Moupiya Maji \inst{1}
\and Anne Verhamme \inst{1}
\and Joakim Rosdahl \inst{2}
\and Thibault Garel \inst{1}
\and Jérémy Blaizot \inst{2}
\and Valentin Mauerhofer \inst{1}
\and Marta Pittavino \inst{3}
\and Maria-Pia Victoria Feser \inst{3}
\and Mathieu Chuniaud \inst{2}
\and Taysun Kimm \inst{4}
\and Harley Katz \inst{5}
\and Martin Haehnelt \inst{6}}

\institute {Observatoire de Genève, Université de Genève, Chemin Pegasi 51, 1290 Versoix, Switzerland
\and Univ Lyon, Univ Lyon1, Ens de Lyon, CNRS, Centre de Recherche Astrophysique de Lyon, Saint-Genis-Laval, France
\and Research Center for Statistics, Université de Genève, 24 rue du Général-Dufour, 1211 Genève 4, Switzerland
\and Yonsei University, 625 Science Hall, 50 Yonsei-ro, Seodaemun-gu, Seoul, 03722, South Korea
\and University of Oxford, Clarendon Laboratory, Parks Road, Oxford, UK
\and University of Cambridge, Madingley Road, Cambridge, UK}


\abstract{} {The primary difficulty in understanding the sources and processes that powered cosmic reionization is that it is not possible to directly probe the ionizing Lyman Continuum (LyC) radiation at that epoch as those photons have been absorbed by the intervening neutral hydrogen. 
It is therefore imperative to build a model to accurately predict LyC emission using other properties of galaxies in the reionization era.}
{In recent years, studies have shown that the LyC emission  from galaxies may be correlated to their Lyman-alpha (\lya) emission. In this paper we study this correlation by analyzing thousands of simulated galaxies at high redshift in the SPHINX cosmological simulation. We post-process these galaxies with the \lya radiative transfer code RASCAS and analyze the \lya - LyC connection.}
{We find that the \lya and LyC luminosities are strongly correlated with each other, although with dispersion. 
There is a positive correlation between the escape fractions of \lya and LyC radiations in the brightest \lya emitters (escaping \lya luminosity $\Llyaout > 10^{41}$ erg/s), similar to that reported by recent observational studies. However, when we also include fainter \lya emitters (LAEs), the correlation disappears, which suggests that the observed relation may be driven by selection effects. We also find that the brighter LAEs are dominant contributors to reionization, with $\Llyaout > 10^{40}$ erg/s galaxies accounting for $> 90\%$ of the total amount of LyC radiation escaping into the inter-galactic medium in the simulation. Finally, we build predictive models using multivariate linear regression where we use the physical and the \lya properties of simulated reionization era galaxies to predict their LyC emission. We build a set of models using different sets of galaxy properties as input parameters and predict their intrinsic and escaping LyC luminosity with a high degree of accuracy (adjusted $R^2$ of these predictions in our fiducial model are 0.89 and 0.85 respectively, where $R^2$ is a measure of how much of the response variance is explained by the model). We find that the most important galaxy properties to predict the escaping LyC luminosity of a galaxy are its \Llyaout, gas mass, gas metallicity and star formation rate.}
{These results and the predictive models can be useful to predict the LyC emission from galaxies using their physical and \lya properties and thus help us identify the sources of reionization. }

\titlerunning{Predicting LyC emission of galaxies}
\maketitle

\section{Introduction}

Cosmic reionization is an important period in the evolution of the Universe, when photons from energetic sources (i.e. first stars, galaxies or quasars) ionized the ubiquitous neutral hydrogen gas in the intergalactic medium. This milestone happened over the first billion years of the Universe, ending around $z \sim 6$, and it holds a key for understanding the formation and evolution of the first galaxies 
\citep{Loeb2001, Stark2016, Ocvirk2016, sphinx, Wise2019}. However, the Epoch of Reionization (EoR) is yet to be fully understood. One of the biggest outstanding question is determining the primary sources of the photons that ionize the Universe. The relative importance of two types of sources that are proposed in the literature, i.e., star-forming galaxies and quasars, is still somewhat debated but recent studies indicate that quasars were likely too rare at these redshifts to reionize the Universe 
\citep{Cowie2009, Fontanot2012,Fontanot2014, Kulkarni2019, Faucher2020, Trebitsch2020}, and that photons from star formation are most probably the primary sources of reionization. Yet, it remains to be understood which types of galaxies are most profusely leaking ionizing radiation (photons with wavelength $\lambda < 912\AA$ ,also called Lyman continuum or LyC) and the properties and environments that can make a galaxy a LyC leaker.

The primary difficulty in understanding the processes and sources that powered cosmic reionization is that it is not possible to directly probe the ionizing radiation at that epoch as those photons are all absorbed by the 
intergalactic medium (IGM) on their way to us \citep{Madau1995, Inoue2014}. Due to this, it is imperative to find indirect tracers for LyC emission to identify the sources of reionization.

In recent years, several methods have been proposed in the literature to indirectly measure LyC emission from galaxies: weak ISM absorption lines \citep[][but see \citealt{Mauerhofer2021}]{Heckman2011, Erb2015, Chisholm2017}, a high [OIII]/[OII] ratio \citep[][but also see \citealp{Bassett2019, Katz2020}]{Jaskot2013, Nakajima2014}, and the \lya line of Hydrogen
\citep{Dijkstra2014, Verhamme2015, Dijkstra2016, Verhamme2017, Izotov2018a}.
Among these, the \lya line is particularly interesting.
Since it is a UV line, \lya\ is observable over a wide range of redshifts, allowing one to probe galaxy formation with the same tool over several Gyrs of evolution. Indeed, over the last 20 years, a great amount of \lya\ emitting galaxies have been observed: from the low redshift Universe, using space-based facilities (Lyman Alpha Reference Sample (LARS) and Extended LARS survey (eLARS), studying 14 and 28 LAEs, respectively, at  $0.03 <z< 0.18$ \citep{Hayes2013, Ostlin2014} and the Green Pea sample of 43 LAEs at z = 0.2; \citealt{Henry2015, Schaerer2016, Yang2017}); from the ground, in optical from z$\sim2$ to z$\sim6$ (several thousands of spectroscopically confirmed LAEs, \citealt{Erb2011, Bacon2015, Trainor2015, Urrutia2019}) and in IR at the highest redshifts (e.g. SILVERRUSH survey using the Hyper Suprime-Cam (HSC) recently observed a large sample of 2230 Lyman Alpha emitters (LAEs) at z = 5.7 - 6.6 with narrow band imaging data; \citealt{Ouchi2018, Shibuya2018}). At even higher redshift, it is increasingly difficult to detect LAEs due to attenuation of \lya by the relatively neutral IGM. However, concentrated efforts with very deep photometric and spectroscopic surveys in recent years have led to detections of some \lya emitting galaxies in the extreme redshift range of z = 6 - 9 \citep{Vanzella2011, Ono2012, Schenker2012, Shibuya2012, Finkelstein2013, Oesch2015, Konno2014, Zitrin2015, Song2016, RobertsBorsani2016, Stark2017, Matthee2017, Songaila2018, Matthee2018, Itoh2018, Jung2019, Matthee2020, Meyer2021}. The upcoming JWST surveys are expected to discover many more such galaxies in the EoR soon.

The possibility of the relatively intense \lya radiation from galaxies being a tracer of LyC emission has become much studied in the past few years. \cite{Verhamme2015} explored the escape of \lya and LyC in idealized galaxy models and found that \lya line profiles show distinct signatures (strong, narrow peak and narrow peak separation if it is double-peaked) if the ISM of galaxies is transparent to the LyC. \cite{Dijkstra2016} found similar results in a theoretical study of a suite of 2500 idealized models of a dusty and clumpy ISM with \lya radiative transfer simulations.

\cite{Verhamme2017} performed an observational study of LyC leakers in the sample of Green Pea galaxies (the local analogs of high-z LAEs) and found that in the 8 galaxies where it is possible to detect LyC emission\footnote{After reionization the Universe has stayed ionized, so that
in the local Universe LyC photons can travel without being absorbed, unlike in high-z where neutral hydrogen atoms can absorb them easily.} in addition to \lya, 
the escape fractions of \lya and LyC are indeed positively correlated. Recently \cite{Izotov2021} observed 9 more galaxies in both LyC and \lya in the redshift range of $\sim 0.30 - 0.45$ and found that similar correlations exist in this sample as well. \cite{Steidel2018} studied the KLCS (Keck Lyman Continuum Spectroscopic Survey) sample which included 15 (out of 124) galaxies detected in LyC at $z \sim 3$ and found that the LyC escape fraction is well correlated with the equivalent width of the \lya emission.

The correlation between \lya and LyC radiation shows great promise, but to use this in the reionization era to estimate LyC from galaxies, we need to statistically analyze a large sample of EoR galaxies. Since LyC cannot be observed in this epoch, we need to explore it with simulations. 
Modelling \lya and LyC radiation from a large sample of galaxies in simulations has been particularly challenging, because it requires simulations to overcome several technical challenges.
Such simulations need to incorporate LyC radiation transfer on the fly, i.e. coupled at each hydrodynamical time step, to describe the ionization state of each cell in the simulation volume accurately. These simulations also need to account for the radiative transfer of \lya which requires a massively parallel resonant scattering code. Finally, the production and scattering or absorption of \lya\ and LyC photons happen at small scales in the ISM of galaxies and their eventual escape or absorption is at galactic and inter-galactic scales, so the simulation needs to sample both small and large scales correctly in order to predict reliable escape fractions and reionization topology. All of these requirements make such undertakings challenging and computationally expensive. Hence simulation studies of this kind so far have generally focused on either analyzing a small volume with high resolution, such as isolated galaxies \citep{Verhamme2012, Behrens2014}, \lya nebulae / blobs \citep{Yajima2013, Trebitsch2016}; molecular clouds \citep{Kimm2019}, zoom-in simulation of individual galaxies \citep{Faucher2010, Smith2019, Laursen2019}, or large volumes but with comparatively poor resolution \citep{Yajima2014, Inoue2018, Gronke2020}.

With this in mind, the SPHINX simulations are an ideal choice for this study, being a state-of-the-art radiation hydrodynamics (RHD) simulation and a good balance of a sufficiently large volume and high resolution hosting a large sample of well resolved galaxies. 
SPHINX is a suite of cosmological radiation-hydrodynamics simulations which reaches a resolution up to 10 pc in 10 co-moving Mpc (cMpc) wide volumes \citep{sphinx}. This allows us to investigate the Lya and LyC properties of thousands of simulated galaxies at $z > 6$ \citep[see][]{sphinx, Garel2021}. 

In this paper, we focus on the following questions:
\begin{itemize}
    \item is there a correlation between \lya and LyC emission at galaxy scale during the EoR? 
    \item is it possible to predict the LyC emission of galaxies, knowing their physical and \lya properties? 
\end{itemize}

The paper is structured as follows. We discuss our methods in $\S$\ref{sec:method}, where we describe the SPHINX simulation and the radiative transfer code that we use for \lya post processing, and present our sample of simulated galaxies. In $\S$\ref{sec:lyalyc_rel_z6} we explore the relationship between LyC and \lya luminosities and escape fractions and analyze the contribution of LAEs to reionization. In $\S$\ref{sec:prediction_z6} we build a multivariate regression model where we use the physical and \lya properties of galaxies to predict their intrinsic and escaping LyC luminosities and escape fractions, determine the most important variables required for each prediction and apply our models to observed data for comparison. In $\S$\ref{sec:discuss} we discuss the limitations of our study and in \S~\ref{sec:summary} we summarize our results.

\section{Methods}
\label{sec:method}
In this section we present the simulation, the selection procedure to build our sample of galaxies, and our methods to calculate LyC and \lya emissions from them.

\subsection{The Reionization Simulation}
\label{subsec:simulations}
SPHINX \citep{sphinx} is a suite of cosmological hydrodynamical simulations of the epoch of reionization. In this study we analyze galaxies in the 10 cMpc wide SPHINX volume previously presented in \cite{sphinx} which uses the binary stellar population model from BPASS \citep{Stanway2016}.


SPHINX is run with the RAMSES-RT code \citep{Teyssier2002, Rosdahl2013}. 
It simulates an average density patch of the Universe. The spatial resolution reaches 10.9 pc at $z = 6$, the dark matter mass resolution is $2.5\times 10^5\Msun$ per particle and the stellar mass resolution is $10^3\,M_{\odot}$ per stellar particle \citep[we refer to][for details of the simulation]{sphinx}. 
Within the simulation the radiation tracked is split into three photon groups, which encompass the ionization energies for HI, HeI, and HeII. These photons interact with hydrogen and helium in the simulation via photo-ionization, heating, and momentum transfer. The simulation is run until $z = 6$ 
and it uses Planck results \citep{Planck2014} for the cosmological parameters, i.e. $\Omega_\Lambda = 0.68, \Omega_m = 0.32, \Omega_b = 0.05, h= 0.67,$ and $\sigma_8=0.83$.

\subsection{Halo and Galaxy samples}
\label{sec:sample}

\begin{figure*}[ht]
    \centering
    \includegraphics[width=0.3\textwidth]{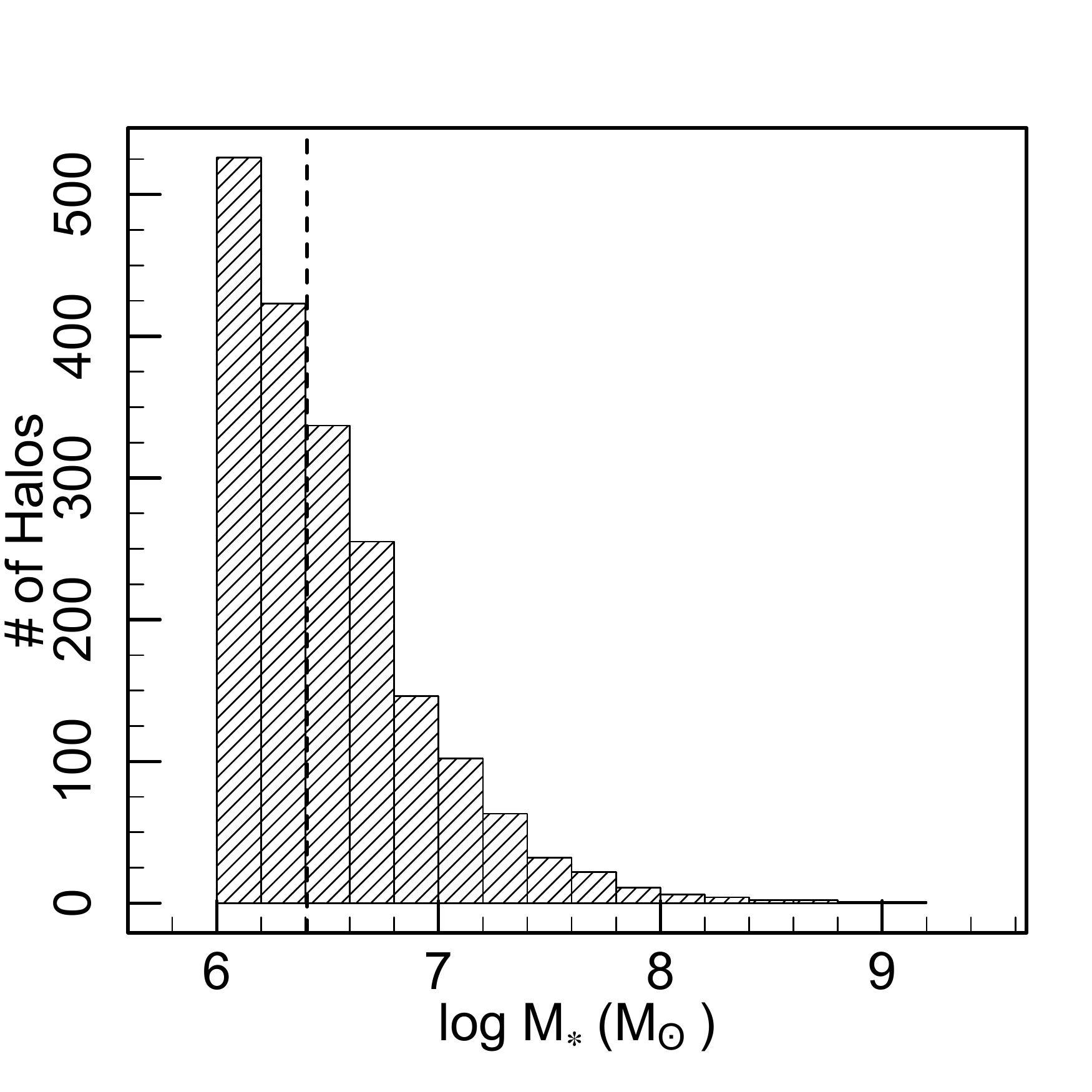}
    \includegraphics[width=0.3\textwidth]{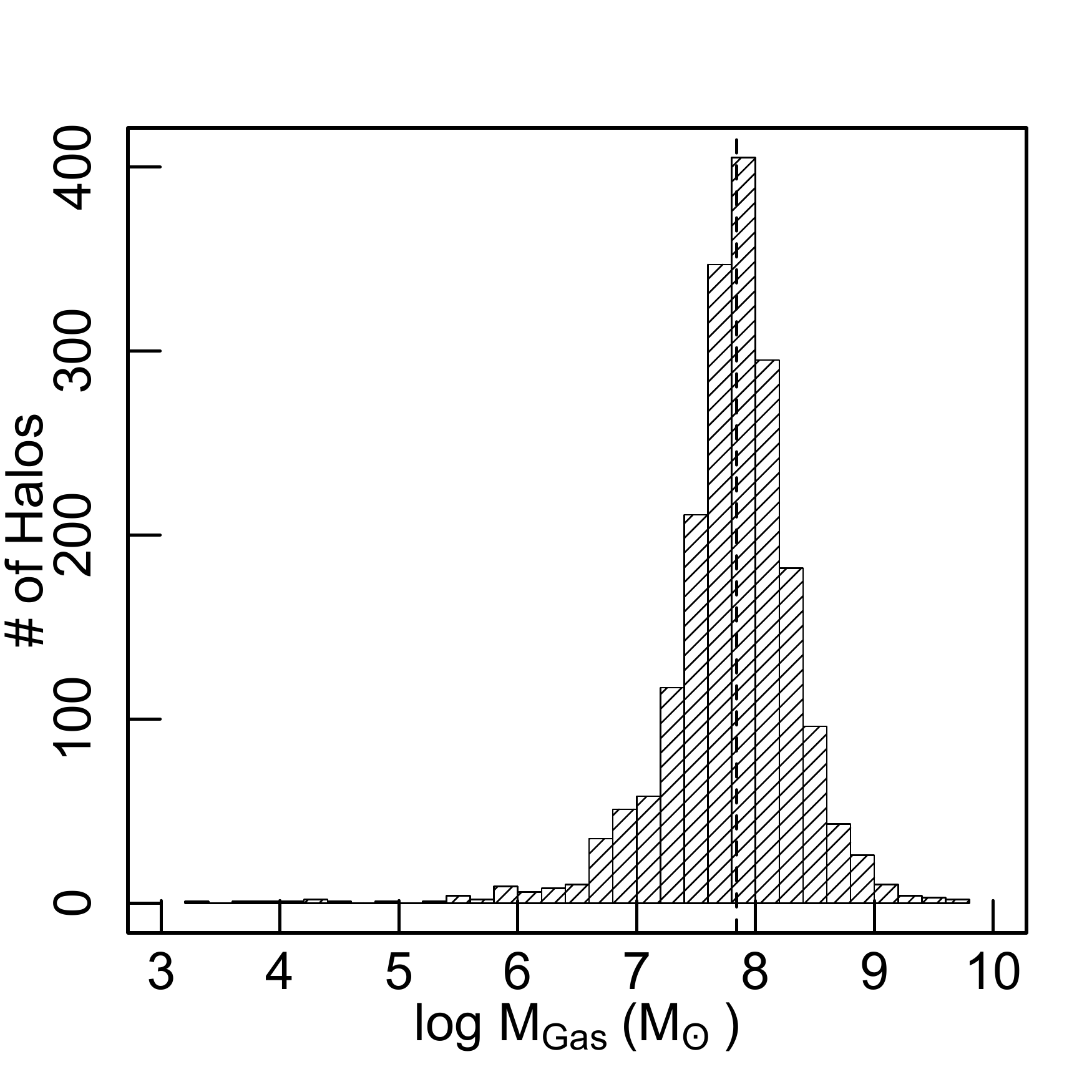}
    \includegraphics[width=0.3\textwidth]{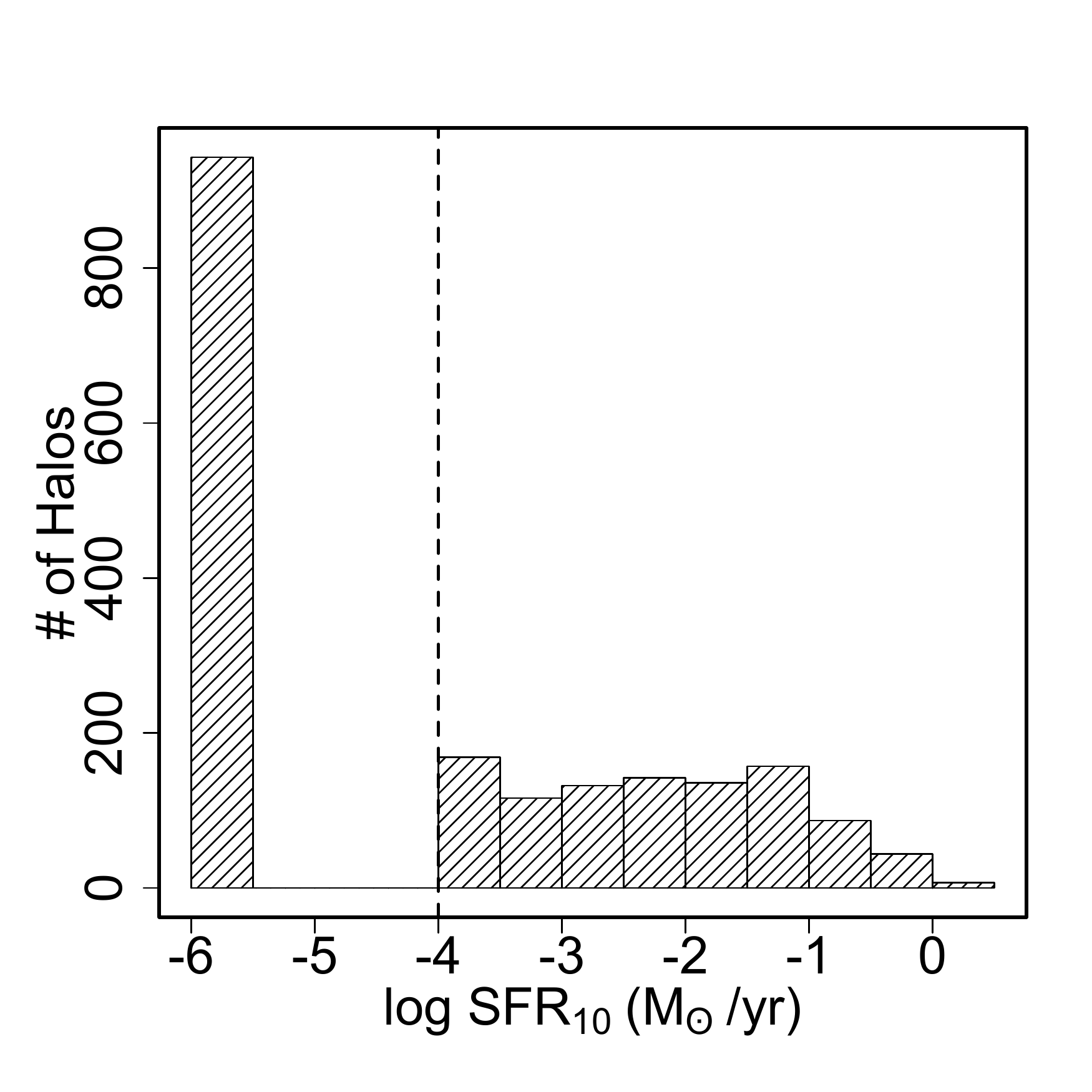} 
  \caption{Histograms of physical properties of the simulated galaxies in our sample. The histograms show the distribution of stellar mass (left), gas mass (middle) and SFR$_{10}$(right) of the galaxies. The median values are shown by dashed black lines. The stellar mass histogram shows 
  $M_{\star}$ within 30$\%$ of the halo virial radius. There are 10 galaxies with $M_{\star}>10^8\Msun$. The gas mass has a peaked distribution with few galaxies having very little gas (further discussion in \S~\ref{sec:discuss}). 
  There are 943 galaxies with zero SFR$_{10}$, these galaxies are represented in the bar at $10^{-6}$ (discussed in \S~\ref{sec:gal_selection}).}
    \label{fig:allsnaps_physical}
\end{figure*}

We use the same halos and galaxies as described and analyzed in \cite{sphinx}. In short, galaxies are detected in two stages. The group finder algorithm ADAPTAHOP \citep{Aubert2004, Tweed2009} is run on the dark matter particles, and the overdense virialized regions are identified as halos (and sub halos, sub-sub-halos etc depending on their level of structure). Halos are considered to be resolved when they have virial masses ($M_{vir}$) greater than 300 times the dark matter particle mass, i.e. $M_{vir} > 7.4\times 10^7\Msun$. Then ADAPTAHOP is run on stellar particles, and it identifies the overdense groups with at least 10 stellar particles as galaxies. Finally, the most massive galaxy within 0.3 $R_{vir}$ is assigned to each halo to build the galaxy-halo catalog.


In our analysis, we select systems which have i.) stellar mass $M_\star > 10^6$ $M_{\odot}$ (this is stellar mass within $0.3 R_{vir}$ of the halo) and ii.) main halo is at level 1, i.e. they are not a substructure of a parent halo. 
We exclude less massive galaxies with $M_\star < 10^6$ $M_{\odot}$ from our sample and focus on bright galaxies that are potentially observable. This stellar mass limit also means that all of our galaxies contain at least $10^3$ stellar particles, which ensures that the selected galaxies are reasonably well resolved.

We analyze snapshots of the SPHINX simulation at 5 different redshifts: $z = 6, 7, 8, 9,$ and 10. We select all galaxies that satisfy our criterion described above. The numbers of selected galaxies at these redshifts are respectively, 674, 509, 362, 236 and 152. Among these galaxies at $6\leq z\leq10$, the maximum galaxy stellar mass is $1.33\times 10^9 \Msun$, and there are 10 galaxies with $M_\star > 10^8 M_{\odot}$. We have compared the properties of the galaxies at these different redshifts and found that there is no significant evolution in terms of physical or radiative (\lya or LyC) properties (more discussion at \S~\ref{appendix:z6z10}). Therefore, we combine our galaxy sample as a larger sample size can give better statistical significance for our understanding. Our final sample comprises of 1933 galaxies. 

Figure~\ref{fig:allsnaps_physical} shows distributions of their stellar mass, gas mass and star formation rate. We recall that the stellar mass distribution in this figure shows the  
stellar mass within 30$\%$ halo virial radius. 
The median stellar mass 
is $10^{6.41}$ \Msun. 

The gas mass shown in Figure~\ref{fig:allsnaps_physical} is the total gas mass of the halo, calculated by summing up the mass of all the gas cells inside $R_{vir}$. 
We find that the gas mass has a normal distribution (median mass $10^{7.84}\Msun$) with some halos containing very small amounts of gas, likely because a recent supernova or starburst has blown the gas away from these small systems. 

The star formation rate, SFR$_{10}$ shows the star formation rate of the galaxy averaged over the last 10 Myrs. 10 Myrs is a typical lifetime of the massive stars,
after which they undergo a supernova (the most massive
stars live for about 3 Myr) and 10 Myr is also the typical
timescale of the production of LyC and \lya. 
In Figure~\ref{fig:allsnaps_physical} we show the distribution of the log SFR$_{10}$. There are 943 galaxies in our sample that have SFR$_{10}$ = 0. We artificially set their SFR values equal to $10^{-6}$\Msun/yr (which is lower than the lowest non-zero SFR) to show them in the histogram.
The median value of SFR$_{10}$ is $10^{-4}$ \rm{\Msun}/yr. 



\subsection{LyC emission from SPHINX galaxies}
\label{sec:lycmethod}
The production and escape of LyC photons in SPHINX has been described in \cite{sphinx}. In short, the instantaneous escape fractions of LyC photons are calculated in post-processing, using RASCAS \citep{rascas2020}. Rays are traced from every stellar particle inside a halo out to its virial radius. Along each ray, the optical depth ($\tau$) is calculated for hydrogen and helium. 
For each stellar particle, the escape fraction is the average of $e^{-\tau}$ calculated with rays in 500 random directions. Then the global escape fraction of the halo (\flyc) is the luminosity-weighted average escape fraction of all the stellar particles inside the halo. 
The LyC photons we consider range from $0 - 912$\AA\, and in the simulation they are described in 3 groups of photons: photons that ionize HI (UV$_{HI}$, 912 - 504\,\AA, 13.6 - 24.59 eV), HeI (UV$_{HeI}$, 504 - 228\,\AA, 24.59 - 54.42 eV) and HeII (UV$_{HeI}$, 228 - 0\,\AA, 54.42 - $\infty $ eV). The distributions of intrinsic (\Llycin) and escaping (\Llycout) LyC luminosities, and escape fractions for our galaxy sample are further described in \S~\ref{sec:lyalyc_distrib}.

On the contrary, observations of LyC usually focus on a small part of the ionizing spectrum, close to the Lyman limit (912 \AA). This observed LyC luminosity known as \Llycnineout, i.e. the escaping LyC luminosity at 900\AA, is defined as $\Llycnineout = \Llycninein \times \flycnine$ (\Llycninein and \flycnine are intrinsic luminosity and escape fraction at 900\AA, respectively). So we perform additional LyC measurements more similar to what is done observationally. We can estimate the intrinsic LyC luminosity of the simulated galaxies at 900\AA\,(\Llycninein), using the BPASS models \citep{Stanway2016} that have been used in modeling the ionizing emission in the SPHINX simulation. Using RASCAS we distribute $10^5$ photon packets with wavelengths between 10 - 912\AA\ among the stellar particles and then transfer them until they are absorbed by HI, HeI, HeII, dust or escape the halo virial radius. Thereafter we have both the intrinsic and escaping spectral energy distribution from 10 - 912\AA\ and this allows us to derive the LyC escaping luminosity and escape fraction over different wavelength ranges, e.g. 890-912 \AA. The average luminosity in this range is the luminosity at 900\AA\, (i.e. \Llycninein and \Llycnineout are in units of erg/s/\AA).

\begin{figure*}
    \centering
    \includegraphics[width=0.3\textwidth]{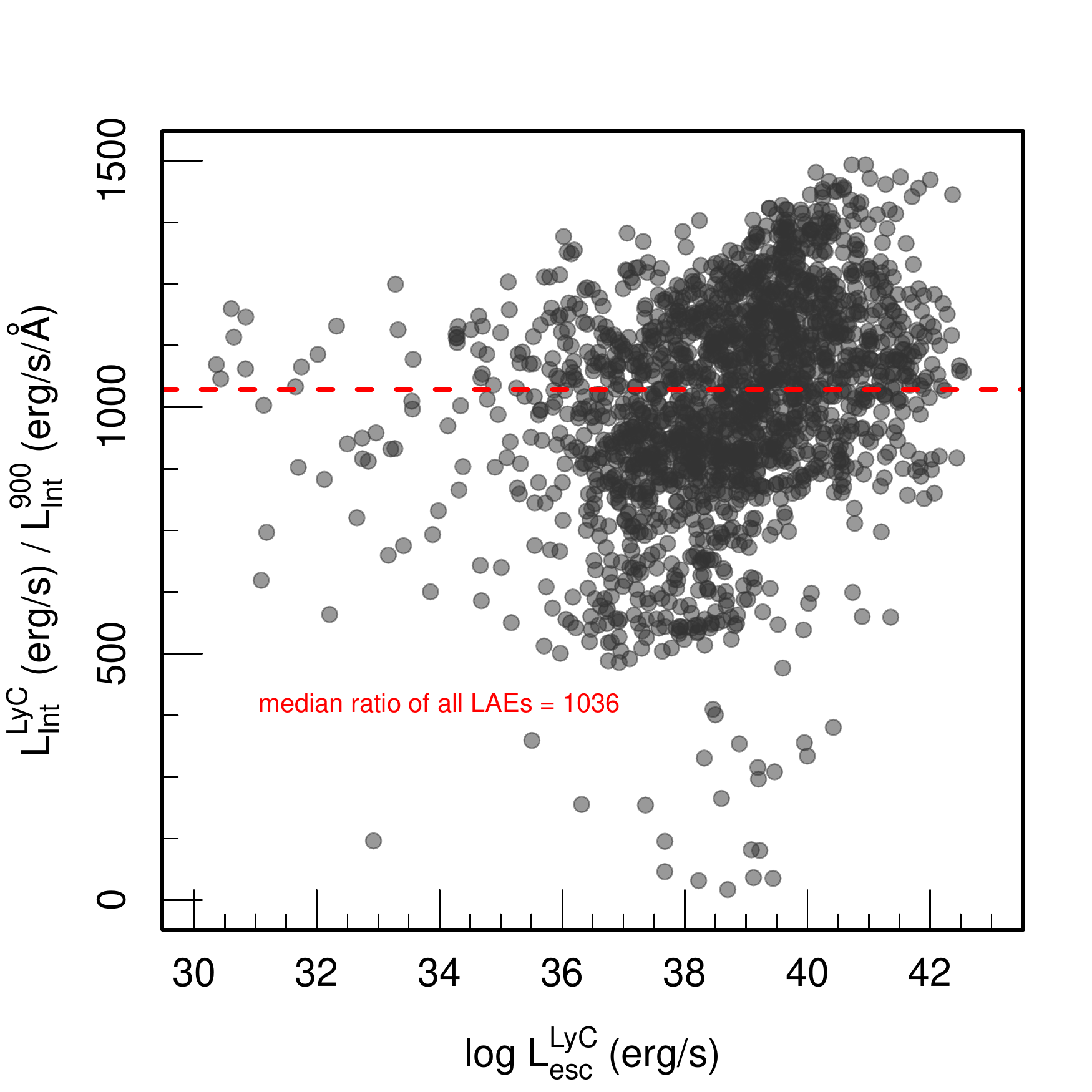}
    \includegraphics[width=0.3\textwidth]{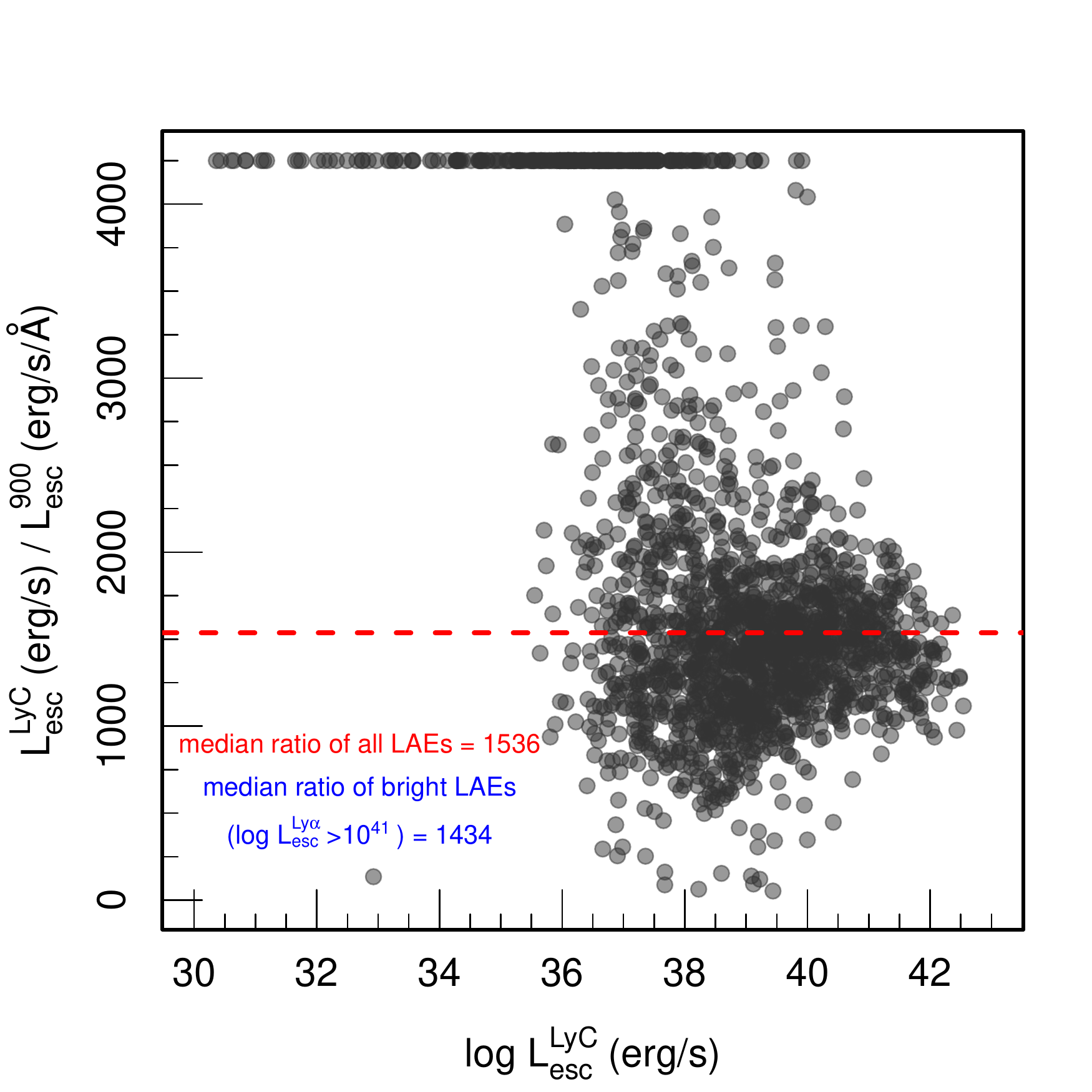}
    \includegraphics[width=0.3\textwidth]{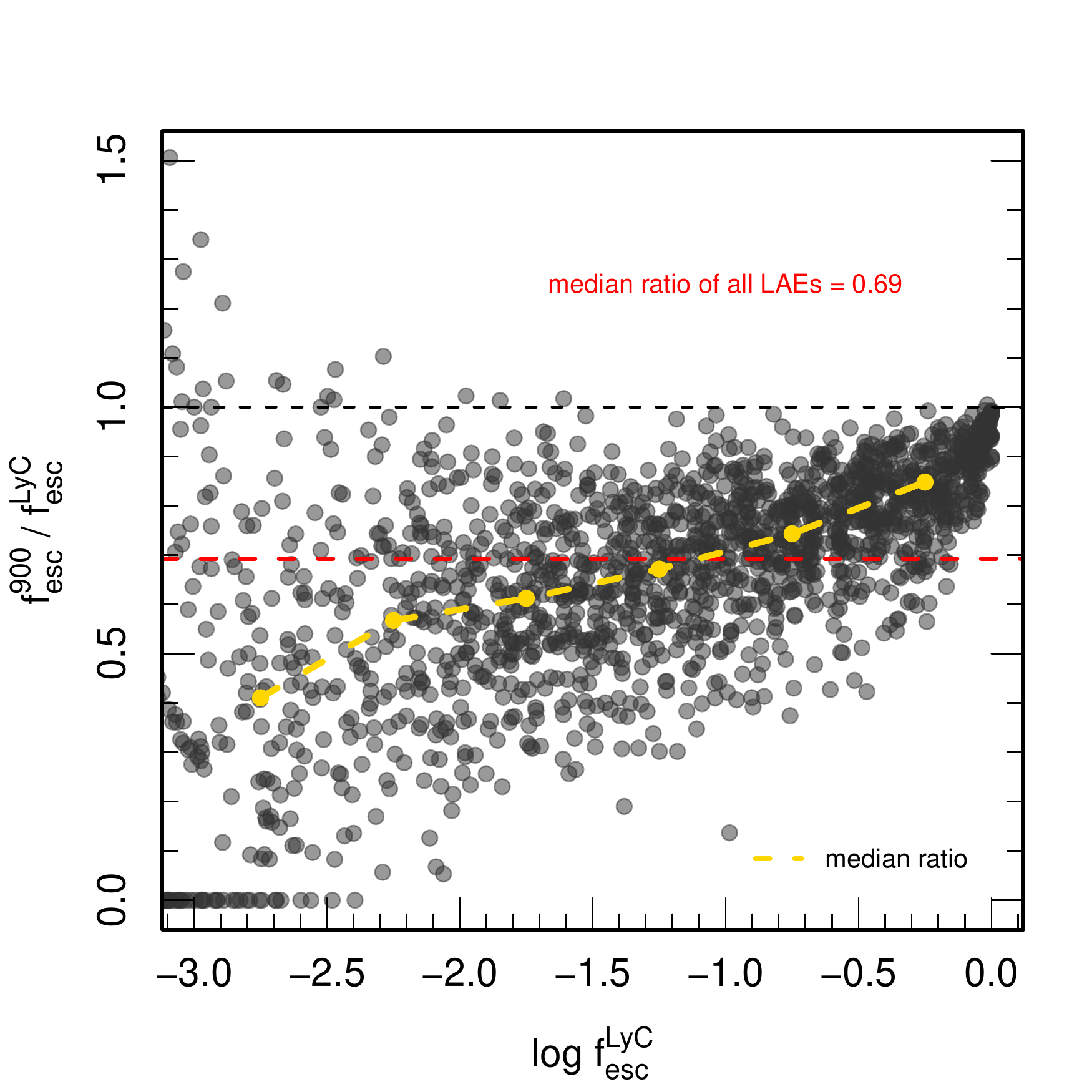}
    \caption{Ratio of total intrinsic (left) / escaping (middle) LyC luminosity emitted over the range of 0 - 912 \AA\, and the intrinsic / emitted LyC at 900 \AA, as a function of their total escaping LyC luminosity. The median ratio calculated with all simulated galaxies is shown in red dashed line. Some of the galaxies have \Llycnineout = 0, i.e. their ratios of \Llycout/\Llycnineout = Inf, these ratios are represented at a value of 4250 in the middle panel. The ratio of \flycnine/\flyc as a function of the \flyc is shown at the right panel (note that the escape fraction ratio is 900\AA\, divided by total and the other two ratios are total divided by 900\AA\, quantities). The dashed yellow line shows the median ratio as a function of \flyc (we divide the log \flyc between -3 to 0 into groups of 0.5 dex each and find the median ratios). }
    \label{fig:ratio_900}
\end{figure*}

Figure~\ref{fig:ratio_900} (left and middle panel) shows the ratio of total LyC (i.e. 0 - 912\AA) emission (intrinsic and escaping luminosities) to the LyC emission at 900\AA\, as a function of their total escaping LyC (\Llycout) luminosities for all simulated galaxies.
Since we integrate over a wavelength range 900 times larger for the total luminosity, we expect a rough ratio of around 900 between the two intrinsic luminosities. The ratio of the escaping luminosities is expected to be higher because the cross-section of Hydrogen photoionisation is approximately proportional to $\lambda^{3}$ at $\lambda < 912\AA$, 
so \Llycnineout could be more attenuated than \Llycout. Indeed, the median ratios for intrinsic and escaping luminosities are 1036 and 1536, respectively. We also find that this ratio for both intrinsic and escaping luminosities has a significant scatter, probably due to the particular star formation history and morphology of each galaxy.

In particular, we find that 242 galaxies, i.e. 12\% of our galaxies, have $\Llycnineout = 0$, i.e. for these galaxies $\Llycout/\Llycnineout=Inf$, these ratios are represnted at a value of 4250 in the middle panel. Almost all of them are also faint in total LyC emission, with only 6 among them having $\Llycout > 10^{39}$ ergs/s. This result suggests that few LyC leakers can be missed by surveys probing only the flux close to the Lyman limit.




By comparing the number of intrinsic photons with the escaping photons, we also obtain the escape fraction at $900\AA$. We show the ratio of \flycnine to \flyc as a function of \flyc in the right panel of figure~\ref{fig:ratio_900} and find that, except very few galaxies (with low luminosities) \flycnine is lower than \flyc. The overall median ratio is 0.69. \cite{Kimm2019} also finds similar ratio while investigating the escape of LyC radiation from turbulent clouds. We divide the (log) \flyc into groups of 0.5 dex each and find that the median ratio increases slightly with \flyc. 

\subsection{\lya emission from SPHINX galaxies}
\label{subsec:rascascode}

In order to investigate the correlation between LyC leakage and the observable Lya properties of galaxies, we now turn our interest to the Lya post-processing of SPHINX galaxies.

A \lya photon (wavelength - 1215.67 \AA, energy - 10.2 eV) is emitted when a hydrogen electron jumps from the 2p to the 1s (ground) state. It is not only the hydrogen line with the largest flux, but also a resonant line. To obtain the \lya properties of galaxies in the SPHINX simulation, we post process them using RASCAS \citep{rascas2020},
which is a fully parallelized 3D radiative transfer code developed to perform the propagation of any resonant line 
in numerical simulations. It performs radiative transfer on an adaptive mesh using the Monte Carlo technique. We describe below the different steps of our implementation.

\textbf{\lya intrinsic luminosities:}
\lya emission can be triggered by two processes, recombination and collisional de-excitation \citep{Dijkstra2014review}. LyC photons from massive stars in galaxies ionize the neutral gas in their ISM and afterwards, the free proton and electron recombine. 
The electron can initially enter into any energy level, and then cascades to ground level with a probability of $\approx$ 0.67 to emit a \lya photon \citep{Partridge1967, Dijkstra2014review}. Alternatively, HI atoms can be excited collisionally, and when the electron returns to the ground state, a \lya photon can be emitted. 
So, for any given halo in our sample, we track both recombinations and collisional excitations from all cells inside the halo virial radius to capture the intrinsic \lya emission. 
For recombinations, the \lya photon emission rate in each cell is \citep{Cantalupo2008} :
\begin{equation}
    N_{\gamma, rec} = n_e n_p \alpha_B(T) \epsilon^B_\lya(T) \times (\Delta x)^3
\end{equation}
where, $n_e$ and $n_p$ are the number density of electrons and protons respectively (these come from the simulation), $\alpha_B(T)$ is the case-B recombination coefficient, $\epsilon^B_\lya(T)$ is the fraction of recombination events that produces a \lya photon eventually (at $T = 10^4 K$, it is 0.67) and $(\Delta x)^3$ is the cell volume. For collisional excitation, the \lya emission rate is given by \citep{Goerdt2010} :
\begin{equation}
    N_{\gamma, col} = n_e n_{HI} C_\lya(T)\times (\Delta x)^3
\end{equation}
where $n_{HI}$ is the number density of neutral hydrogen, and $C_\lya(T)$ is the rate of collisionally induced 1S-to-2P level transitions (we do not consider higher order transitions). We refer to \cite{rascas2020} for a detailed description of how we fit each of the coefficients $\alpha_B(T)$, $\epsilon^B_\lya(T)$ and $C_\lya(T)$.
Once these luminosities are known in each cell,  we emit a total of $10^5$ photon packets from the cells inside a galactic halo with the probability of a cell emitting a photon packet proportional to its luminosity. The number of photon packets has been chosen so as to minimise the computational cost while preserving the accuracy of the \lya angle-averaged escape fraction and luminosity. Performing convergence tests on  the ten most massive galaxies in our sample, we find that these quantities are well converged using $10^5$ photon packets.

\textbf{\lya propagation and escape: }
In each cell, we cast \lya photons isotropically and propagate them through the halo with RASCAS code. 
Each \lya photon can be scattered, i.e. absorbed and re-emitted, numerous times whenever they encounter HI atoms in the ISM, until they finally escape the halo or are absorbed by dust. 
The dust is modelled by specifying a cross section per hydrogen atom and a pseudo dust number density dependent on HI and HII density and metallicity \citep{rascas2020}. The dust absorption coefficient in each cell is given by $(n_{\rm HI} + f_{\rm ion}n_{\rm HII})\,\sigma_{\rm dust}(\lambda)\rm{Z/Z_{0}}$, where $f_{ion} = 0.01$ (abundance of dust in ionized gas), $Z$ is the gas metallicity in that cell, the effective dust cross-section $\sigma_{\rm dust}$ and $Z_0$ (= 0.005) are normalized to the Small Magellanic Cloud (SMC) models following \cite{Laursen2009}.

The boundary beyond which a \lya photon can be considered as having escaped is not an obvious choice. At $z \geq 6$ where reionization takes place, configuration of galaxies are complex, partly because in many cases galaxies are interacting or colliding with each other. So we perform convergence tests on the ten most massive galaxies in our sample. In each of them, we set the boundary at $R_{vir}$, 2$R_{vir}$, and 3$R_{vir}$ where $R_{vir}$ is the corresponding halo virial radius, and run \lya radiative transfer in each case. We find that beyond $R_{vir}$, the escape fraction converges, with only small increments in accuracy. So we fix $R_{vir}$ to be the boundary of \lya escape. Both the production and the propagation of photons are allowed within this radius, which encompass the main galaxy and in many cases, its satellites.


We use the core-skipping method to speed up the calculation \citep{rascas2020}. We have tested the core-skipping method by simulating the \lya radiation transfer in the 10 most massive galaxies in our simulation with and without core-skipping and found that the \lya results, e.g. luminosities and escape fraction, are very similar (median $0.6\%$ difference) and we gain significant (up to a factor of 100) speedup in the calculation. The distributions of intrinsic and escaping \lya luminosities, and escape fractions, for our galaxy sample, are further described in \S~\ref{sec:lyalyc_distrib}.

\section{LyC - \lya relationship}
\label{sec:lyalyc_rel_z6}

The goal of our study is to investigate the connection between the \lya and LyC properties of galaxies in order to investigate if, or how, \lya can trace the total ionizing radiation escaping from galaxies at EoR.
To that end, in this section we first discuss the relationship between their intrinsic and escaping luminosities and then analyze their escape fractions.

\subsection{Observed LyC emitters}
\label{sec:obs_lces}
Before we explore the relationship between the various \lya and LyC properties, we review existing observed sample of LyC emitters (LCEs) in order to facilitate the comparison of our simulated galaxies with observed ones. 

Although there are many observations of \lya at different redshifts, 
it is difficult to observe LyC even at low redshift galaxies because earth's atmosphere blocks UV radiation, so no ground based observations are possible. However, in recent years it has become possible to obtain direct observations of LyC leakers using space-based facilities, e.g. HST \citep{Verhamme2017, Izotov2016a, Izotov2016b, Izotov2018a, Izotov2018b, Izotov2021}. We compile these observations (23 galaxies) in Table~\ref{tab:observations}, where we note their redshift, available physical properties, i.e. stellar mass, SFR, surface SFR density, escaping luminosity and escape fraction in LyC and \lya. The SFR is derived from H$\beta$ observations and therefore correspond to SFR on a short time scale.  They can thus be considered similar to the SFR$_{10}$ in our simulated galaxies.

\begin{figure*}
    \centering
    \includegraphics[width=0.3\textwidth]{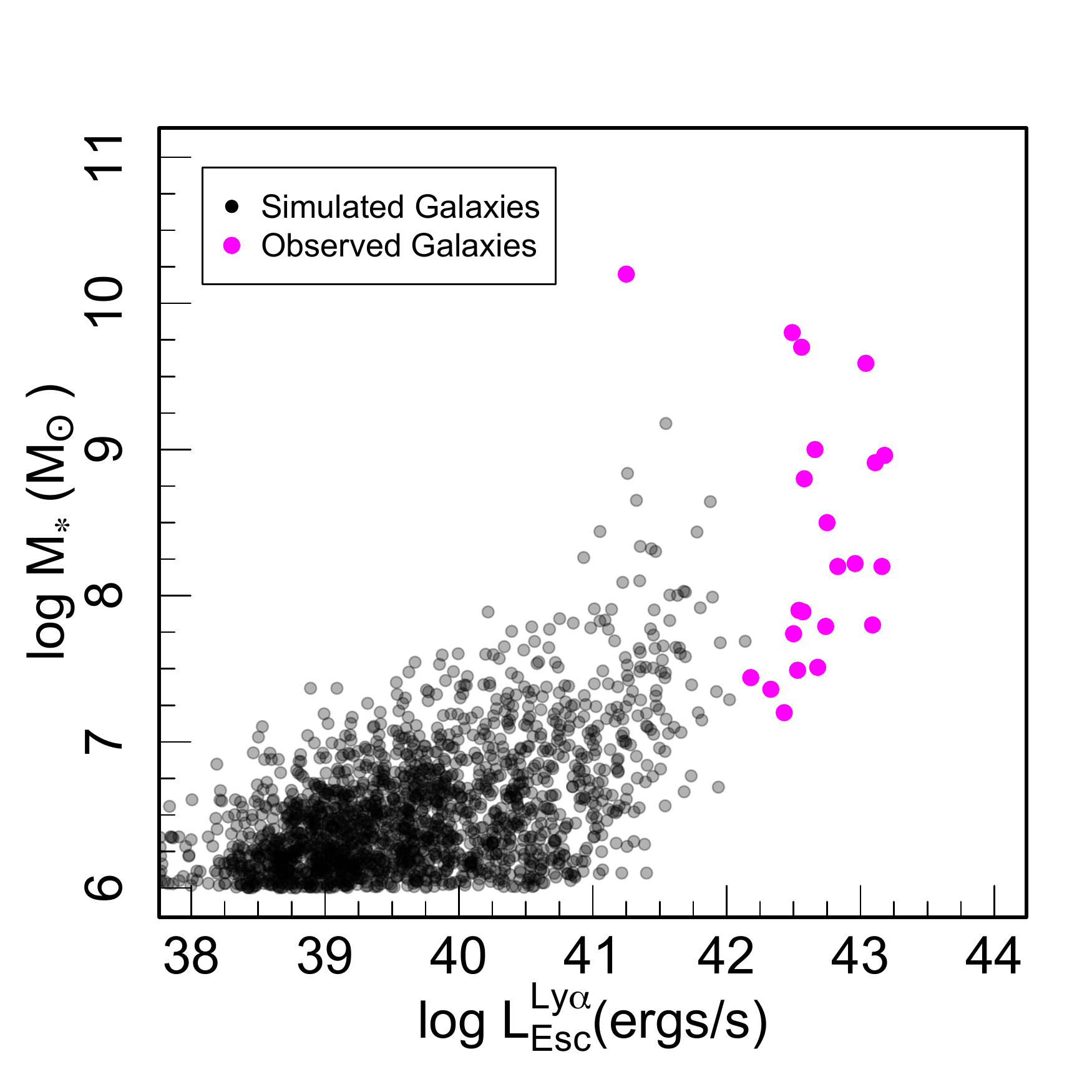}
    \includegraphics[width=0.3\textwidth]{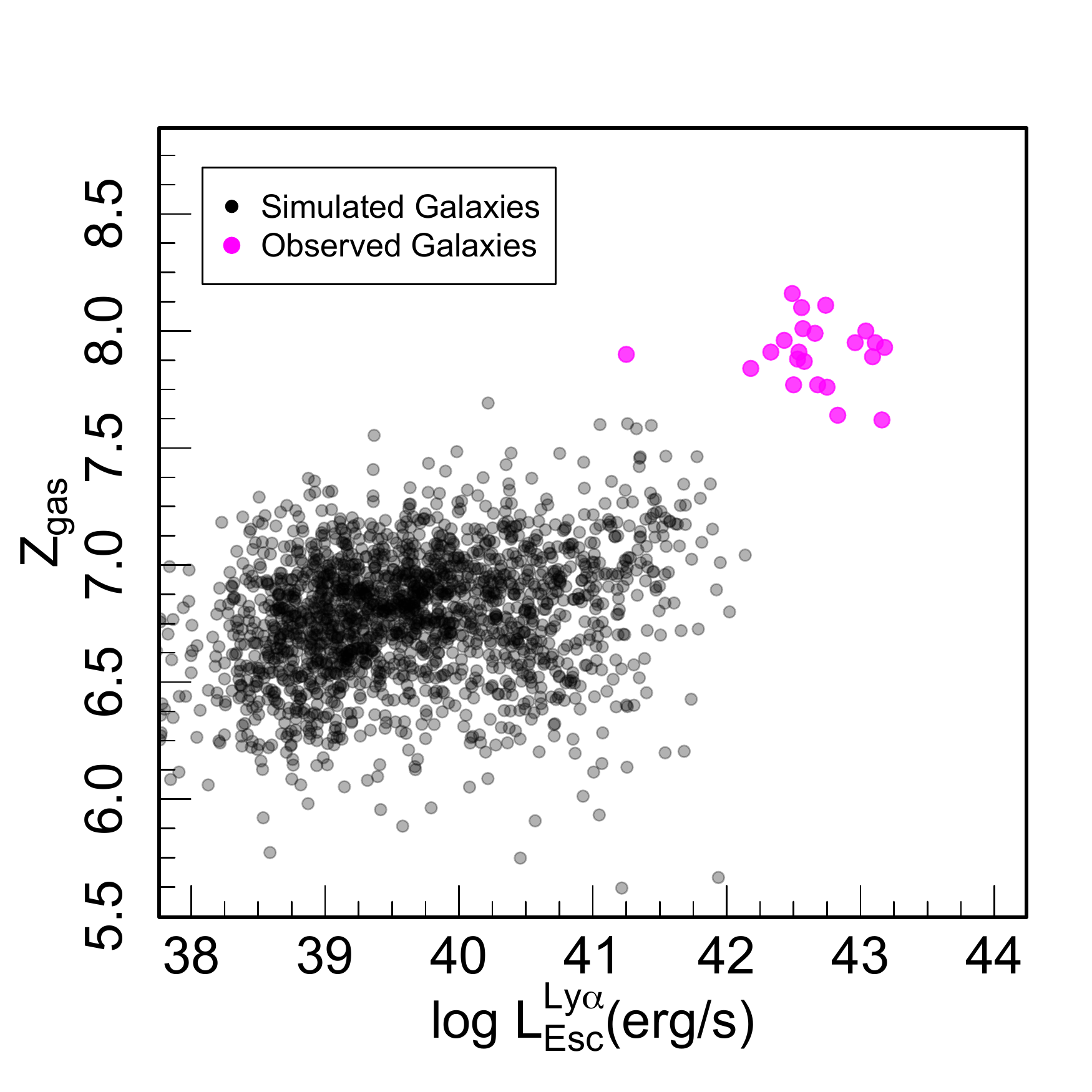}
    \includegraphics[width=0.3\textwidth]{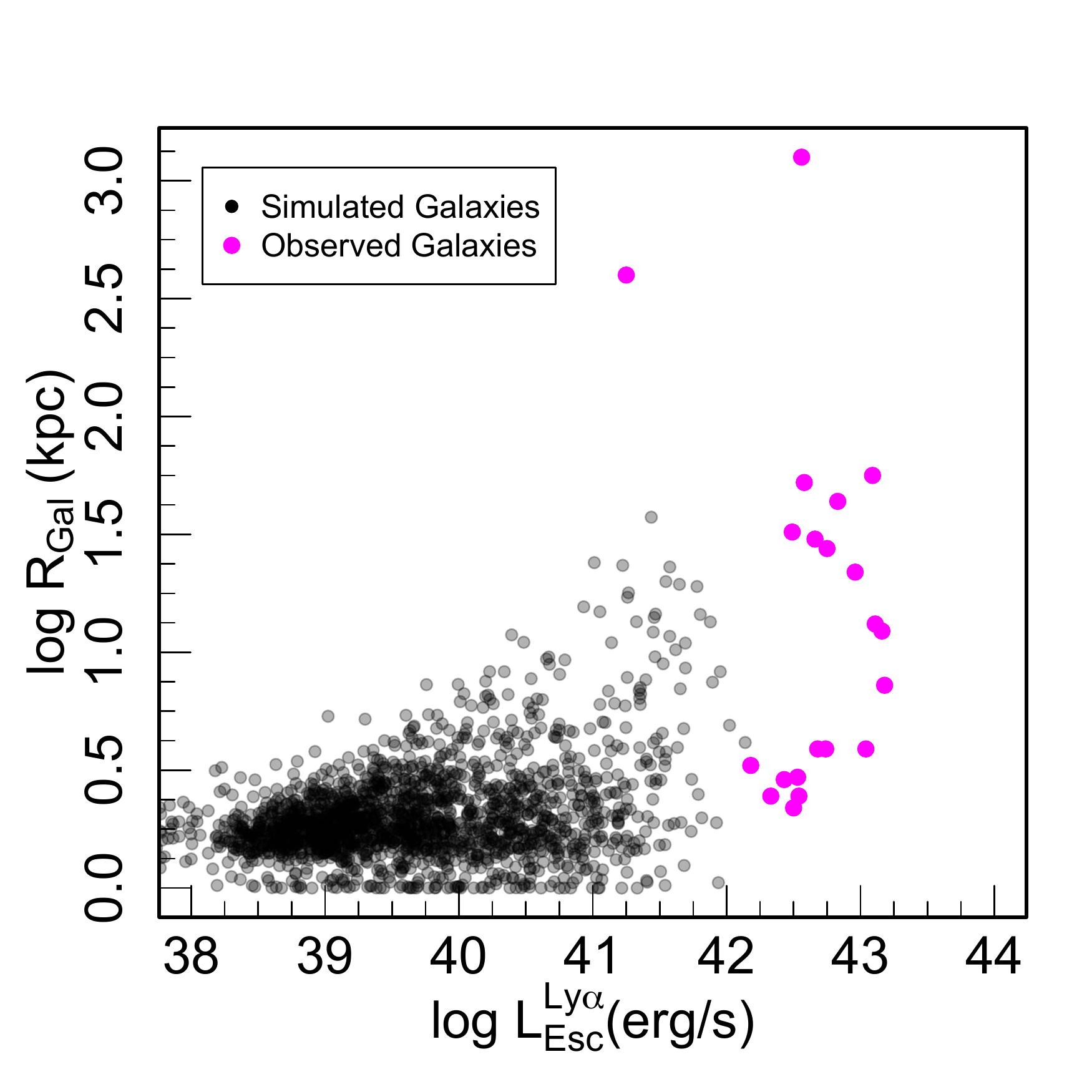}\\
    \includegraphics[width=0.3\textwidth]{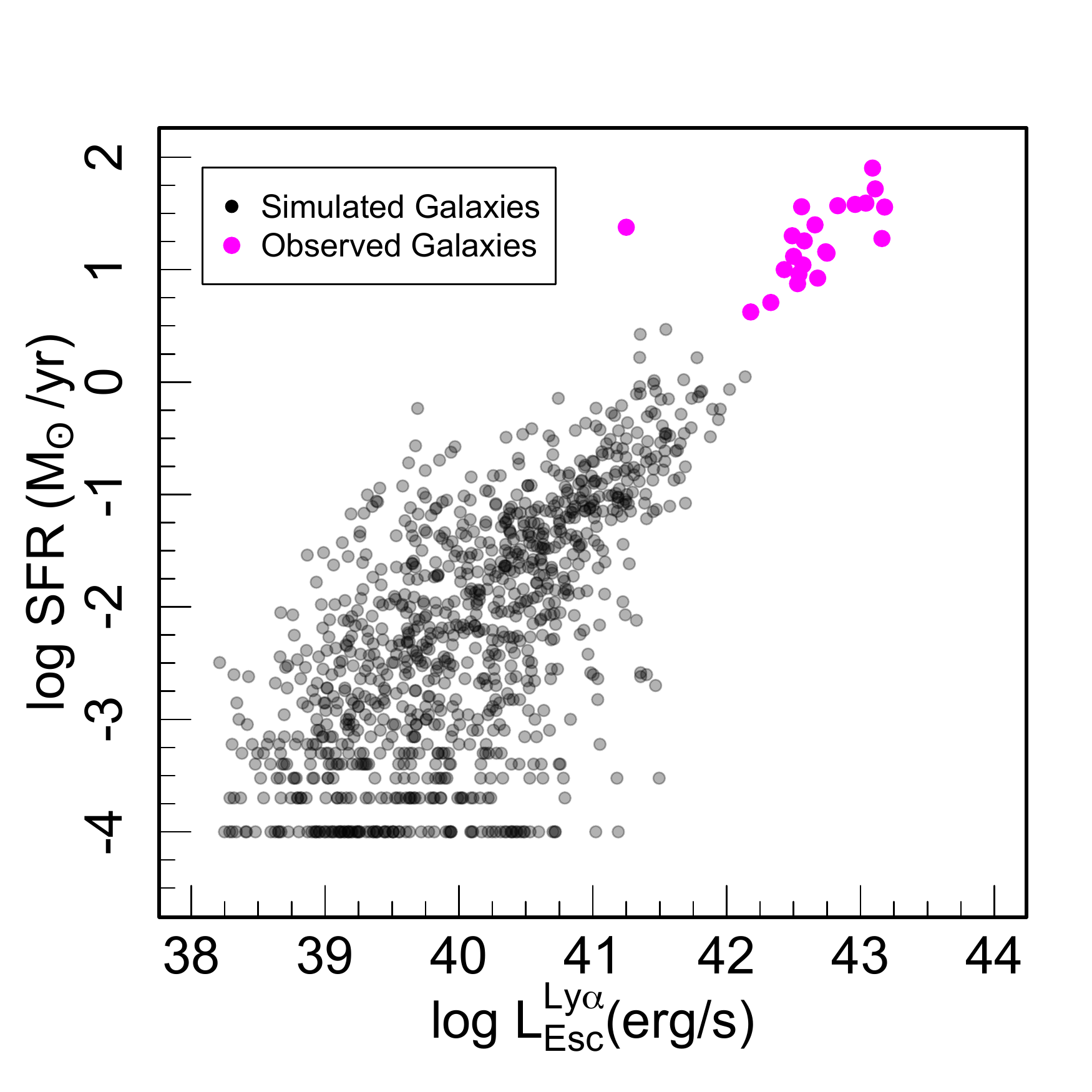}
    \includegraphics[width=0.3\textwidth]{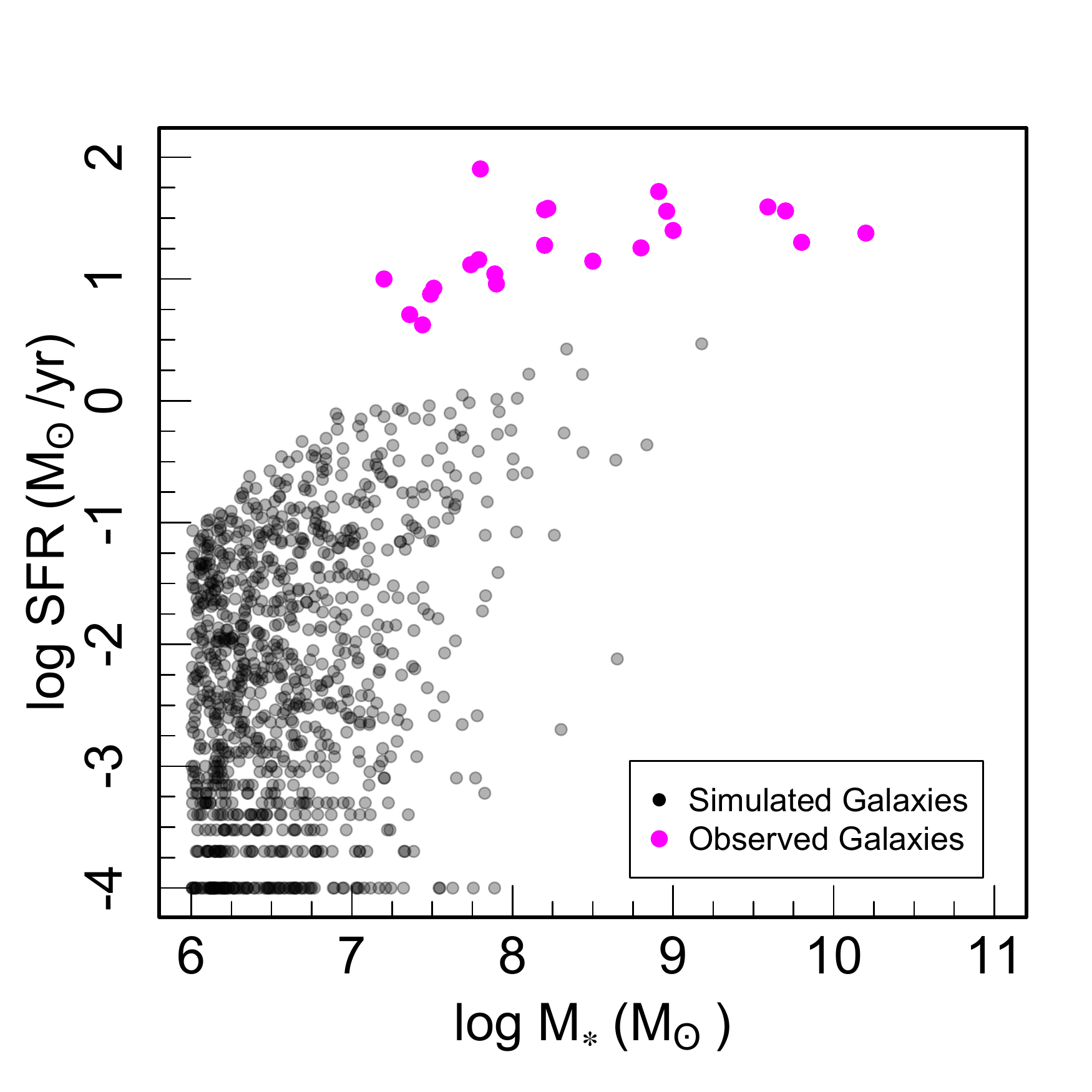}
    \caption{Comparison of physical properties of observed LCEs (magenta points) and the simulated galaxies (black points). Here we show the stellar mass (top left), gas metallicity (oxygen abundances i.e. 12+log10(O/H) for observed galaxies, top middle), galaxy radius (galaxy virial radius for simulated ones and exponential disc scale length for observed ones, top right) and SFR (SFR$_{10}$) as a function of their escaping \lya luminosities. The \lya luminosities of the SPHINX galaxies do not scale with stellar masses, metallicities or galaxy sizes, whereas they correlate with recent star formation, as expected. In the top right panel we show SFR (SFR$_{10}$ for simulated galaxies) as a function of the stellar mass of galaxies. The properties of the observed LCEs are listed in Table~\ref{tab:observations} and more details can be found in their corresponding reference papers.}
    \label{fig:compare_obs_prop}
\end{figure*}

Figure~\ref{fig:compare_obs_prop} shows a comparison of the physical properties of these observed LCEs with our simulated sample. We find that \lya luminosities of the SPHINX galaxies do not scale well with stellar masses, gas metallicities or galaxy sizes, whereas they correlate with recent star formation, as expected since higher SFR means more energetic photons are being produced which can be reprocessed in the ISM as \lya. The SFR$_{10}$ of the galaxies correlates weakly with the stellar mass and has a large scatter. Because of the finite volume of our simulation, our sample is restricted to relatively faint and low-mass galaxies such that most of the observed objects considered here are brighter, slightly more massive, slightly bigger and have higher star formation than our simulated sources. The observed galaxies also have higher metallicities compared to the simulated ones, which is perhaps not surprising as the observed sample is at a much lower redshift ($z\sim0.3$ compared to $z\sim6$), hence they can be more metal enriched. While this certainly represents a limitation of our study, investigating the LyC-\lya connection in our sample can still be used to interpret available observational data and guide future surveys that will target galaxies more similar to our sample.

It is important to note that the LyC luminosity in the Table~\ref{tab:observations} is \Llycnineout, i.e. the LyC luminosity at 900\AA. However, the escaping LyC luminosity that counts for reionization is the total luminosity of all photons that can ionize HI, i.e. all photons with $\lambda = 0 - 912$ \AA\, so we consider this total LyC throughout the paper. These two measures of LyC luminosities can be very different as discussed in \S~\ref{sec:lycmethod}, and the contribution of the highly ionizing spectrum for observed LCEs ($< 900$\AA) is still largely unknown. Since the observed LCEs are brighter in \lya ($>10^{41}$ erg/s) than the bulk of our galaxies, we recalculate the median of the \Llycout / \Llycnineout ratio for bright LAEs and find the ratio to be 1434 (Figure~\ref{fig:ratio_900}, middle panel). This ratio can be used to convert observed 900\AA\, luminosities to total LyC luminosities, if needed.

As we are interested in investigating the global theoretical connection between \lya and the ionizing radiation of galaxies in EoR, hereafter throughout this paper we consider the global \lya\ and LyC photon budgets from galaxies, i.e. summed over all directions and relevant wavelengths (i.e. 0 - 912\AA\, for LyC luminosities and \flyc) unless otherwise specified.

\begin{table*}[ht]
\centering
   \caption{Observed data. The columns here denote the name of the galaxy, its redshift, \lya and LyC escape fraction, stellar mass, star formation rate, star formation rate, oxygen abundance, \lya and LyC luminosity (at 900\AA) and the reference respectively. The uncertainties of the escape fractions are noted in the table. The typical uncertainty of luminosities is $\sim 10\%$.}
     \begin{threeparttable}
        \begin{tabular}{|c|c|c|c|c|c|c|c|c|c|c|c|}
        \hline
        Galaxy & z & \flya & \flyc & 
        \specialcell{log $M_\star$\\ (log $\Msun$)}& 
        \specialcell{SFR\\(\Msun/yr)}&
        \specialcell{12+log(O/H)}&
        \specialcell{log \Llyaout\\(erg/s)}&  
        \specialcell{log $\Llycnineout$\\(erg/s/\AA)}&  
        Ref  \\
        \hline
        \hline
J0901 + 2119&	0.2993&	0.14$\pm{0.01}$&	0.027$\pm{0.007}$&	9.8&    20&        8.16&     42.49&  39.20&  \textit{a, n} \\
J0925 + 1409&	0.3013&	0.29$\pm{0.03}$&	0.078$\pm{0.011}$&	8.91&   52.2&            7.95&   43.11& 39.84&  \textit{c, n}\\
J1011 + 1947&	0.3322&	0.18$\pm{0.01}$&	0.114$\pm{0.018}$&  9.0&    25&        7.99&      42.66&  39.73&  \textit{a, n}\\
J1152 + 3400&	0.3419&	0.34$\pm{0.07}$&	0.132$\pm{0.011}$&	9.59&   39&           8.00&    43.04& 40.23&  \textit{d, n} \\
J1154 + 2443&	0.3690&	0.61$\pm{0.03}$&	0.46$\pm{0.02}$&	8.2&    18.9&      7.62&     43.16&  40.26&  \textit{e, n}\\
J1243 + 4646&	0.4317&	0.52$\pm{0.04}$&	0.726$\pm{0.097}$&  7.8&    80&       7.89&      43.09&  40.78&  \textit{a, n}\\
J1248 + 4259&	0.3629&	0.17$\pm{0.01}$&	0.022$\pm{0.007}$&	8.2&    37&       7.64&      42.83&  39.26&  \textit{a, n}\\
J1256 + 4509&	0.3530&	0.32$\pm{0.03}$&	0.380$\pm{0.057}$&	8.8&    18&       7.87&      42.58&  40.27&  \textit{a, n}\\
J1333 + 6246&	0.3181&	0.51$\pm{0.09}$&	0.056$\pm{0.015}$&	8.50&   14&          7.76&     42.75&   39.44&  \textit{d, n}\\
J1442 - 0209&	0.2937&	0.54$\pm{0.05}$&	0.074$\pm{0.01}$& 	8.96&   36&          7.93&     43.18&   39.74&  \textit{d, n}\\
J1503 + 3644&	0.3557&	0.30$\pm{0.04}$&	0.058$\pm{0.006}$&	8.22&   38&          7.95&     42.96&  39.84&  \textit{d}\\
Tol1247 - 232&	0.0488& 0.10$\pm{0.02}$&	0.045$\pm{0.012}$&	9.7&    36.2&       8.1&       42.56&     40.4&  \textit{f,g,i,n}\\
Haro 11&    	0.021&	0.04&	0.032$\pm{0.012}$&	10.2&   23.8&       7.9&        41.25&     39.60&  \textit{f,h,j,l,o}\\
J0232 - 0426& 0.45236&  0.425$\pm{0.053}$&   $<0.04$&   7.49&   7.5&       7.88&      42.53& 38.73&  \textit{m}\\
J0919 + 4906& 0.40512&  0.687$\pm{0.089}$&   0.162$\pm{0.059}$&   7.51&     8.4&      7.77&      42.68& 39.63&   \textit{m}\\
J1046 + 5827& 0.39677&  0.318$\pm{0.043}$&   $<0.02$&   7.89&   11.0&        8.01&      42.57& 38.69&  \textit{m}\\
J1121 + 3806& 0.31788&  0.432$\pm{0.052}$&   0.35$\pm{0.056}$&   7.20&      10.0&      7.96&      42.43& 39.85&   \textit{m}\\
J1127 + 4610& 0.32230&  0.397$\pm{0.085}$&   0.111$\pm{0.040}$&   7.44&     4.2&      7.84&      42.18& 39.05&   \textit{m}\\
J1233 + 4959& 0.42194&  0.412$\pm{0.039}$&   0.121$\pm{0.034}$&   7.79&     14.4&     8.11&     42.74& 39.72&  \textit{m}\\
J1349 + 5631& 0.36366&  0.403$\pm{0.044}$&   $<0.07$&   7.36&   5.1&      7.91&     42.33& 38.72&  \textit{m}\\
J1355 + 1457& 0.36513&  0.231$\pm{0.028}$&   $<0.01$&   7.74&   13.1&    7.77&     42.50& 38.72&  \textit{m}\\
J1455 + 6107& 0.36793&  0.365$\pm{0.045}$&   $<0.01$&   7.90&   9.1&    7.91&     42.54& 38.68&  \textit{m}\\

          \hline
    \end{tabular}
        \begin{tablenotes}[online, flushleft]
            \item[] a)\cite{Izotov2018b},
            \item[] b)\cite{Borthakur2014},
            \item[] c)\cite{Izotov2016a},
            \item[] d)\cite{Izotov2016b},
            \item[] e)\cite{Izotov2018a},
            \item[] f)\cite{Leitet2013},
            \item[] g)\cite{Verhamme2017},
            \item[] h)\cite{Leitet2011},
            \item[] i)\cite{Puschnig2017},
            \item[] j)\cite{Pardy2016},
            \item[] k)\cite{Heckman2015},
            \item[] l)\textit{http://lasd.lyman-alpha.com/},
            \item[] m)\cite{Izotov2021},
            \item[] n)\cite{Gazagnes2020}
            \item[] o)\cite{Micheva2010}
        \end{tablenotes}
     \end{threeparttable}
     \label{tab:observations}
    \end{table*}

\subsection{Distributions of \lya and LyC properties}
\label{sec:lyalyc_distrib}

\begin{figure*}
    \centering
    \includegraphics[width=0.3\textwidth]{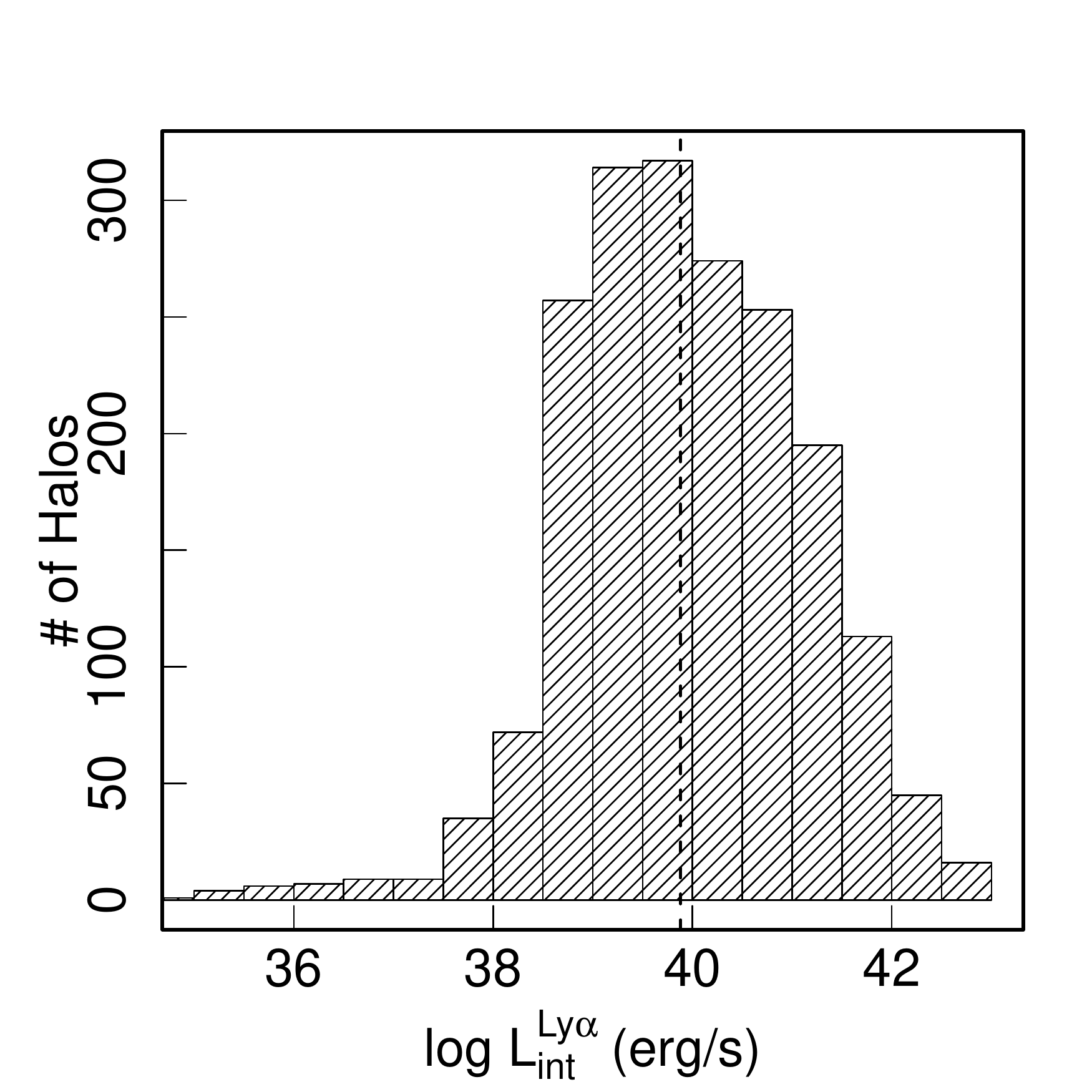}
    \includegraphics[width=0.3\textwidth]{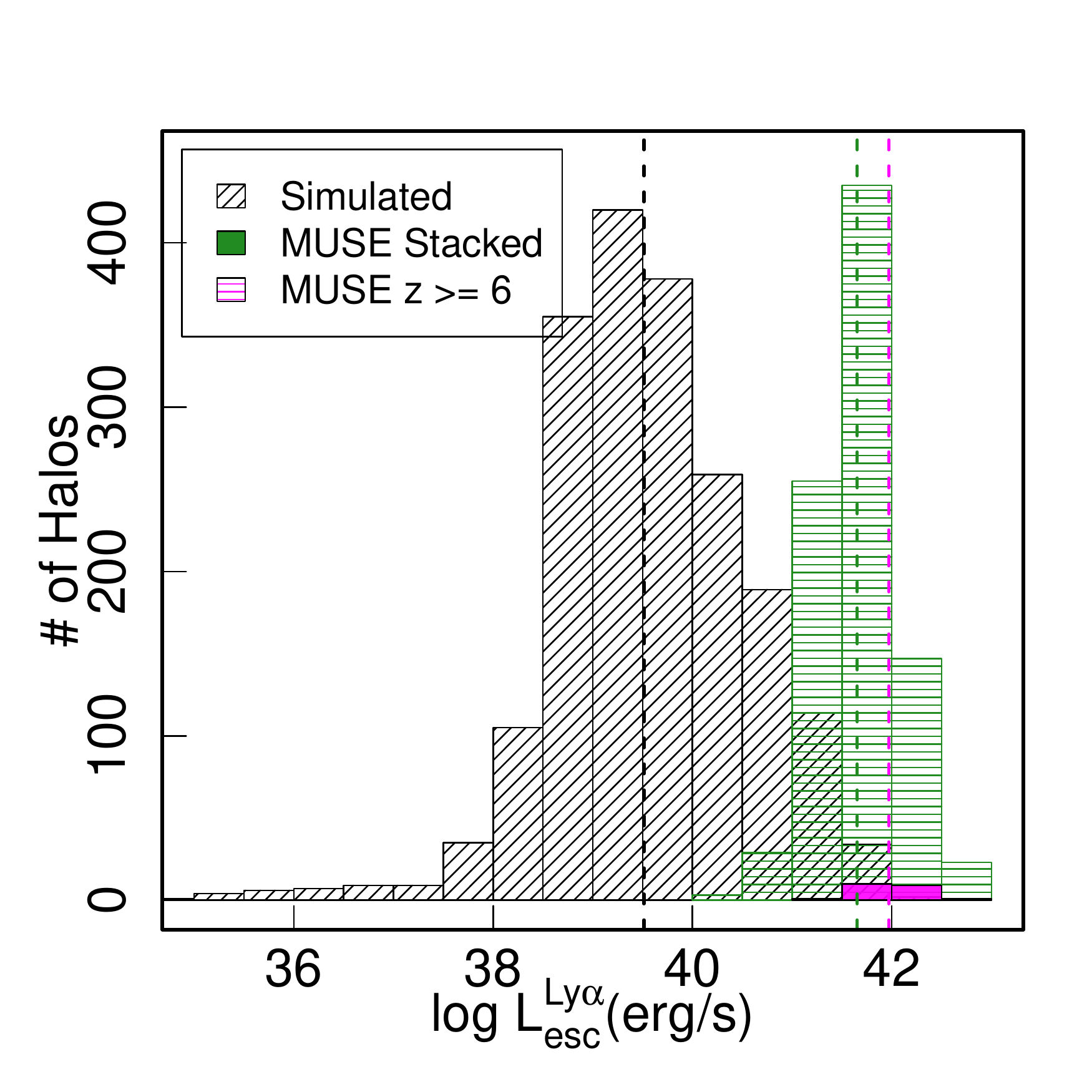}
    \includegraphics[width=0.3\textwidth]{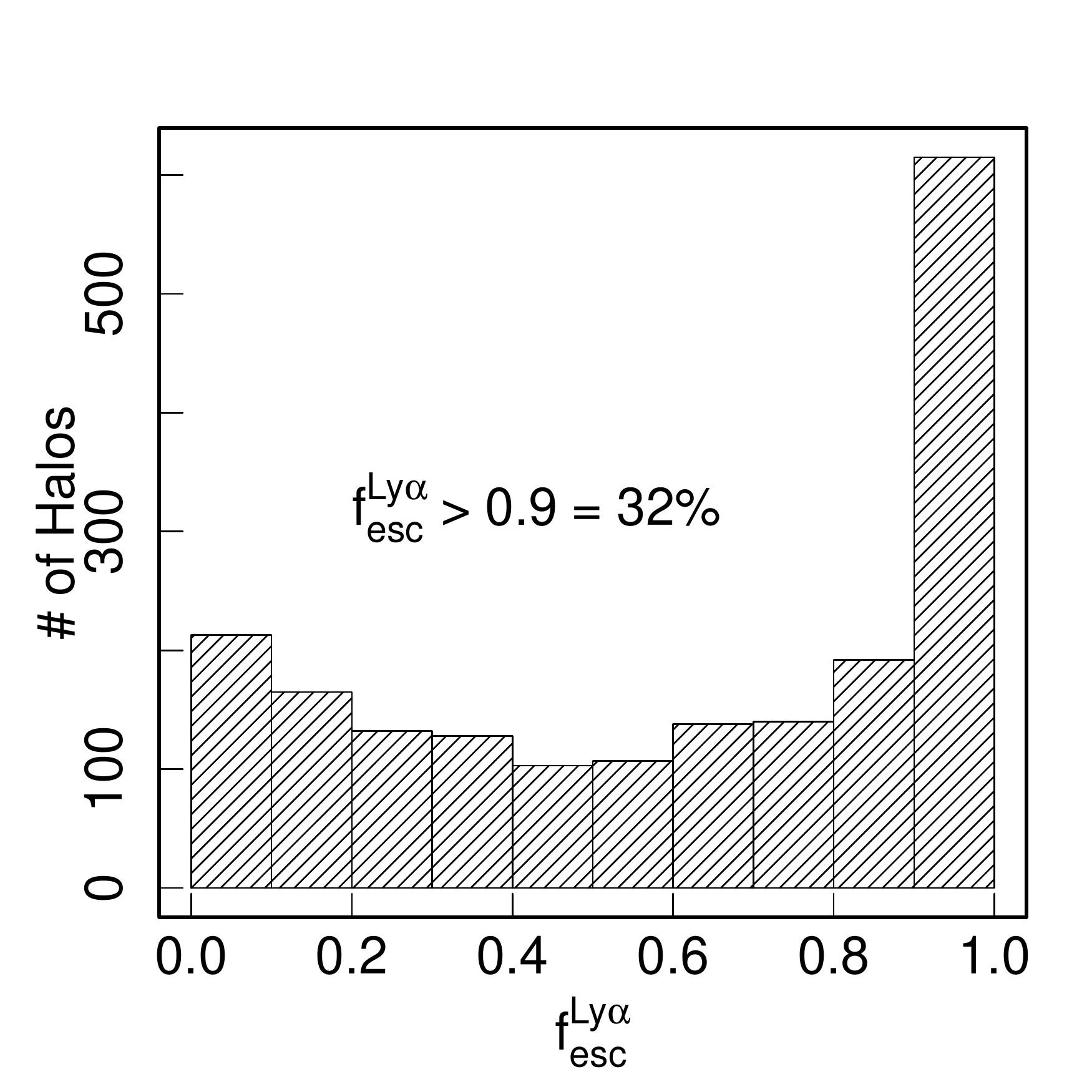}\\
    \includegraphics[width=0.3\textwidth]{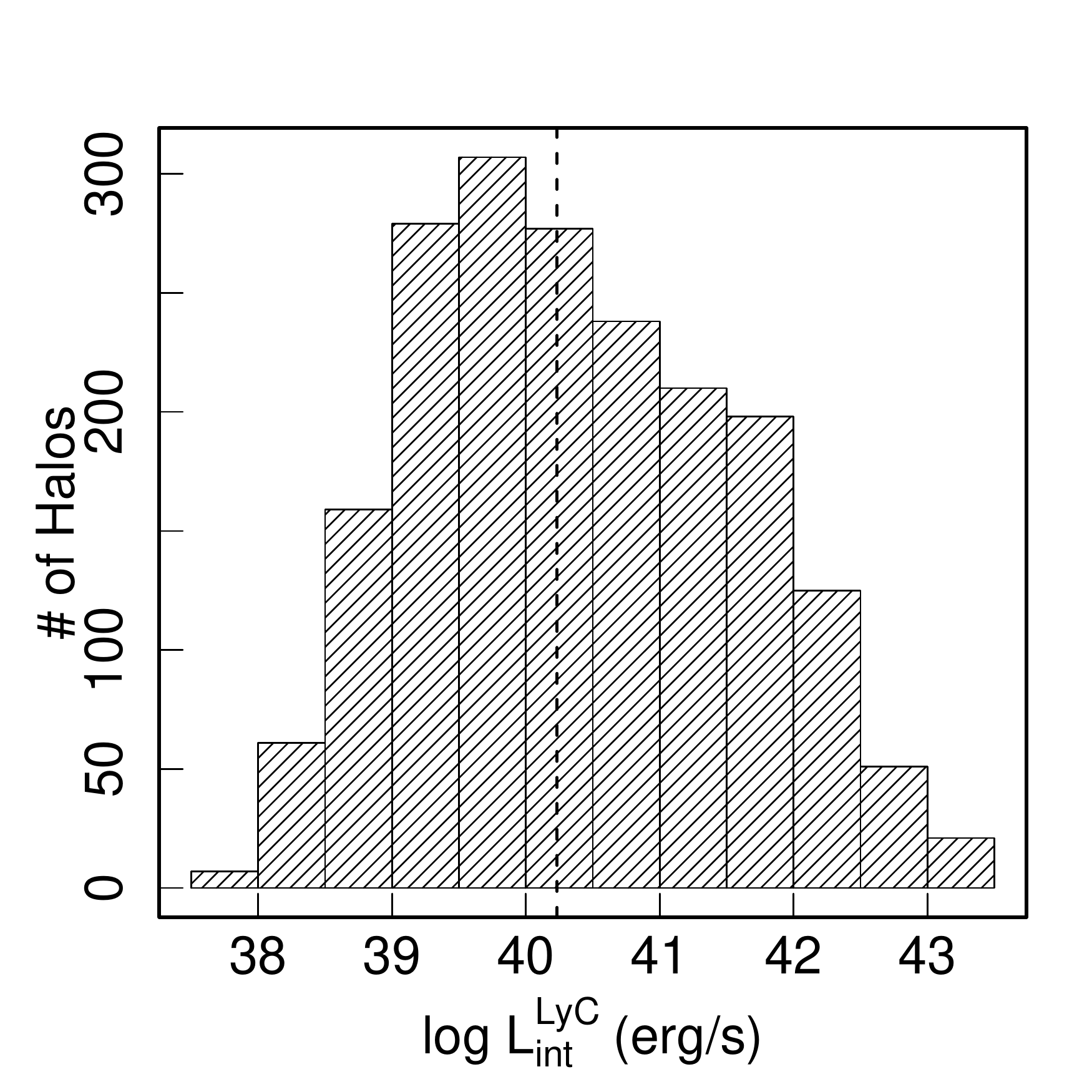}
    \includegraphics[width=0.3\textwidth]{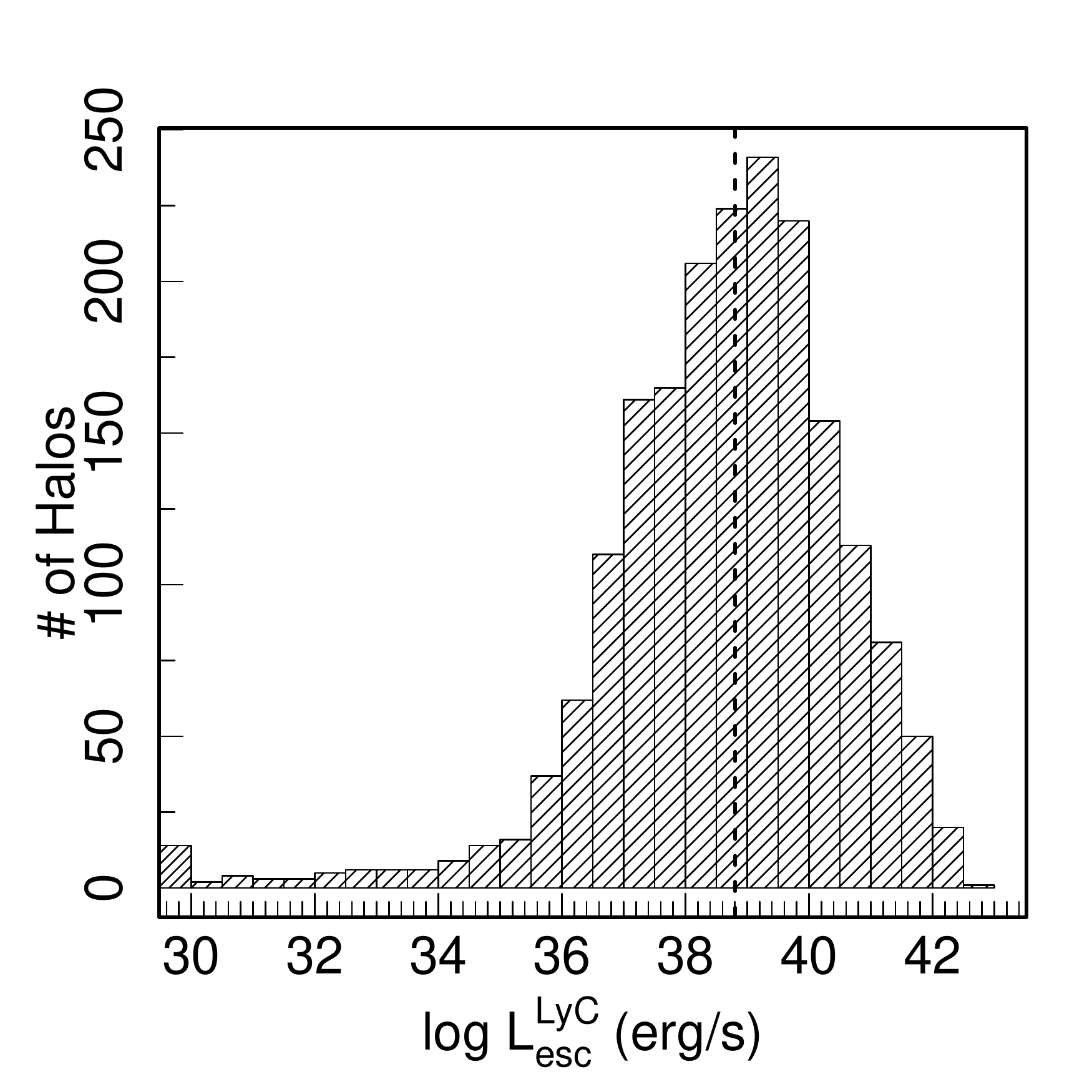}
    \includegraphics[width=0.3\textwidth]{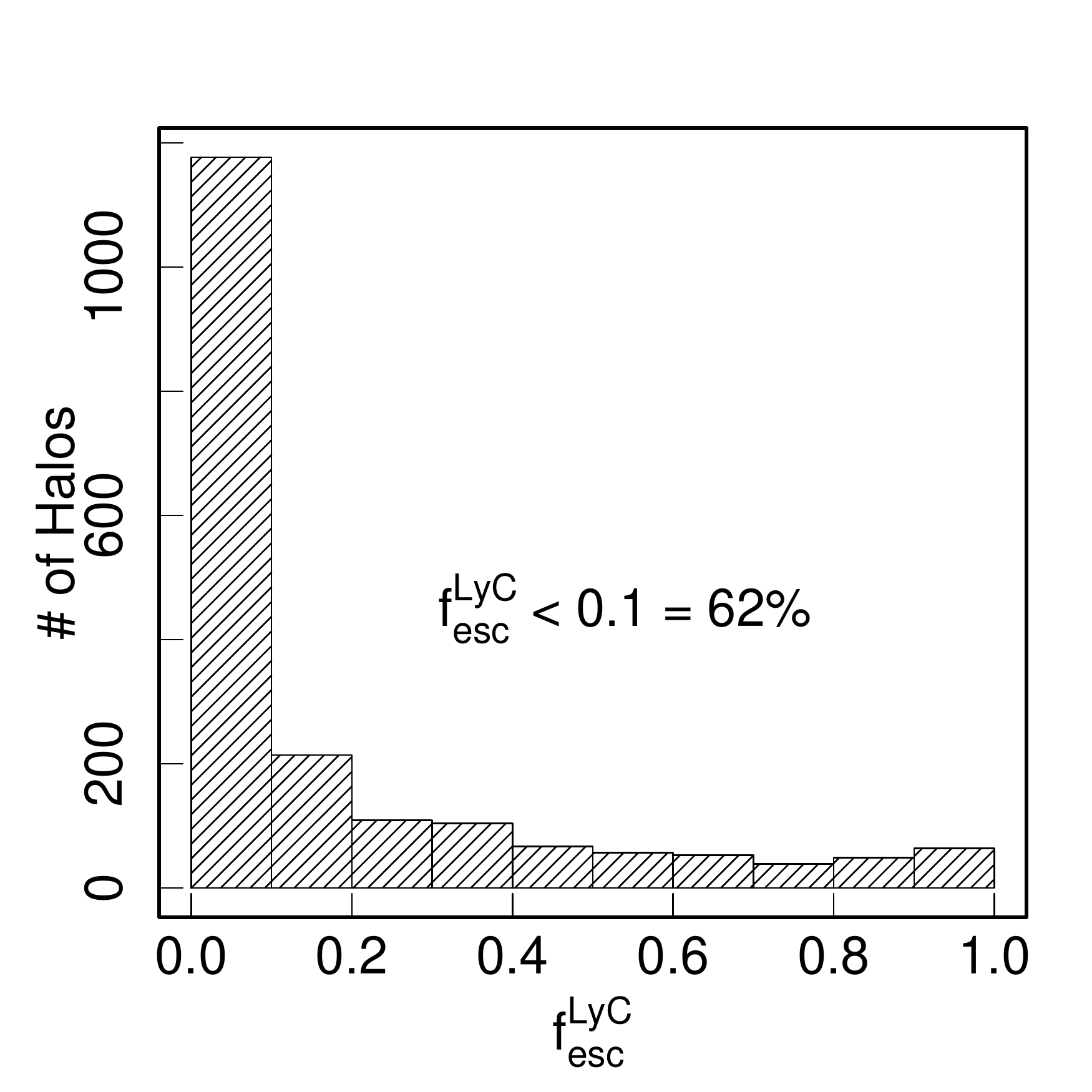}
    \caption{Histograms of \lya and LyC emission of our sample of 1933 galaxies. The top row shows the \lya properties of our sample with intrinsic luminosity (left), escaping luminosity (middle) and escape fraction (right). The bottom row shows the same properties but for LyC radiation. In the middle panel of the top row, we also show the distribution of \Llyaout of galaxies observed in MUSE GTO surveys (MUSE galaxies are shown in green shade, among these, galaxies that are at z > 6 are shown in pink). The dashed lines show their respective median values. The luminosities have a peaked distribution, \flya has a bi-modal distribution with a strong peak at 1, and in most galaxies \flyc is very low.}
    \label{fig:lya_lyc_prop_hist}
\end{figure*}

Figure~\ref{fig:lya_lyc_prop_hist} shows the distribution of the \lya and LyC properties of our simulated galaxy sample, namely their intrinsic luminosities, escaping luminosities and their escape fractions.  In all four cases (intrinsic and escaping for \lya and LyC), the luminosities have a peaked distribution. The median values of \Llyain and \Llyaout are 39.88 and 39.51 erg/s (in log scale) respectively, with the maximum escaping luminosity at $1.375\times10^{42}$ erg/s. The LyC luminosities show a similarly peaked distribution with median (log) values at 40.23 (\Llycin) and 38.80 (\Llycout). We note that the maximum luminosities of simulated galaxies are a consequence of the finite volume of the simulation box and the low end of the luminosities are affected by galaxy mass selection and the mass resolution of the simulation \citep{Garel2021}.

In contrast, \flya shows a bi-modal distribution with the major peak at 1 (the minor peak is at 0). We find that $32\%$ of the sample has $\flya > 0.9$. The distribution of \flyc shows that most galaxies have low \flyc, with $62\%$ of galaxies with $\flyc < 0.1$. Since LyC can be absorbed by HI and HI is plentiful in the ISM, it is very hard for LyC to escape, resulting in very low \flyc in most galaxies. \lya on the other hand is absorbed only by dust, so has a easier time to escape, which results in the peak around \flya=1.

Among these 6 quantities, only the escaping \lya luminosity is observable at the EoR. Our sample is fainter than most available LAE data but it can still be compared with the faint LAEs from MUSE surveys. Therefore in the histogram of \Llyaout in Figure~\ref{fig:lya_lyc_prop_hist} we also show the distribution of \Llyaout from galaxies in MUSE GTO surveys. The 
MUSE data are taken from the MUSE-Deep survey \citep{Drake2017} and MUSE Extremely Deep Field (MXDF) (Bacon et al. in prep).
In total there are 892 MUSE galaxies in the redshift range of $z = 2.92 - 6.64$ with luminosities $10^{40.33 - 43}$ erg/s. Among these, 21 galaxies are at $z > 6$. We see that there is overlap between the most luminous end of our simulated galaxies and the faint end from MUSE, in the luminosity range of $\sim 10^{40 - 42}$ erg/s.
Our simulated luminosities are the total \lya output of the galaxy in all directions, before IGM attenuation. The observed data is, of course, directional measurement after IGM attenuation. 
We discuss the potential observational biases towards bright galaxies, and the lack of very bright LAEs in our sample due to the simulation box size limit in \S~\ref{sec:discuss}.

\subsection{Investigating the \lya-LyC luminosity relationship}
\label{sec:lyalyc_lum}
To assess possible correlations between the 
LyC and \lya radiation in 
galaxies, in the first step we analyze their intrinsic and escaping luminosities.

\begin{figure*}[ht]
    \centering
    \includegraphics[width=0.45\textwidth]{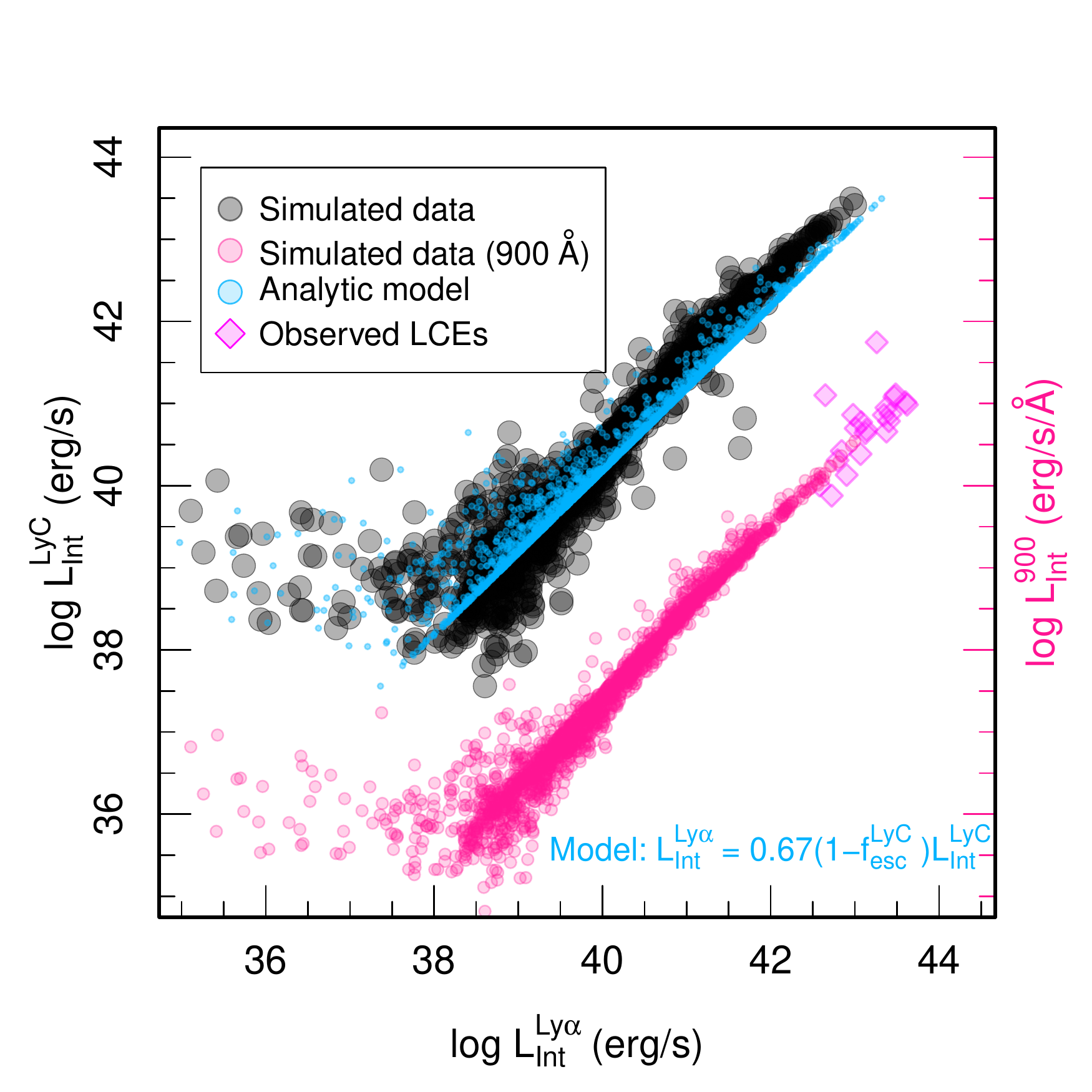}
    \includegraphics[width=0.45\textwidth]{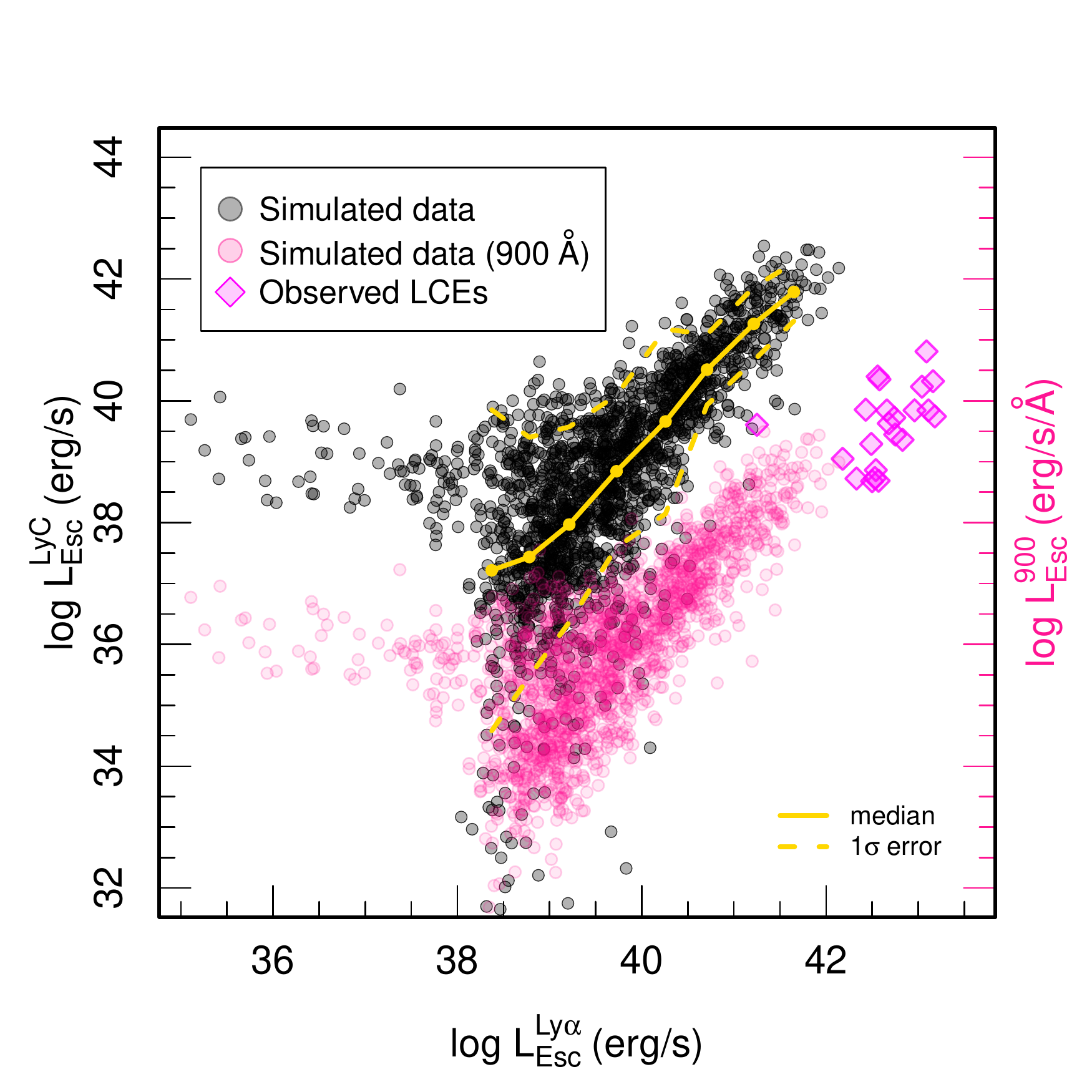}
    \caption{Left- Intrinsic LyC luminosity of galaxies as a function their intrinsic \lya luminosity. The black and the pink points show the total LyC luminosity (0 - 912\AA) and the 900\AA\, luminosity of the simulated galaxies respectively (\S~\ref{sec:obs_lces}). The diamond shaped magenta points show the observed LCEs described in Table~\ref{tab:observations}. 
    The sky blue points show the intrinsic luminosities derived from the analytic model described in the left figure. Right- Escaping LyC luminosity of galaxies as a function of their escaping \lya luminosity. The solid yellow line shows the median \Llycout as a function of \Llyaout (we divide the log \Llyaout between $38 - 42$ into groups of 0.5 dex each and find the median luminosities). The dashed line show $1\sigma$ deviation from this, which illustrates the typical dispersion of the escaping luminosities.}
    \label{fig:lya_lyc_lum_inout}
\end{figure*}

In Figure \ref{fig:lya_lyc_lum_inout} we 
show the LyC luminosities of galaxies as a function of their \lya luminosities. 
We find that for intrinsic luminosities, \lya and LyC have a fairly tight positive correlation. The production of both LyC and \lya is strongly related to the star formation rate of the galaxy because massive stars directly emit LyC photons and these same photons generate \lya by photo-ionizing the HI in the ISM, which then can produce \lya through recombination. 

Furthermore, we also show their intrinsic LyC luminosity at 900\AA\ (as discussed in \S~\ref{sec:lycmethod}), and find that the intrinsic luminosities of observed LCEs (derived as observed luminosity/escape fraction) 
also fall on the same tight correlation, though extending to higher luminosities. This suggests that the correlation between intrinsic \lya and LyC luminosities is valid over a large range of \lya luminosities.

In the same figure we also show predictions for intrinsic \lya luminosities from a simple model based on case B recombination \citep{Spitzer1978} given by, $\Llyain = 0.67(1-\flyc)\Llycin$. This model assumes that all LyC photons that do not escape the galaxy will ionize the neutral hydrogen gas in the ISM. It also assumes that $67\%$ of them will be reprocessed as \lya photons through recombinations. 

We find that the simulated data is generally matched well by this model. Some galaxies, especially among lower \lya luminosity galaxies, lie below the analytical relationship, 
implying that the contribution of collisions is increasingly important for faint and low mass \lya emitters. For example, we find that in galaxies where $\Llyain > 10^{42}$ erg/s, collisional emission contributes only $\sim$ a few percent of the total \lya production, but it can rise to $\sim 50\%$ in galaxies $10^{38}\leq \Llyain \leq 10^{40}$ erg/s (see discussion and figure in \ref{appendix:rec_col}, also \citealt{Rosdahl2012}). 
We also find that in all luminosity ranges, some galaxies fall above the analytical relationship, i.e. there are some galaxies which have less \lya production than estimated by the analytical equation. This is mainly due to the fact that a fraction of the most energetic photons go towards ionising He or HeI, rather than HI, and as a result they cannot be reprocessed as \lya.
We also note that galaxies at the very faint end of \lya ($\Llyain < 10^{38}$ erg/s) have LyC luminosity in the range of $10^{38} - 10^{40}$ erg/s. These galaxies are extremely gas deficient, so they produce very little \lya and the stars in them continue to produce LyC for a long time (further discussed in $\S\ref{sec:discuss}$).

Furthermore, from the right panel of Figure~\ref{fig:lya_lyc_lum_inout} we find that the escaping luminosity of \lya and LyC is also well correlated. The escaping \lya and LyC luminosities of the observed LCEs are also shown in this figure along with the \Llycnineout of the simulated galaxies and these LCEs seem to follow the similar trend. We note that the correlation is tight at higher luminosities, although the scatter is overall larger compared to the correlation between the intrinsic luminosities. The scatter increases as the galaxies become fainter in \lya (or LyC). This is mostly due to the fact that 
the faint LAEs have a very wide range of \lya and LyC escape fractions (discussed further in \S~\ref{sec:lyalyc_fesc}, see also Fig~\ref{fig:fesc_lya_lyc}, \ref{fig:median_fesc}). Hence galaxies with similar intrinsic luminosities can end up with very different escaping luminosities, which scatters the points horizontally and vertically.
The escape fractions of the galaxies depend on the structure of the ISM, in particular on the possibility of having holes or low HI column density channels in the ISM which can facilitate the escape of LyC. We discuss the escape fractions in more detail in the next section. We also show this figure color-coded with \flya and \flyc in Figure~\ref{fig:lumout_color} and further discuss the relationship of escaping luminosities with escape fractions in \S~\ref{appendix:colorbyfesc}. Moreover, we note that there are no galaxies with simultaneously very low \lya and LyC luminosities. This is an effect of the stellar mass limit we imposed on our galaxies. We recall from \S~\ref{sec:sample} that we analyze here all galaxies with $M_\star >10^6\Msun$. We checked that if we do include less massive galaxies in our sample, they start to fill up this faint section of the plot, as they are very faint in both \lya and LyC. The few extremely faint LAEs we do have in our sample are extremely gas deficient as we discussed in the previous paragraph, so it is easy for the LyC emission to escape from these systems, hence their intrinsic and the escaping LyC luminosities remain almost same. 




\begin{figure}
    \centering
    \includegraphics[width=0.45\textwidth]{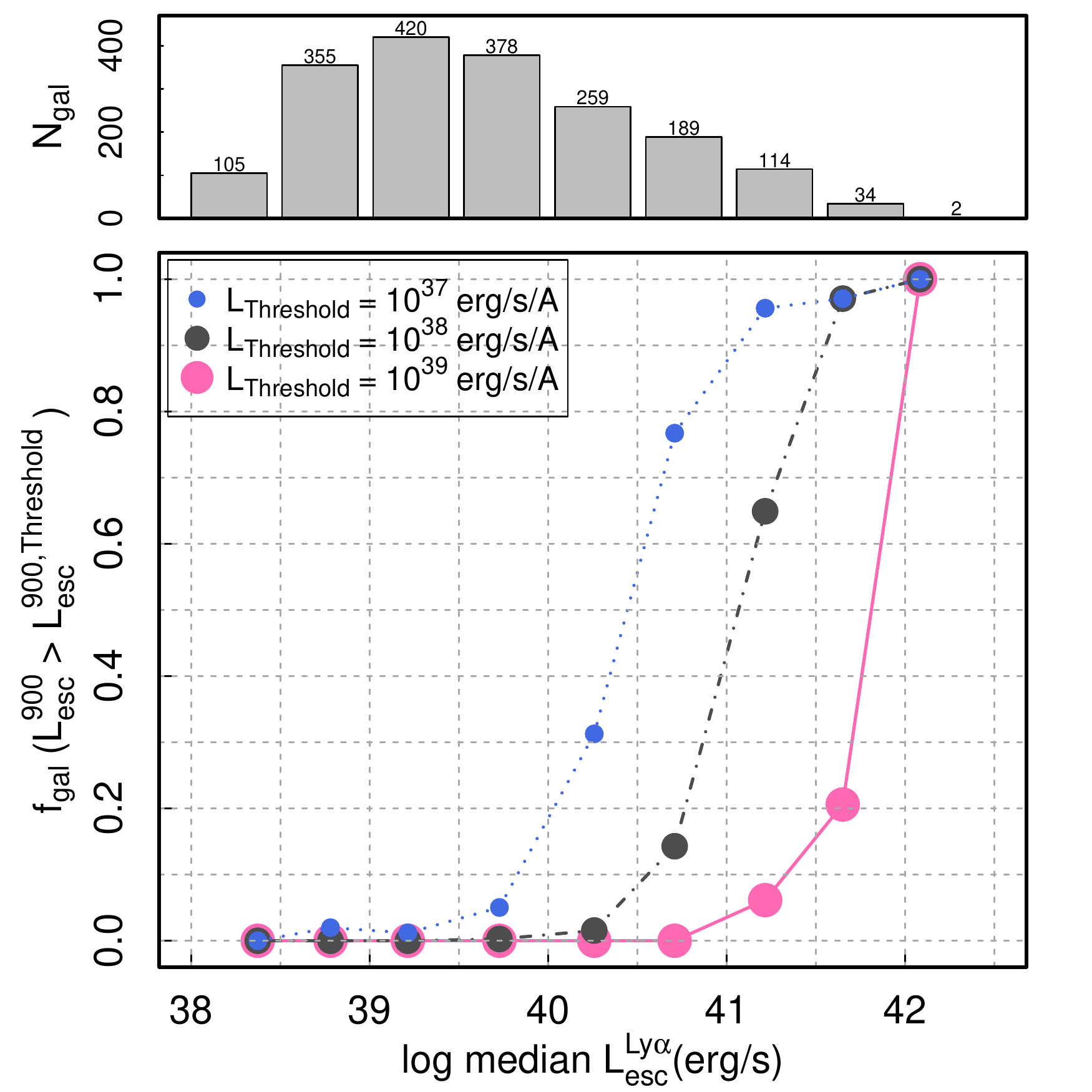}
    \caption{Fraction of galaxies with \Llycnineout luminosity above a threshold value against their median escaping \lya Luminosity.}
    \label{fig:median_lya}
\end{figure}


\subsection{Fraction of LyC leakers in LAE samples}
As shown in the previous sections, \lya and LyC luminosities are correlated with one another. Hence, we can wonder what fraction of LAEs would be detectable as LyC leakers, assuming typical LyC and \lya detection limits. 
To answer this question, we divide galaxies in our sample with \Llyaout between $10^{38}$ to $10^{42.5}$ erg/s into 9 equally logarithmically-spaced bins (bin width 0.5 dex). In each group we calculate the median \lya luminosity and the fraction of galaxies that have their \Llycnineout luminosity higher than a given threshold value and report these fractions against their median \Llyaout in Figure~\ref{fig:median_lya}. We do this exercise for three different threshold values of escaping LyC luminosity, $L_{\rm{Threshold}} = 10^{37}, 10^{38}$and $10^{39}$ erg/s and we find that as galaxies become brighter in \lya, the fraction of galaxies with $\Llycnineout > L_{\rm{Threshold}}$ increases. For example, given a threshold LyC luminosity of $10^{38}$ erg/s, $65\%$ of LAEs with luminosity $\Llyaout = 10^{41 - 41.5}$ erg/s and $97\%$ of LAEs with luminosity $\Llyaout = 10^{41.5 - 42}$ erg/s are bright in LyC emission. Granted, our simulated galaxies are at high redshift (z = 6 - 10) but these results could be useful at lower redshifts, where LyC emission can be detected. \cite{Katz2019, Katz2020} have shown that low metallicity LyC leakers at $z\sim3$ are good analogues of EoR galaxies. The observed \Llycnineout limit around z = 3 is $\sim 1.61\times10^{39}$ erg/s (flux limit $2\times 10^{-20}$ erg/s/cm$^2$/\AA\, or $5.5\times 10^{-4}\,\mu$Jy, Kerutt et al, in prep). At this threshold LyC luminosity, our analysis highlights that among LAEs with luminosity $10^{41.5 - 42}$  erg/s, $\sim 15\%$ of  galaxies will be detected as LyC emitters.

\subsection{Escape fraction}
\label{sec:lyalyc_fesc}

\begin{figure*}[ht]
    \centering
    \includegraphics[width=0.45\textwidth]{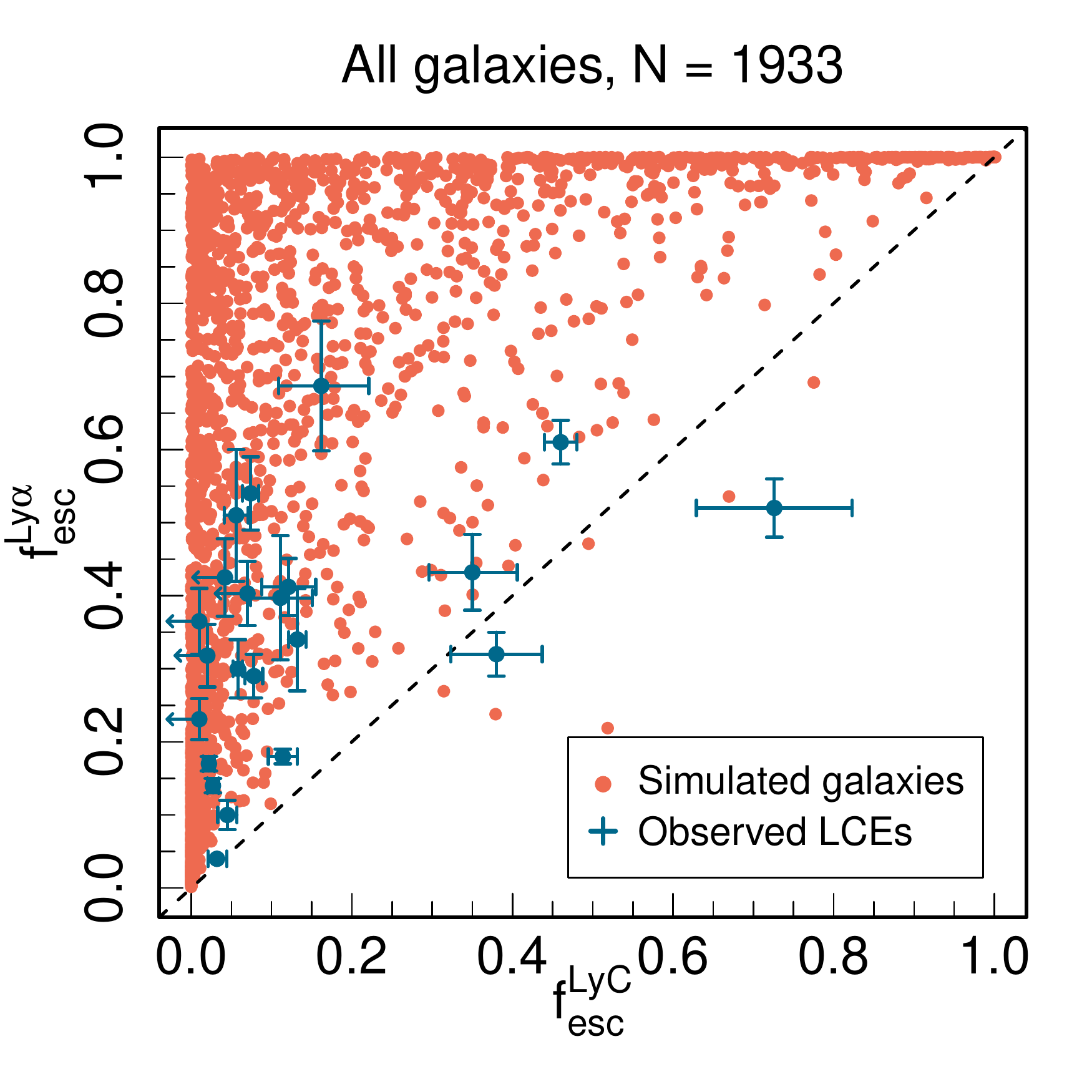}
    \includegraphics[width=0.45\textwidth]{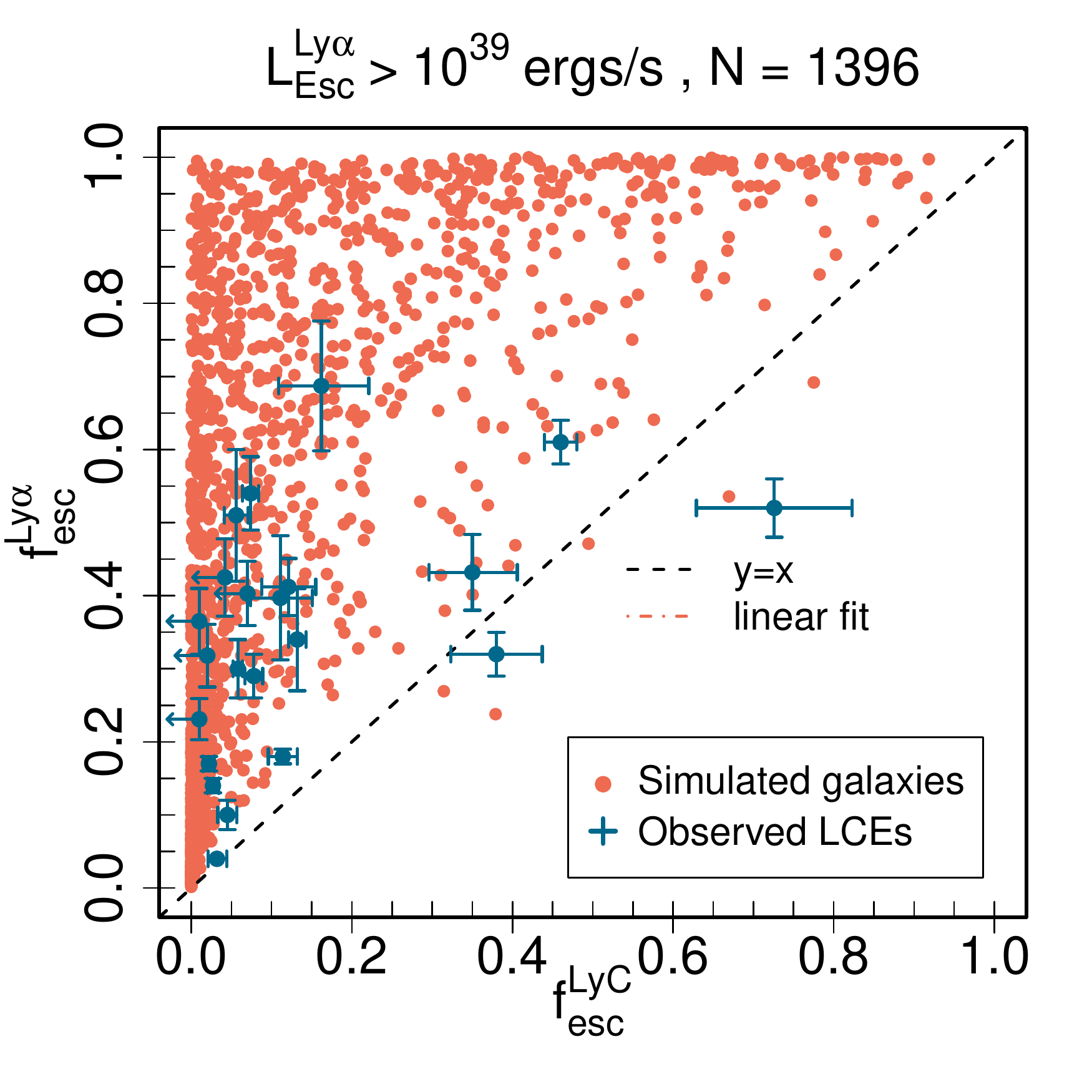}
    \includegraphics[width=0.45\textwidth]{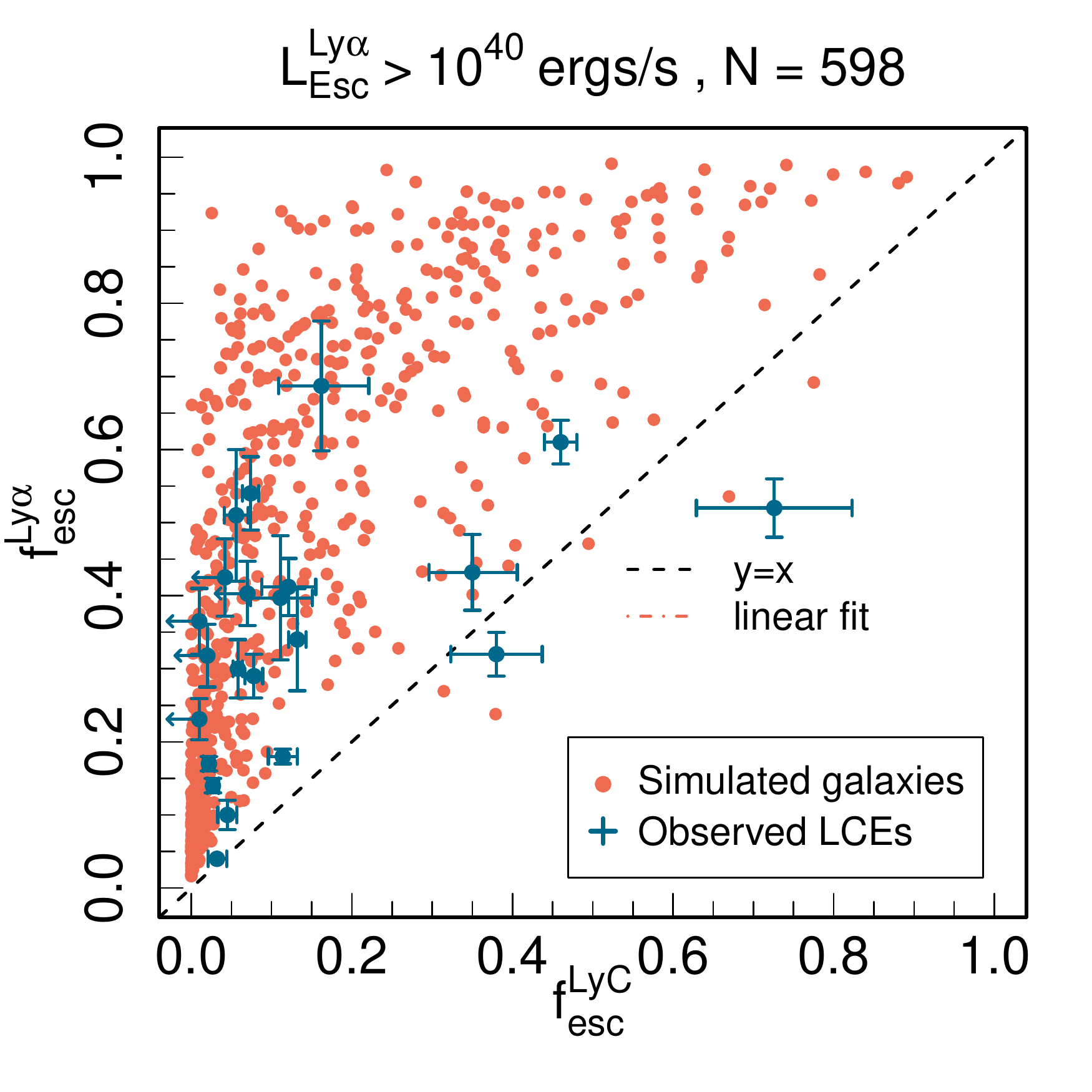}
    \includegraphics[width=0.45\textwidth]{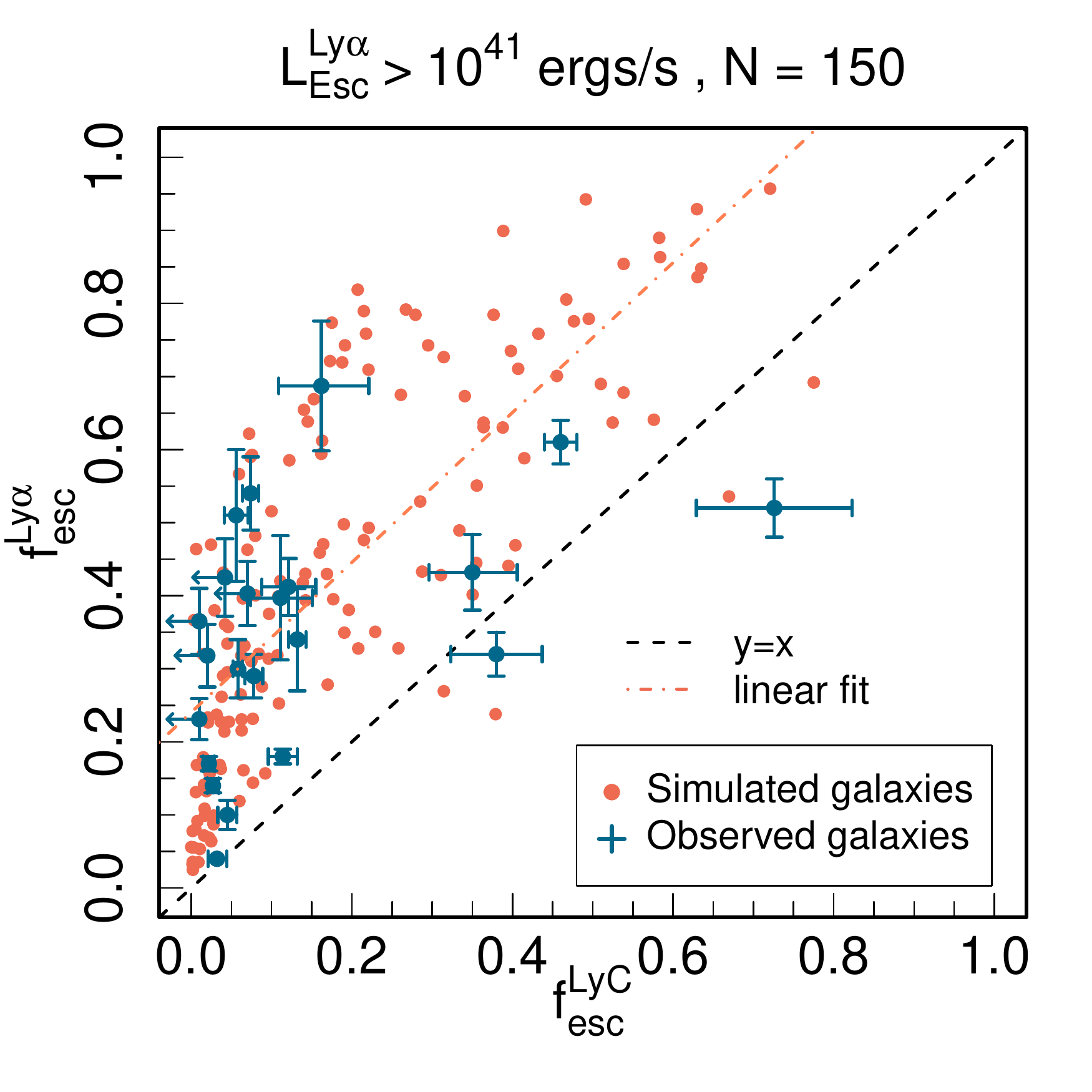}
   \caption{Escape fractions of \lya vs LyC. The plots here show progressively brighter sample selection, all galaxies (top left), and galaxies with $\Llyaout > 10^{39}$ erg (top right), $\Llyaout > 10^{40}$ erg (bottom left), and $\Llyaout > 10^{41}$ erg (bottom right) respectively. In each plot we include the \flya and \flyc of observed LCEs from Table 1 (blue points) with their error bars. The observed LCEs are all bright in \lya, with \Llyaout $>10^{41}$ erg/s. For few galaxies the observed \flyc is an upper limit, these are marked by blue arrows. The dashed black line shows the y=x or equality line. The orange dashed line in the bottom right plot shows a linear fit of the simulated galaxies, which yields a slope of 1.02. This plot shows that if we include all galaxies, including very faint ones, \flya and \flyc are very scattered and not correlated, but as we restrict our sample to progressively brighter LAEs, a correlation emerges. In the last panel, the simulated galaxies are in the same luminosity range as the observed ones ($>10^{41}$ erg/s), and they both show similar correlation.}
    \label{fig:fesc_lya_lyc}
\end{figure*}

Figure~\ref{fig:fesc_lya_lyc} shows the \flya - \flyc relationship of our simulated galaxies. 
Here we have plotted galaxies with progressively brighter sample selections: all galaxies (N = 1933), galaxies with $\Llyaout > 10^{39}$ erg/s (N = 1396), $> 10^{40}$ erg/s (N = 598), and finally $> 10^{41}$ erg/s (N = 150). We find that if we consider all 1933 galaxies, including the very faint ones, the escape fractions of \lya - LyC are very scattered and not correlated. The escape fractions occupy the whole space above the equality line, with only a few galaxies with $\flya < \flyc$. However, if we limit our sample to only \lya bright galaxies, the dispersion decreases. If we include only the brightest galaxies with $\Llyaout > 10^{41}$ erg/s, a positive correlation emerges between the two escape fractions. 
A linear regression of these bright galaxies yields the following model (with standard errors), $\flya = (1.02\pm{0.07})\flyc + (0.24\pm{0.02})$. We find that the observed LCEs (Table~\ref{tab:observations}) which are all bright LAEs ($>10^{41}$ erg/s), fall in the same escape fraction range as the simulated galaxies, which is an encouraging indication that escape fractions of our simulated galaxies are not significantly different from the escape fractions calculated from observed local LCEs.
The correlation between \flya and \flyc in the simulated bright galaxies and the observed ones is also very similar. This analysis indicates that the linear positive correlation of \flya and \flyc that we find in observed LyC emitters \citep{Verhamme2017} may be a selection bias which holds true only when we consider the brightest LAEs. 

Additionally, in Figure~\ref{fig:fesc_lya_lyc} we find that in galaxies with very low \flyc, the \flya can take any value between 0 to 1, but in galaxies with high \flyc, the \flya is always very high. Conversely, galaxies with low \flya always have low \flyc, but in galaxies with high \flya, \flyc can range from 0 to 1. \cite{Dijkstra2016} also found similar distributions using idealized models. 
We also note that \flya is always greater than \flyc, except for a few outlier galaxies in our simulated sample where $\flya < \flyc$. Theoretically it is expected that the \lya escape fraction is greater than LyC because \lya is only destroyed by dust while LyC can also be killed by HI atoms in the ISM. \lya photons can scatter numerous times and have a greater possibility to find channels in the ISM with low HI column density through which they can escape the galaxy \citep{Dijkstra2016}.
However in 
6 out of 1933 (or $0.3\%$) of our galaxies we find that this is not the case. Similarly for observed LCEs, although most of them have higher \flya, in 2 out of 23 galaxies (~8.7$\%$), \flya is less than \flyc. It is possible that in these systems there are dusty escape channels with low HI column density (the dust model allows for dust in ionized gas) such that it is optically thin to LyC photons but not to \lya. We looked into these 6 simulated galaxies and found that these systems are comprised of interacting galaxies with complex configurations. The distribution of \lya and LyC sources differ and they have escape channels of low density gas columns very close to the center where LyC production happens which can greatly aid its escape. 


\subsubsection{Median escape fraction at different \lya luminosities}
\label{sec:median}

\begin{figure}[ht]
    \centering
    \includegraphics[width=0.45\textwidth]{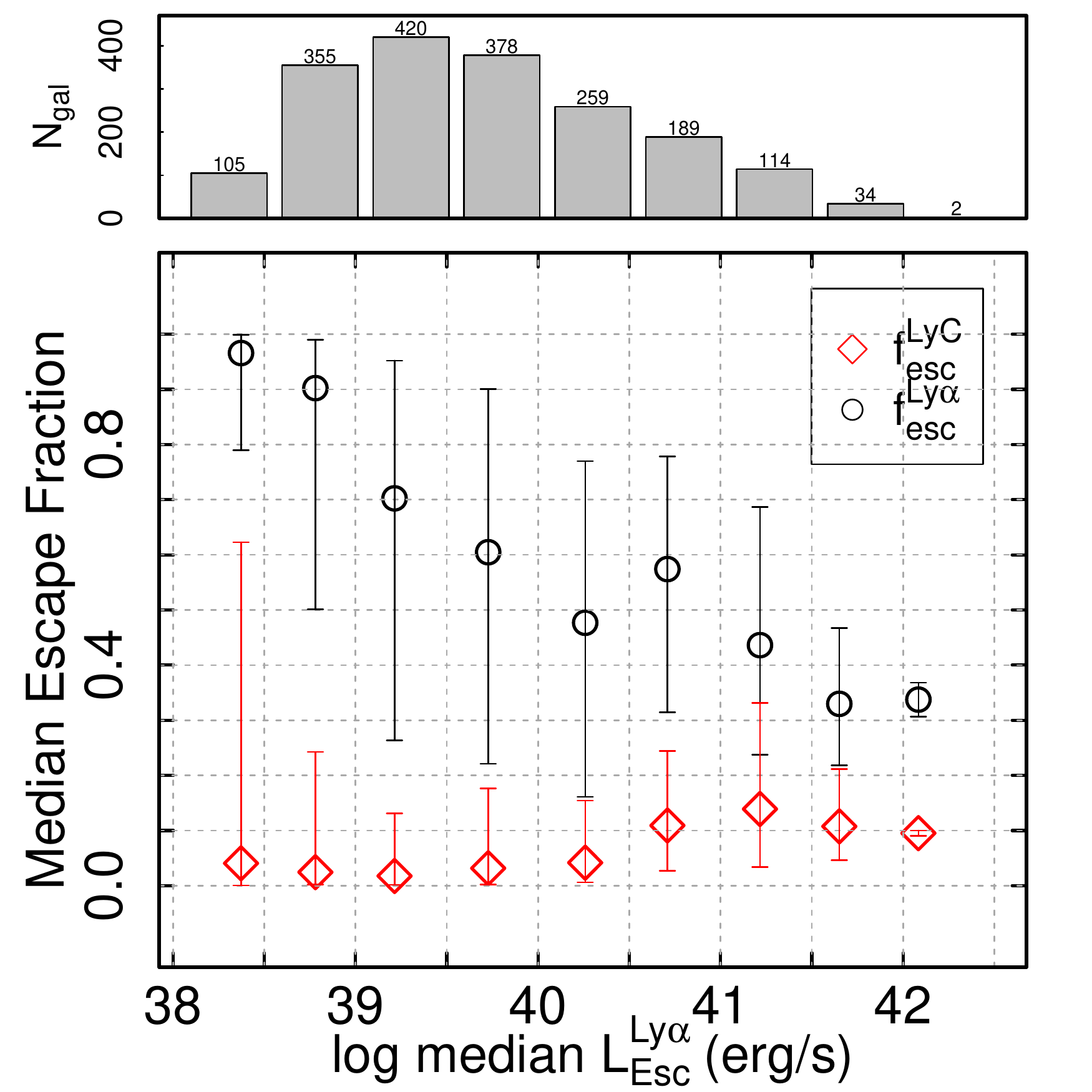}
    \caption{Median \flya and \flyc of galaxies in different escaping \lya luminosity groups, plotted against the median \lya escaping luminosity of the group. The vertical lines through each median point connect the 1st quantile ($25\%$) and the 3rd quantile($75\%$) of the escape fraction distribution in that luminosity bin. The histogram shown on the top indicates the number of galaxies in the respective luminosity bins.}
    \label{fig:median_fesc}
\end{figure}

Since we have a large sample of galaxies with both \lya and LyC radiative transfer, it is instructive to study how \flya and \flyc correlate with the \lya luminosity of galaxies. To analyze this, we have taken all galaxies in our sample with \Llyaout from $10^{38}$ to $10^{42.5}$ erg/s and divided the luminosities into 9 equally logarithmically-spaced bins (bin width 0.5 dex). We show the median escape fractions against median luminosities in Figure~\ref{fig:median_fesc} and find that as the luminosity increases \flya decreases.
The drop in \flya is fairly gradual and in our highest luminosity bins, $10^{41.5 - 42.5}$ erg/s, the median value of \flya is $\sim 0.3$. Brighter galaxies have higher mass in all components, including dust mass, and as dust content increases, more \lya is absorbed by dust which reduces \flya. At the bright end, $\Llyaout \approx 10^{42}$erg/s, our sample size decreases to only a couple of galaxies, owing to the limited simulation volume. Therefore, although the flattening of the median curves in bright LAEs suggest a similar value for even brighter galaxies, we cannot make any concrete prediction for much brighter LAEs. 

The median \flyc is low 
for all \lya luminosities. In galaxies with $\Llyaout < 10^{40.5}$ erg/s median \flyc is very low ($\sim 0.02$), and in brighter galaxies it rises to $\sim 0.1$. A large fraction of faint LAEs have zero or very low \flyc as ionizing photons are absorbed by HI gas in the surrounding ISM which drives the median low (Chuniaud et al 2021, in prep).


The median \lya luminosity of MUSE LAEs is around $10^{41.6}$ erg/s, as shown in Figure~\ref{fig:lya_lyc_prop_hist}. Our simulation predicts that the typical \flya and \flyc of galaxies at this luminosity are around 0.3 and 0.1, respectively. Here we note that the \lya luminosities of MUSE galaxies are what we observe after \lya  has gone through IGM attenuation. The escaping \lya luminosity of galaxies can be affected adversely by IGM attenuation, especially at $z > 6$. In our simulation, we have not considered the effects of IGM. Along the same lines, the observed luminosities of MUSE galaxies are what we measure along our line of sight whereas the simulated luminosities and escape fractions quoted here are global ones. We provide further discussion on the effects of IGM attenuation and line of sight variability in \S~\ref{sec:discuss}.

\subsection{Contribution of LAEs to reionization}
\label{sec:lyalyc_reion}

\begin{figure*}
    \centering
    \includegraphics[width=0.45\textwidth]{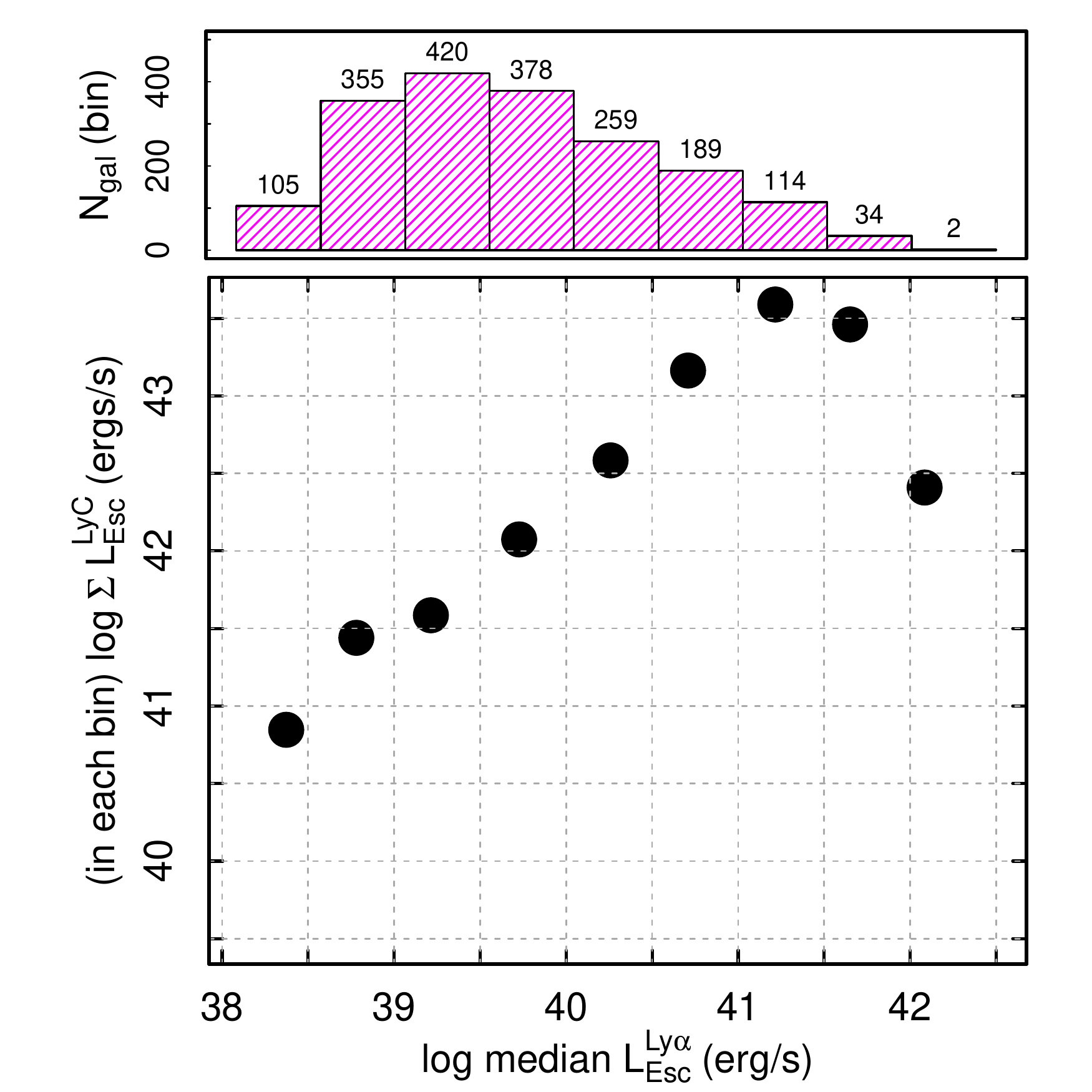}
    \includegraphics[width=0.45\textwidth]{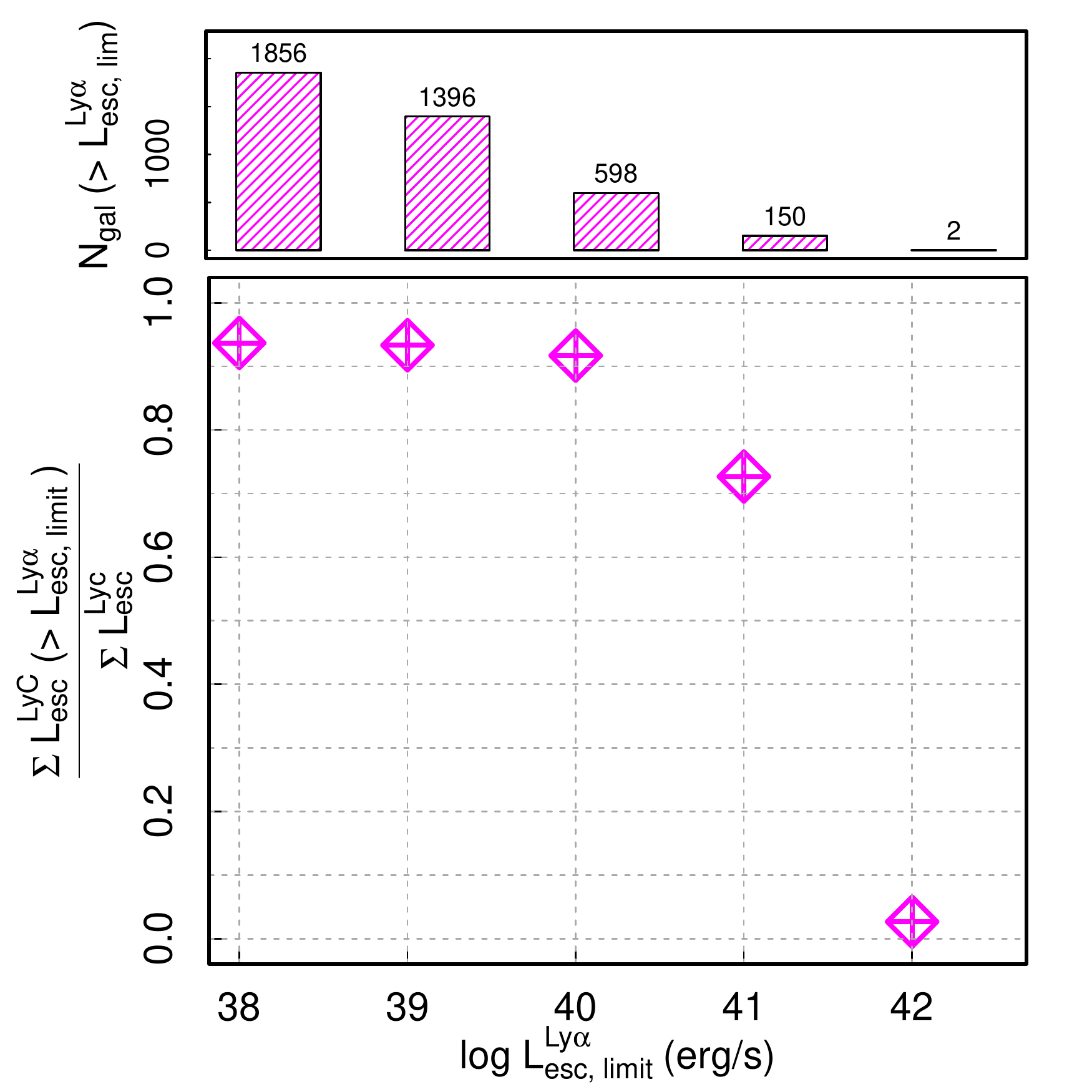}
    \caption{Left : Total escaping LyC luminosity of galaxies grouped
    by their \lya luminosities as a function of their median \lya luminosity. The histogram above shows the number of galaxies in the corresponding bins below. Right : Conditional total escaping LyC luminosity of galaxies brighter than a given \lya luminosity limit as a function of the \lya luminosity limit. The histogram above indicates the number of galaxies with \Llyaout $>$ the respective \lya luminosity limits below, so these are numbers of galaxies that have been used to calculate the respective fractions.}
    \label{fig:cont_to_reion}
\end{figure*}


In our analysis, we have both ionizing or LyC luminosities and the \lya luminosities for a large sample of simulated galaxies in EoR, so we can investigate the role of LAEs as sources of cosmic reionization. 
Similar to the previous section, we take our sample of galaxies that have \lya luminosities in the range $10^{38} - 10^{42.5}$ erg/s range and divide them into 9 equally logarithmically-spaced bins (bin width 0.5 dex). 
For each group of galaxies we calculate their total escaping ionizing luminosities and plot it as a function of their median escaping \lya luminosity in Figure~\ref{fig:cont_to_reion} (left panel). We find that as galaxies become brighter, their total escaping LyC luminosity in each group increases. Since our 10 Mpc$^3$ simulation volume does not contain  galaxies brighter than $1.35\times10^{42}$ erg/s, \citep[see also the \lya luminosity functions, e.g. Fig 5, in][]{Garel2021}, there is a downward trend at the extreme bright end of our sample ($10^{41.5 - 42}$). Therefore our sample size is too small to be conclusive about a peak at $10^{41}$ erg/s. Nevertheless, the luminosity range of $10^{38} - 10^{41}$ erg/s is well sampled, and we find that in this luminosity range, the brighter LAEs have higher total \Llycout.  



Now we calculate the total ionizing luminosity in the whole simulation box. We recall that our galaxy sample consists of galaxies with selection criterion provided in \S~\ref{sec:sample}, i.e. galaxies at level 1 and with $M_{\star} > 10^6\Msun$. Then, to be consistent in our comparisons, we estimate the total ionizing luminosity in the box by summing up the LyC luminosity ($\Sigma\,\Llycout$) of all galaxies at level 1, i.e. from a total of 8783 such galaxies in our simulation.

We compute the contribution of galaxies with \lya luminosity brighter than some limit to the total ionizing luminosity emitted by all simulated galaxies. The result of this is shown in the right panel of  Figure~\ref{fig:cont_to_reion}. We find that simulated LAEs brighter than $10^{40}$ erg/s (N = 598) contribute more than $90\%$ to the total ionizing luminosity of the box, even though the number of faint LAEs is much larger than bright ones. So $6.8\%$ (598 out of 8783 galaxies) of the galaxies, which hosts $37\%$ of total stellar mass, are responsible for more than $90\%$ of the escaping ionizing radiation. Including all LAEs brighter than $10^{38}$ erg/s (N = 1856) can account for $\approx 95\%$ of the total LyC luminosity.

In the MUSE Ultra Deep Field survey \citep[Figure 5]{Drake2017} at z=3 the \lya luminosity limit is $10^{41.25}$ erg/s (50$\%$ completeness). Our analysis suggests that the LAEs with  $\Llyaout > 10^{41.25}$ erg/s at EoR could have contributed $\sim 57\%$ of the ionizing radiation budget.

The faint LAEs produce a small amount of LyC intrinsically, compared to the bright LAEs (\S~\ref{sec:lyalyc_lum}). From our analysis of escape fractions in the previous section we know that the median \flyc of all galaxies is rather low. Consequently the escaping LyC luminosities of faint LAEs is generally low. Therefore, we find that although faint LAEs are far more numerous, brighter LAEs as a group contribute more to the escaping ionizing luminosity. 
We have also explored the effect of the lower mass limit of the galaxies (discussed further in \S~\ref{appedix:reion}) on this reionization study and found that if we lower the mass limit of our galaxies from $10^6$ to $10^5$ \Msun, LAEs brighter than $10^{40}$ contribute $97\%$ of the total ionizing radiation (Fig~\ref{fig:reion_105_106}). This shows that although lowering the mass limit slightly increase these fractions, the differences are very small, our results thus converge. Therefore, we conclude that the primary sources of reionization are likely bright LAEs with $\Llyaout > 10^{40}$ erg/s.


\section{Predicting LyC luminosities and escape fractions}
\label{sec:prediction_z6}

The major goal towards studying the connection between \lya and LyC emission from galaxies is to discover a correlation or develop a model that can estimate the LyC emission of EoR galaxies using the observable properties of galaxies, as the ionizing photons themselves cannot be observed. 

In the previous section (\S~\ref{sec:lyalyc_rel_z6}) we have found that the escape fractions of \lya and LyC are correlated in bright LAEs during the EoR, as observations have suggested, but when we include all LAEs in our sample, including the fainter ones, there is no correlation, which implies that the observed relation may be due to a selection bias. We also found that the intrinsic luminosities in \lya and LyC are well correlated, whereas the escaping luminosities have a positive correlation but with much more dispersion, especially at the faint end. Thus in the quest for predicting the LyC emission, it is important to explore beyond the simple 1:1 correlation. Since we have a large dataset of galaxies with a number of their physical, \lya and LyC properties we now investigate if it is possible to construct a statistical model that predicts the LyC emission using other properties, e.g. mass, SFR and \lya. 


Our galaxy sample is generally fainter (highest $\Llyaout \sim  1.37\times 10^{42}$ erg/s) and less massive (highest stellar mass $M_\star \sim 1.33\times 10^9 \Msun$) than typical observed LAEs. The model that we can build with this data can be best applied to galaxies with properties similar to SPHINX galaxies. Whether this model can be applied to more massive or more luminous galaxies cannot be conclusively determined based on this study alone. Nevertheless, building such a predictive model for LyC using our data is an important first step towards a quantitative understanding of the contribution of galaxies to reionization. This analysis will also identify which galaxy properties are the main predictors of LyC emission and this can help identify strong LyC emitters among observed samples of EoR galaxies and guide future surveys.


\subsection{Multivariate model: A general framework}
\label{sec:multivariate_model}
In our simulation we have a large data-set of hundreds of galaxies each with several physical and radiative properties that can be measured in their real-world counterparts. Given the large number of variables available, we aim to build a model that can be interpreted easily. 
Multivariate linear regression is a common statistical method for building such models, it is also straightforward to interpret and gain insights from the final model. 

Recently \cite{Runnholm2020} did an analysis where they applied multivariate linear regression to observed galaxies at low-redshift to predict escaping \lya luminosities using observed galaxy properties. In this study, they have analyzed galaxies in the Lyman Alpha Reference Sample (LARS) and extended LARS (e-LARS) containing 14 and 28 galaxies respectively, within a redshift range of $0.028 \leq z \leq 0.18$ and found that using either observed or derived physical quantities it is possible to predict \lya luminosities of galaxies accurately with their multivariate regression method. Keeping these considerations in mind, we choose to use multivariate linear regression for predicting LyC and \lya properties of $z\geq 6$ galaxies.

A multivariate linear regression model can be written as follows:
\begin{equation}
    y = \beta_0 + \beta_1 x_1 + \beta_2 x_2 + ... + \beta_n x_n 
    \label{eq:multivar}
\end{equation}
where $x_1, x_2,..x_n$ are independent variables or predictor variables (which would be a set of known properties of the galaxy) and $y$ is the dependent variable or response variable that we want to predict, which in our case are LyC luminosities (intrinsic and escaping) and LyC escape fraction. The resulting model is characterized by the values of the coefficients in the equation, i.e., $\beta_0, \beta_1, \beta_2,..\beta_n$.




\subsubsection{Variables in the model}
\label{sec:var_selection}
For our model building purpose, we explore various galaxy properties and we feed different combinations of them into the linear regression method. 

Here we list the properties of galaxies that can be considered as \textit{x-}variables or known variables and ones that are response or \textit{y} variables. 
\begin{enumerate}
    \item $M_{Gas}$ - Total gas mass of the halo. The gas mass is calculated by summing up the mass of all the gas cells inside halo radius. In our sample, $M_{Gas}$ values ranges from $10^{3.2}$ - $10^{9.7} \Msun$.
    
    \item $M_{\star}$ - Total stellar mass within $0.3 R_{vir}$ of the halo. 
    

    \item Galaxy $R_{vir}$ - Virial radius of the main galaxy associated with halo. The median radius is $\sim 0.3$ kpc (median halo $R_{vir}$ is 3.9 kpc).
    
    \item SFR$_{10}$ - Star formation rate of the halo averaged over last 10 Myr. 
    
    \item SFR$_{100}$ - SFR of the halo averaged over last 100 Myr.
    
    \item $\tau_\star$ - Mass-weighted mean stellar age of all stellar populations within $30\%$ of the halo virial radius (median age $\sim 102$ Myr).
    
    \item $Z_{\star}$ - Mass-weighted metallicity of stars within $30\%$ of the halo virial radius. 
    
    \item $Z_{gas}$ - Mass-weighted metallicity of gas within the halo virial radius.
    
    \item \Llyain - Intrinsic \lya Luminosity. 
    
    \item \Llyaout - Escaping \lya Luminosity. 
    
    \item \flya - \lya Escape fraction, defined as the ratio of the escaping and intrinsic \lya luminosity.
    
    \item \Llycin - Intrinsic ionizing Luminosity. 
    
    \item \Llycout - Escaping ionizing Luminosity. 
    
    \item \flyc - LyC Escape fraction.
\end{enumerate}

We show the histogram of these variables for our sample of galaxies used in building multivariate models in Figure ~\ref{fig:hist_pred_prop}.

\subsubsection{Preparing the data} 
\label{sec:preparing_data}
When we use multivariate methods for constructing a predictive model, it is important that all variables involved in the model have the same order of magnitude. 
However, standardizing the measurement scales has no impact on the validation and interpretation of the models. Data standardizing comprises of various techniques, for example, z-score standardization where if the data is Gaussian it is shifted so that the new dataset is centered around 0 and has a standard deviation of 1 ($z = \frac{x - \mu}{\sigma}$, where z = new data, \textit{x} = old data, $\mu=<x>$ and $\sigma$ = standard deviation of x) or min-max standardization where the data is scaled between 0  to 1 ($z = \frac{x - x_{min}}{x_{max} - x_{min}}$). In our analysis, not all of the galaxy properties have a Gaussian distribution (as can also be seen from Figure~\ref{fig:compare_stacked_z6}). More importantly, our variables typically cover many orders of magnitudes in range. So for standardizing our data, we first take logarithmic values of all variables and then subtract the median value from them to center them. So for any variable $x$ we scale it to $x_{\rm scaled}$ or $x_s$ by,
\begin{equation}
 x_{s} = \mathrm{log}(x) - \mathrm{median}(\mathrm{log}(x)) 
 \label{eq:standardization}
\end{equation}

The next steps for constructing the model are carried out with these scaled variables (equation \ref{eq:multivar} will be applied on scaled variables for building the models). The variables we have plotted in Figure~\ref{fig:lycout_vs_all} (and \ref{fig:Lin_vs_everything}) and discussed in $\S$\ref{sec:appl_to_sphinx} are these scaled variables.

\subsubsection{Estimating the quality of the fit}
\label{sec:r2def}
There are several metrics that can be used to quantify how suitable the model is or how well it fits the data. A popular statistical metric for the multivariate regression model is the $R^2$ which is a measure of how much of the response variance is explained by the model, i.e. the linear combination of the predictors. It is mathematically defined as:
\begin{equation}
    R^2 = 1 - \frac{\Sigma(y_i - f_i)^2}{\Sigma(y_i - \overline{y})^2}
\end{equation}
where $y_i$ is the actual \textit{y} value, i.e. \textit{y} value from our simulation of i-th halo, $\overline{y}$ is the mean value of these \textit{y} values, and $f_i$ is the predicted value for the i-th halo computed using the model. $R^2 = 0$ means that the model explains no response variance and $R^2 = 1$ means that the model explains all the response variance, i.e. it can predict \textit{y} exactly. So the closer the $R^2$ value is to 1, the better the model. 

Although $R^2$ is a widely used metric of model performance, it should be noted that the value of $R^2$ always increases, however slightly, when more and more variables are added to the model. Therefore, in models where the number of \textit{x-}variables is large, $R^2$ may slightly overestimate the model performance. To ensure that our metric does not depend on the number of x-variables, we define the adjusted \radj as,
\begin{equation}
    \radj = 1 - (1-R^2)\frac{n-1}{n-p-1}
\end{equation}
where $n$ is the number of data points (galaxies) and $p$ is the number of \textit{x-}variables in the model \citep[see e.g.][]{feigelson_babu_2012}.
The adjusted $R^2$ increases only when the addition of a \textit{x-}variable increases the $R^2$ more than it would just by chance.
The value of \radj will always be equal to or less than $R^2$.  
From here onward, whenever we mention $R^2$ and its values, either in text or in figures, we mean the \radj, unless otherwise specified. 

\subsubsection{Finding the most important predictors}
\label{sec:rankingmethod}
We perform a stepwise forward and backward selection method to determine which $x$ variables are the most important for predicting y. In forward selection, the model takes the \textit{x-}variables one by one, and inspects which of them lead to the largest value of $R^2$ by itself and classifies that as the most important \textit{x-}variable (rank 1). Then the model adds each of the remaining \textit{x-}variables one by one to rank 1, and the variable that produces the largest increase in $R^2$ value is the second most important \textit{x-}variable (rank 2). This continues until all the variables have been added and a ranked choice of \textit{x-}variables has been made. In the backward selection method, the model starts with all \textit{x-}variables and then determines which one variable removal decreases the value of $R^2$ the least, this is least important variable. The process continues until all but one variable have been removed and a ranking has been generated. We use both methods on our data-set.


\subsubsection{Validating the models}
\label{sec:cross_valid_method}
After building the regression models it is important to estimate the performance of the model on various datasets. 
To do so, we use repeated k-fold cross validation method to test the model performance. First the entire dataset is randomly divided in k subsets, where the number k (typically 5 or 10) can be specified. Then we reserve one subset as test data and estimate the model using the rest of the subsets which act as training data. We then use this estimated model on the test data in order to calculate the fit/error indicators (which can be \radj or root-mean-square-estimate or mean absolute error). We repeat this process k-times and ensure that each of the subset has acted as the test dataset once. Then we calculate the average of these indicators from these k measurements of errors. This whole process of dividing into test-train datasets and computing the average indicators is then performed multiple times and finally we average all the indicators corresponding to each model and compare it with the \radj of the full model.

\subsection{Application to SPHINX galaxies}
\label{sec:appl_to_sphinx}



\begin{figure*}[h]
    \centering
    \includegraphics[width=0.9\textwidth]{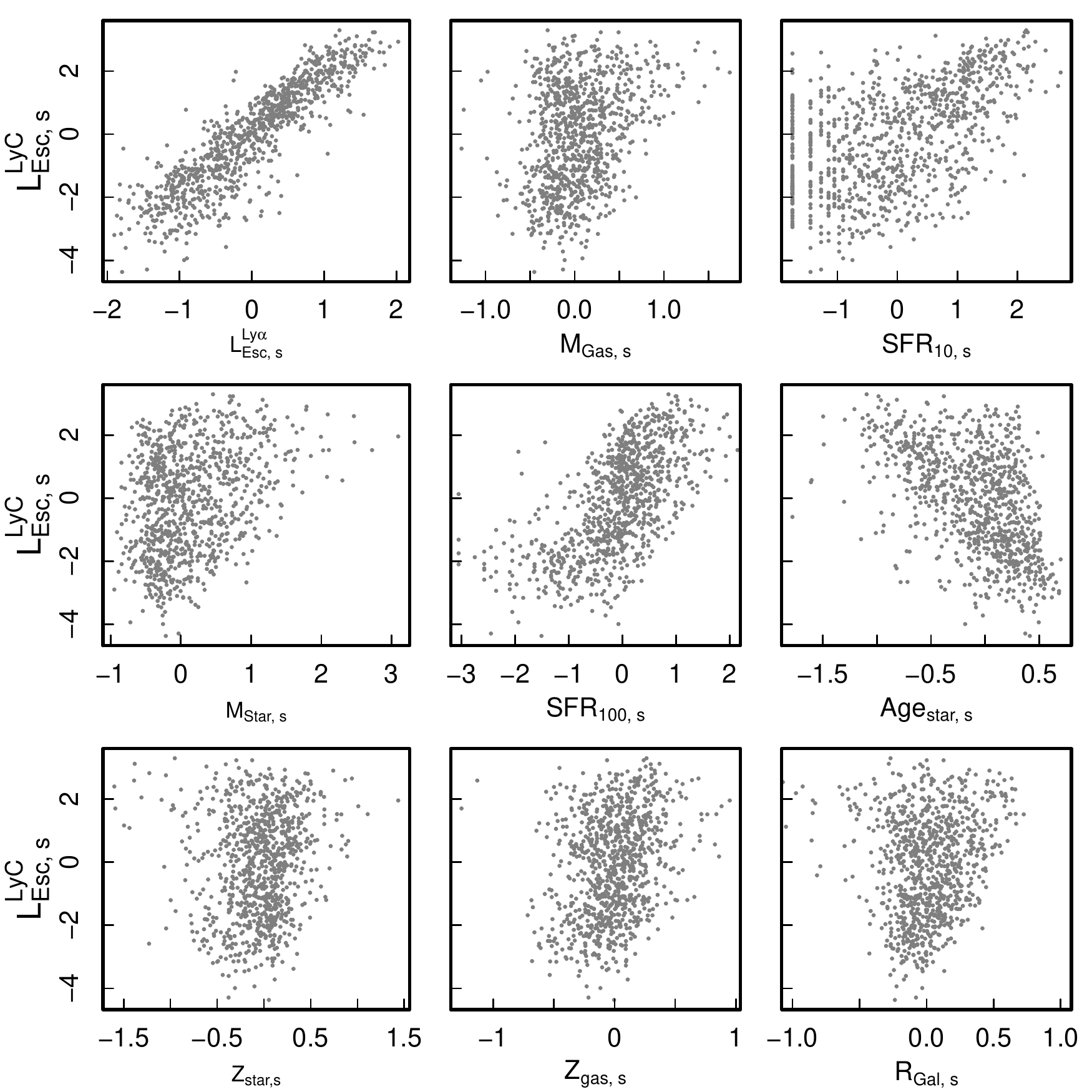}
    \caption{LyC escaping luminosity vs. each of the 9 galaxy variables. All variables plotted here are scaled (as denoted by the subscript \textit{s}) using equation \ref{eq:standardization}, as described in $\S$~\ref{sec:preparing_data}. The particular definitions of the parameters are as follows: $L_{\rm{Esc,s}}^{\rm{LyC}} = \rm{log}(\Llycout/\rm{erg\,s^{-1}}) - 39.25$, 
    $L_{\rm{Esc,s}}^{\lya} = \rm{log}(\Llyaout/\rm{erg\,s^{-1}}) - 40.11$,
    $M_{\rm{Gas,s}} = \rm{log}(M_{\rm{Gas}}/\Msun) - 8.01$, 
    $SFR_{\rm{10,s}} = \rm{log}(\rm SFR_{\rm{10}}/\Msun\,yr^{-1}) + 2.25$, 
    $M_{\rm{Star,s}} = \rm{log}(M_{\rm{Star}}/\Msun) - 6.03$, 
    $SFR_{\rm{100,s}} = \rm{log}(\rm SFR_{\rm{100}}/\Msun\,yr^{-1}) - 0.01$, 
    ${\rm{Age}_{\rm{Star,s}} = \rm{log}(Age_{\rm{Star}}/Myr) - 2.01}$, 
    ${\rm{Z}_{\rm{star,s}} = \rm{log}(Z_{\rm{star}}/Z_{\odot}) + 3.59}$, 
    ${\rm{Z}_{\rm{gas,s}} = \rm{log}(Z_{\rm{gas}}/Z_{\odot}) + 3.56}$, and 
    ${R_{\rm{Gal,s}} = \rm{log}(R_{vir}/kpc)- 0.29}$.}
    \label{fig:lycout_vs_all}
\end{figure*}

From equation~\ref{eq:multivar},  we can deduce that the multivariate linear model is suitable if some (or all) \textit{x-}variables individually vary linearly with y, i.e. at least for some variables $y \propto x_n$. If none of the \textit{x-}variables have any linear correlation with y, it is unlikely that a linear combination of them can determine y. Therefore, 
we first explore if individual correlation between \textit{y} and any \textit{x-}variable exists. 


Such an exploratory plot in shown in Figure~\ref{fig:lycout_vs_all} where we plot the response variable \Llycout vs each of the galaxy properties. 
From this figure we find that \Llycout correlates well with \Llyaout and SFR, along with some other weaker correlations. Similar plot for \Llycin and \flyc is provided in appendix (Figure~\ref{fig:Lin_vs_everything} and \ref{fig:flyc_allx}) where we see that \Llycin is strongly correlated to SFR and \Llyaout and weakly correlated to mass and stellar age and \flyc is correlated to \Llyaout.  This preliminary inspection shows that a multivariate linear regression can be a good model for predicting LyC.

\subsubsection{Sample Selection}
\label{sec:gal_selection}
Before we delve into regression modelling, we examine the galaxy dataset to select a galaxy sample that can be used for building the model. The initial dataset contains 1933 galaxies, which is the sample of all galaxies with stellar mass $\geq 10^6\Msun$ at z = 6, 7, 8, 9, and 10 ($\S$\ref{sec:sample}). 



As discussed above and from Figure~\ref{fig:Lin_vs_everything} it is clear that the star formation rate, especially recent (over last 10 Myr) SFR or SFR$_{10}$, has a strong linear correlation with intrinsic LyC luminosity, and it is also correlated well with the escaping luminosity of LyC (Figure~\ref{fig:lycout_vs_all}). This is also expected from theoretical studies \citep{Stanway2016, Raiter2010, Schaerer2003, Partridge1967} that show star formation is the main driver for the production of both \lya and LyC photons. In our dataset, there are some galaxies (943 out of 1933, most of which are faint LAEs) which have no recent star formation, i.e., the average star formation rate over the last 10 Myr, SFR$_{10} = 0$ (Figure~\ref{fig:lya_lyc_prop_hist}, right panel). So for building our models, we exclude these non star-forming galaxies and with this criterion, there are 990 galaxies left in our dataset.

Next we investigate this modified dataset (galaxies with non-zero SFR) for any significant outliers. 
We find that there are some clear outliers in the distribution of \flyc with values as low as $10^{-20}$. 
We remove galaxies with $\flyc < 10^{-6}$ from the dataset, after which the \flyc distribution is free of outliers. This leads to a dataset of 940 galaxies. 
We find no significant outliers in other galaxy properties.
Incidentally, we note that all of the galaxies in this final dataset of 940 galaxies have both intrinsic and escaping \lya luminosities $> 10^{38}$ erg/s.

\begin{table*}[!htb]
\caption{\radj for predicting different variables with different models. Here Galaxy Properties (GP) refers to physical properties of the galaxies described in \S~\ref{sec:var_selection} (i.e. items 1 - 7).}
   \vspace{-0.5 cm}
   \caption*{}
    \centering
    \begin{tabular}{|c|c|c|c|c|c|c|}
    \hline
      Model & 
      \Llyain & \Llyaout & \flya &
      \Llycin & \Llycout & \flyc \\
      \hline
      \hline

       \makecell[l]{1. GP (Galaxy Properties)} 
       & 0.8758	&	0.7061	&	0.2812  &	0.8665	&	0.5336	&	0.2631\\
       \makecell[l]{2. GP + \Llyaout} 
       & 0.9031	&	NA		&   0.6886	&   0.8969	&	0.8516	&	0.6561\\
        \hline
    \end{tabular}
    \label{tab:r2table}
\end{table*}



\subsubsection{Building the models and the most important variables}
\label{sec:model_details}

Our main goal is to predict the LyC luminosities and \flyc using other properties. However, for many galaxies at high redshift the observation of \lya luminosity can also be difficult, owing to increasing IGM opacity. Moreover, estimation of the intrinsic \lya luminosity and hence \flya is also challenging at all redshifts, as these are not observables and must be derived using stellar models which can have many underlying assumptions. So it can be useful to also build models for predicting these \lya emissivities which may complement existing methods.

Therefore, we explore the full predictive power of multiple linear regression models with our dataset and we aim to build models to predict the following 6 quantities:
\begin{itemize}
    \item \Llycin, \Llycout, and \flyc,
    \item \Llyain, \Llyaout and \flya.
\end{itemize}

We investigate several combinations of physical parameters that we can access in the simulation to build a good predictive model. We calculate the performances of these models using the metric \radj and our most relevant model results are summarized in Table~\ref{tab:r2table}.

\textbf{Model 1} - In Model 1, as predictors we supply all physical galaxy properties (GP), i.e. items 1 - 8 from our list in $\S$ \ref{sec:var_selection}, namely gas mass, stellar mass, galaxy radius, SFR$_{10}$, SFR$_{100}$, stellar age, stellar and gas metallicity. We find that given only the physical properties of galaxies, we can predict the  intrinsic LyC luminosity quite accurately ($R^2 = 0.87$) but the emerging luminosity and the escape fraction cannot be modeled very well ($R^2 = 0.53\, \rm and\, 0.26$ respectively). Conversely, both \lya intrinsic ($R^2=0.88$) and escaping ($R^2=0.71$) luminosities can be predicted quite well with galaxy properties.

\begin{figure*}[h]
    \centering
    \includegraphics[width=0.45\textwidth, trim= {0 0 0 40}, clip ]{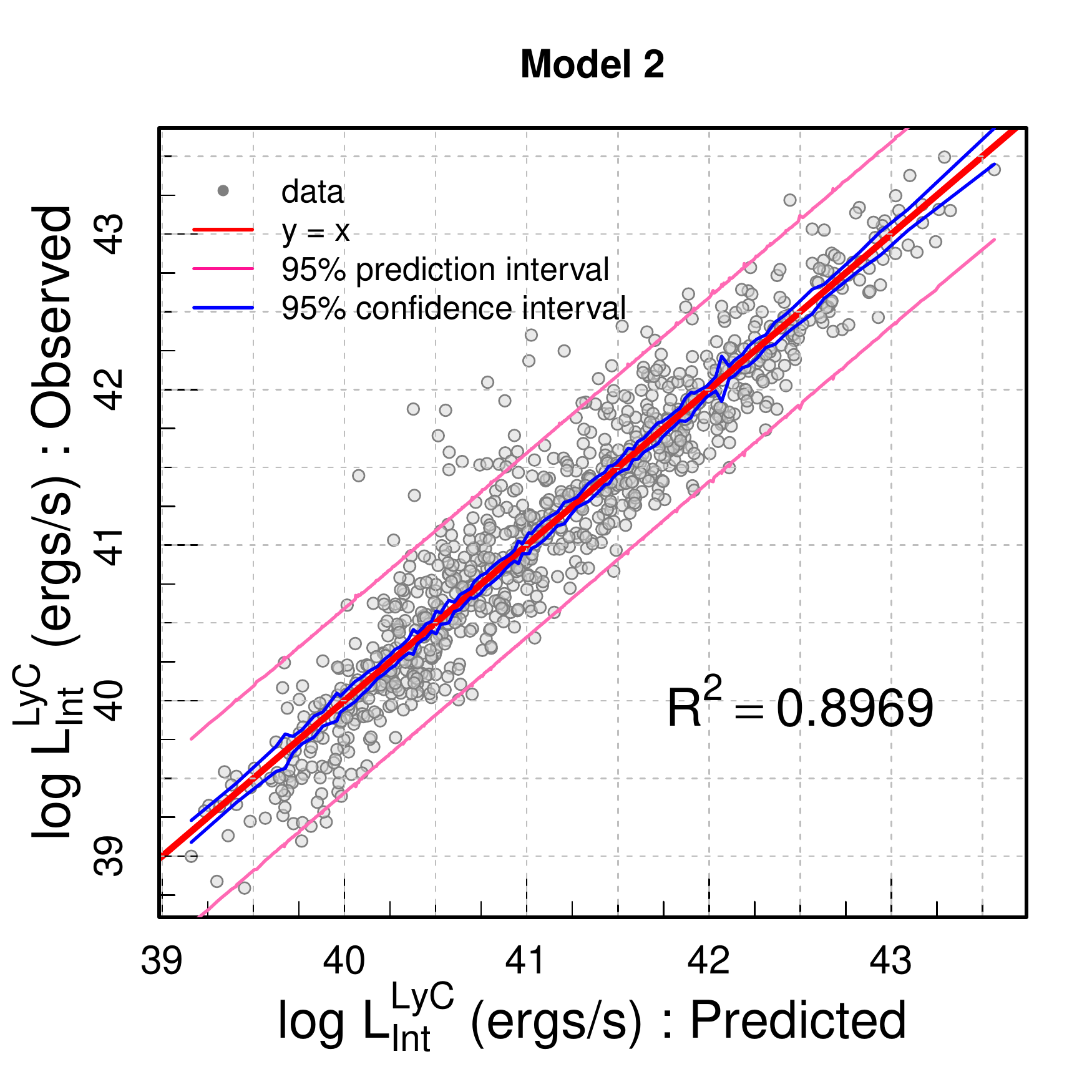}
    \includegraphics[width=0.45\textwidth, trim= {0 0 0 40}, clip]{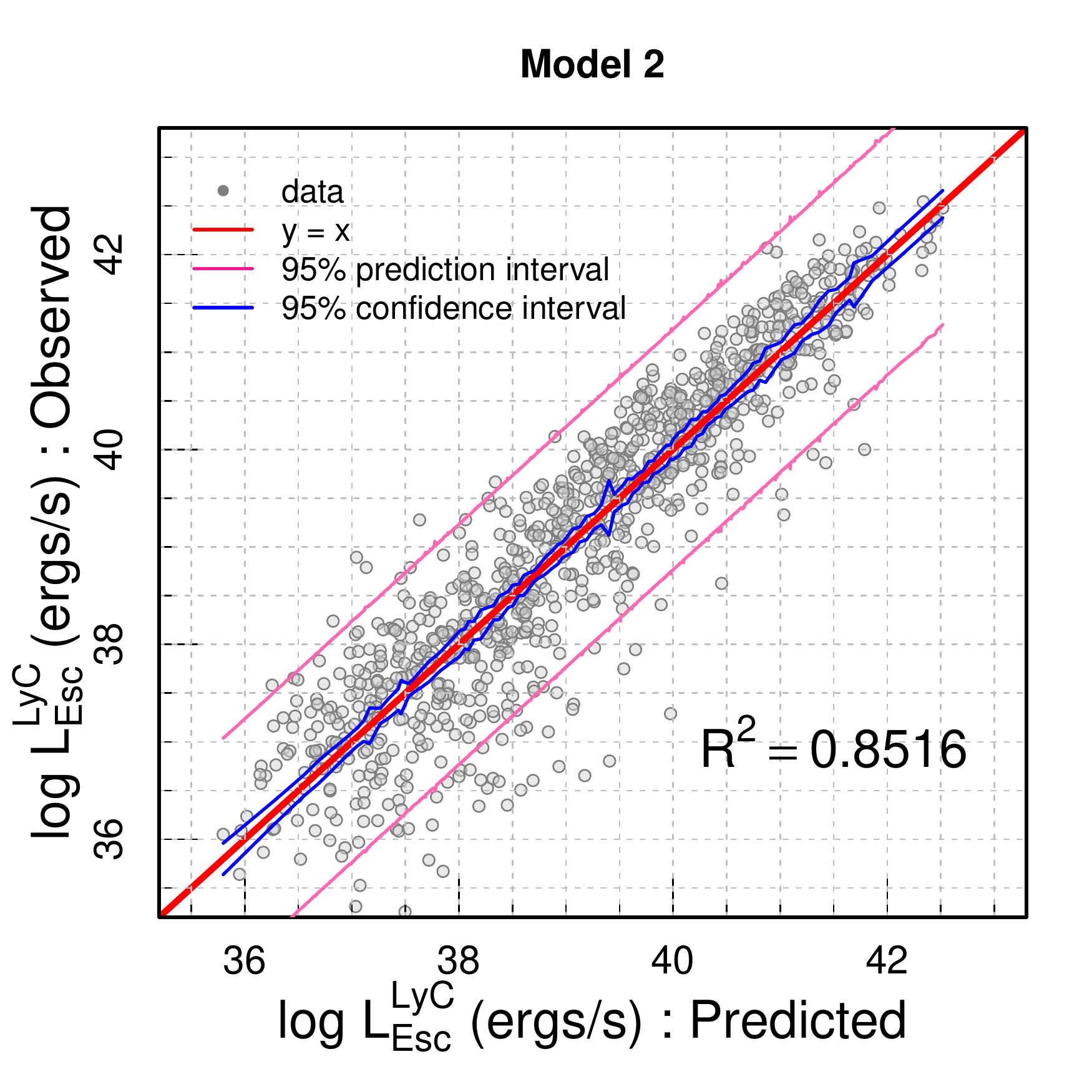}\\
    \includegraphics[width=0.45\textwidth, trim= {0 0 0 40}, clip]{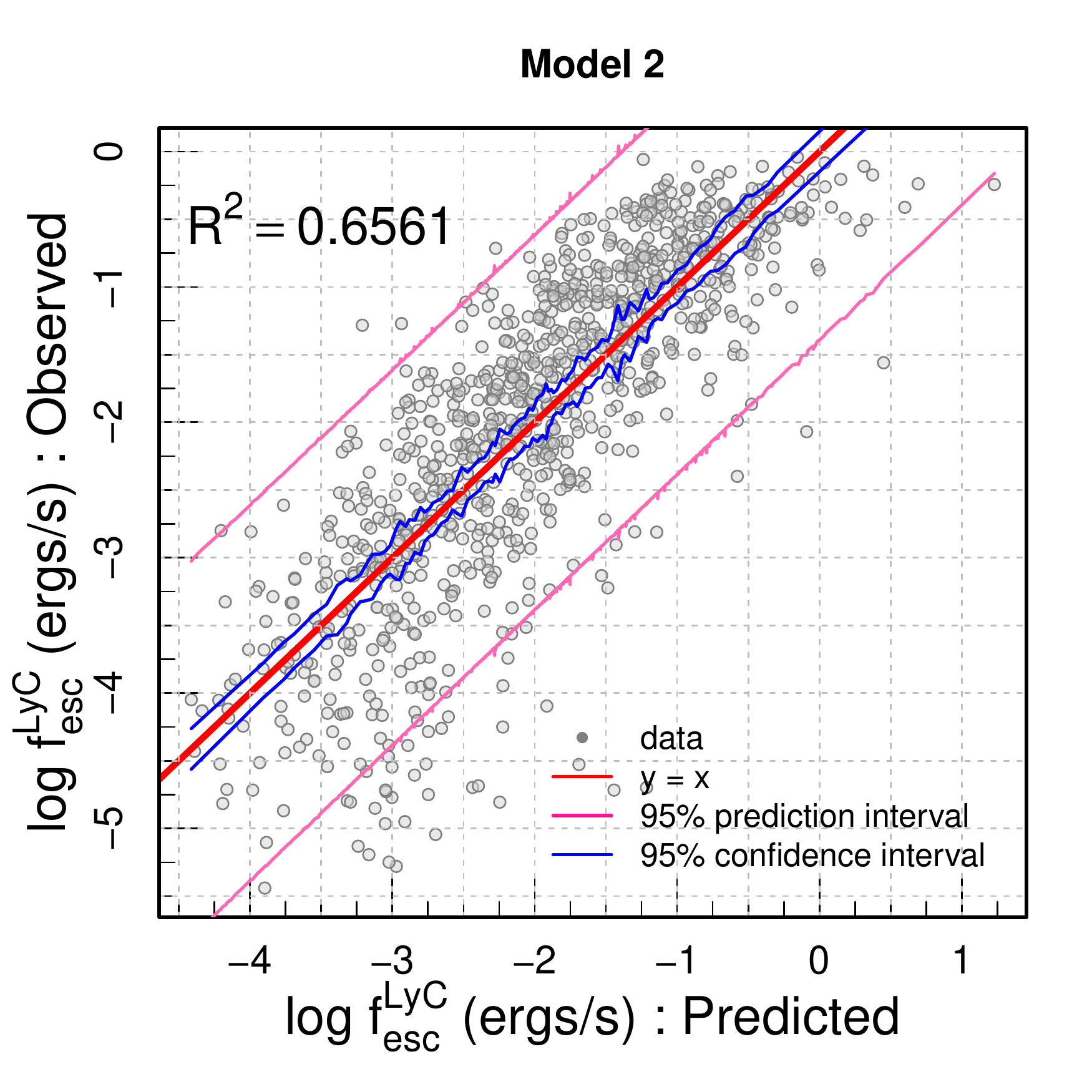}
    \caption{Prediction of intrinsic luminosity, escaping luminosity and escape fraction of LyC from Model 2, where the input variables are the physical galaxy properties and the escaping \lya luminosity.
    The $R^2$ value for each fitting is noted in the plots. The red lines show the 1:1 correlation or y=x line. The pink lines show the $95\%$ prediction interval and the blue lines show the $95\%$ confidence interval.}
    \label{fig:predict_lyc}
\end{figure*}

\textbf{Model 2} - When we add \lya escaping luminosity to our input list of predictors, (Model 2 in Table~\ref{tab:r2table}) we find that in addition to the intrinsic luminosities, now the LyC escaping luminosity is also predicted with high accuracy, with $R^2 = 0.85$. The average error (root mean square error, RMSE, is the average difference between the predicted and actual value) in predicting the \Llycout is  a factor of $\sim 4$ (RMSE = 0.62 in log scale, Figure~\ref{fig:predict_lyc}). Both \flya and \flyc are also fairly well predicted with this model, with an $R^2$ value of 0.69 and 0.64 respectively.

We consider Model 2 as our fiducial model and we show the predicted intrinsic and escaping LyC luminosities and \flyc from Model 2 in Figure~\ref{fig:predict_lyc} against the observed values from the simulation. 
In each of these plots we also show the $95\%$ confidence interval and the $95\%$ prediction interval. The confidence interval signifies that given a set of predictor values, i.e. \textit{x-}values, the mean of the response variable or y, will fall within this interval with $95\%$ confidence. On the other hand, the prediction interval tells us where the next individual \textit{y} value will fall. Given a set of \textit{x-}values, an individual \textit{y} value will fall within the predictor interval with $95\%$ confidence. The prediction interval accounts for both the uncertainty of the estimation of population mean as well as the variation of the individual y-values. Hence, the predictor interval is always wider than the confidence interval. 
We see in Figure~\ref{fig:predict_lyc} that most of the observed (in our simulation) values of \textit{y} do indeed lie within the $95\%$ predictor interval of our model. 


Figure~\ref{fig:predict_lyc} shows both intrinsic and escaping luminosities are well predicted. 
We give here the equation for predicting \Llycout obtained using this model:


\begin{equation}
    \begin{array}{l}
        \lten \Llycout = 39.30
        + 2.08\,\lten(\Llyaout/10^{40}) \\
        - 1.11\,\lten(M_{Gas}/10^{8}) 
        + 0.85\,\lten(Z_{Gas}\times10^{3}) \\
        - 0.20\,\lten(SFR_{10}\times10^{2}) 
        - 0.21\,\lten(Z_{\star}\times10^{3}) \\
        + 0.16\,\lten(M_\star/10^{6}) 
        - 0.16\,\lten(\mathrm{Age_{\star}}/10^{2})
    \end{array}
    \label{eq:lyc_model2}
\end{equation}

Here the luminosity is in erg/s, mass is in \Msun, SFR unit is \Msun/yr, stellar age is in Myr and metallicity unit is solar metallicity.

Here we note that in our models we have included both the gas mass and the gas metallicity. Since the dust content is modelled by using these factors (as described in \S~\ref{subsec:rascascode}), including the dust in addition to the other parameters does not give us additional information (we tested this and found that inclusion of dust changes the \radj by less than 0.01$\%$).

\textbf{Most important variables:}
\label{sec:most_imp}
In the models described above, we have supplied 7 or 8 galaxy properties for predicting various \lya and LyC quantities. However, observing and determining many galaxy properties at high redshift can be extremely challenging. Thus it is necessary to identify which of the \textit{x-}variables is the most important in predicting $y$. Here we discuss the ranking of most important predictors in the context of Model 2 and the response variables \Llycin, \Llycout and \flyc. 

We present the results of the ranking process described in \S~\ref{sec:rankingmethod} in Table~\ref{table:ranking}, listing the most important variables with their ranks and their \radj value. 
The \radj value associated with the n-th rank variable is the \radj the model produces including the first to n-th rank variable. The adjusted $R^2$ increases only when the addition of a \textit{x-}variable increases the $R^2$ more than it would be just by chance, otherwise it actually decreases with variable addition ($\S$\ref{sec:r2def}). In a ranking table, such as Table~\ref{table:ranking}, when the adjusted $R^2$ reaches a peak value, the model has reached its best predictive power. We perform both stepwise forward selection and backward selection for the ranking ($\S$\ref{sec:rankingmethod}), and find that both processes give the same ranking in all cases, which suggests that our ranking is stable.



We find that 88\% of the variance in \Llycin can be explained if we only use SFR$_{10}$ and \Llyaout, such that these two values alone can provide a reliable prediction of the intrinsic ionsing power of galaxies. For escaping LyC, knowing the escaping \lya luminosity is the most important factor and combining this with gas mass, gas metallicity and SFR$_{10}$ can account for 85$\%$ of the variance. Lastly, the most important three predictors of \flyc are \Llyaout, SFR$_{10}$ and gas mass, as these three can explain $63\%$ of the response variance. In the case of \flyc, variables with rank 1 - 6 increase the \radj (up to 0.6569), but the addition of more properties decreases the model performance. Similarly, we find that in models for predicting \Llycin, galaxy radius (rank 9) and for predicting \Llycout, SFR$_{100}$ and radius (rank 8 and 9) are not important.

\subsubsection{Minimal model}
\label{sec:minimal_model}

Going one step further, we note that it is extremely difficult to observe gas properties in reionization era galaxies. Among the rest of the predictors used in our models so far, the observed LCEs we have discussed in \S~\ref{sec:obs_lces} and listed in Table~\ref{tab:observations} have only 3 predictors available, namely \Llyaout, SFR$_{10}$ and M$_\star$. It is now interesting to explore if a model built with only these 3 predictors can predict LyC quantities. We build a minimal model with these three predictors only (Model 3) and list the resulting model performances in Table~\ref{tab:r2table_model3}. We find that here also \Llycout is predicted with a high accuracy, $R^2 = 0.80$ and the average error is a factor of $\rm RMSE \sim$ 5.24. The intrinsic luminosities are also predicted very well, with fair performances for escape fractions. The equation for \Llycout we get with this model is:


\begin{equation}
    \begin{array}{l}
        \lten \Llycout = 38.94 \\
        + 2.03\,\lten(\Llyaout/10^{40}) \\
        - 0.15\,\lten(M_\star/10^{6}) \\
        - 0.23\,\lten(SFR_{10}\times10^{2}) \\
    \end{array}
    \label{eq:lyc_model_minimal}
\end{equation}


The equation for \Llycin from this model can be written as:

\begin{equation}
    \begin{array}{l}
        \lten \Llycin = 40.96 \\
        + 0.49\,\lten(\Llyaout/10^{40}) \\
        - 0.08\,\lten(M_\star/10^{6}) \\
        - 0.49\,\lten(SFR_{10}\times10^{2}) \\
    \end{array}
    \label{eq:lyc_int_model_minimal}
\end{equation}


The units of the quantities are the same as described in \S~\ref{sec:model_details}. The ranking of the most important predictors for \Llycout with this minimal model is shown in Table~\ref{table:ranking_model3}, where we find that \Llyaout is has rank 1, followed by SFR$_{10}$ and $M_\star$, same results as we found with Model 2 (\S~\ref{sec:most_imp}) also. 

\subsection{Fitting the model to observed data}
\label{sec:fit_to_obs}

Finally, we explore if such a model can be fitted to real observed data. In \S~\ref{sec:lyalyc_rel_z6} we have listed the properties of known \lya and LyC emitters in Table~\ref{tab:observations}. These galaxies have observations of their stellar mass, star formation rate, \lya luminosity and their LyC luminosity at 900\AA. As discussed in \S~\ref{sec:lyalyc_lum} and shown in Figure~\ref{fig:compare_obs_prop} these observed LCEs are more massive, have higher SFR and they are brighter in \lya and LyC than the SPHINX galaxies. They are also observed at low redshifts, $z \leq 0.45$ whereas the SPHINX galaxies are at z = 6 - 10. The simulated luminosities and escape fractions are angle-averaged quantities whereas observations are, of course, directional (further discussion in \S~\ref{sec:discuss}). Nevertheless, this is the only observed sample we currently have with both LyC and \lya observations, so we evaluate our predictive model on these galaxies.

\begin{figure}[h]
    \centering
    \includegraphics[width=0.45\textwidth]{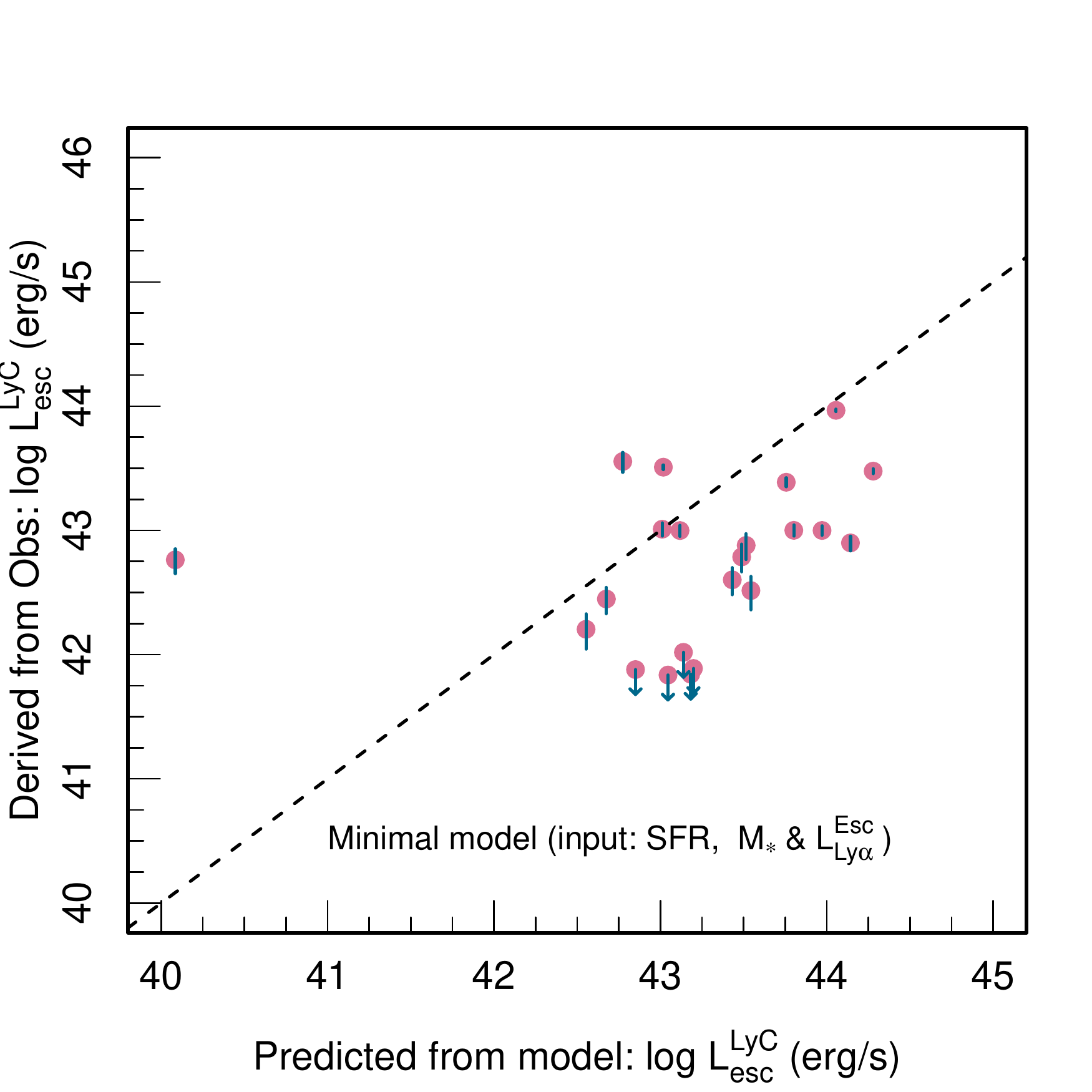}
    \caption{Predicted \Llycout from the minimal model (model 3, described in \S~\ref{sec:minimal_model})
    vs the \Llycout derived from observations for the observed LCEs listed in Table~\ref{tab:observations}. 
    The values in the \textit{y} axis are derived by multiplying the observed $L_{esc}^{LyC,900}$ with the ratio \Llycout/\Llycnineout derived from our simulation (\S~\ref{sec:lycmethod}). The observational luminosity error bars are shown in skyblue. For few galaxies the observed \Llycout is an upper limit, these are marked by skyblue arrows.}
    \label{fig:lyclumout_obs_predic}
\end{figure}

Since the observed galaxies have only 3 predictors available, we use our our minimal model (model 3) described in \S~\ref{sec:minimal_model} and use Equation~\ref{eq:lyc_model_minimal} for predicting the \Llycout of these observed LCEs.
In Figure~\ref{fig:lyclumout_obs_predic} we show the predicted \Llycout from this model vs. the LyC luminosity that is derived from observations of \Llycnineout (by multiplying the observed \Llycnineout with a factor of 1434, the median value of the ratios \Llycout/\Llycnineout derived from our simulation, \S~\ref{sec:lyalyc_lum}). We find that the predicted luminosities are generally close to the observed values.  
In some cases the model over predicts the escaping luminosity, probably due to the differences in the physical properties between these observations and the SPHINX galaxy sample. Models performs best when the given input properties are inside the range of the training data (the ranges of properties for our SPHINX sample are shown in Figure~\ref{fig:hist_pred_prop}), otherwise it needs to be extrapolated. The outlier with low predicted LyC is the galaxy Haro 11 which is located at z = 0.021, much closer than other observations which may affect the galaxy properties.

%





\begin{table*}[!htb]
    \caption{Most important variables for predicting LyC luminosities and escape fractions using Model 2 (${GP + \Llyaout}$)}
    \vspace{-0.5 cm}
    \begin{minipage}{.32\linewidth}
      \centering
        \caption*{}
        \begin{tabular}{|c|c|c|}
        \hline
        \multicolumn{3}{|c|}{{\Llycin}} \\
        \hline
        Rank  & Variable &  Adjusted ${R^2}$  \\
        \hline
          1 & SFR${_{10}}$ & 0.7762 \\
          2 & {\Llyaout} & 0.8856 \\
          3 & SFR${_{100}}$ & 0.8911 \\
          4 & Z${_{Gas}}$ & 0.8923 \\
          5 & ${M_{Gas}}$ & 0.8937 \\
          6 & Stellar Age & 0.8954 \\
          7 & ${M_\star}$ & 0.8968 \\
          8 & Z${_{\star}}$ & 0.8969 \\
          9 & ${R_{Gal}}$ & 0.8969 \\
          \hline
          \end{tabular}
    \end{minipage} 
     \begin{minipage}{.32\linewidth}
      \centering
        \caption*{}
        \begin{tabular}{|c|c|c|}
        \hline
        \multicolumn{3}{|c|}{{\Llycout}} \\
        \hline
        Rank  & Variable &  Adjusted ${R^2}$  \\
        \hline
          1 & {\Llyaout} & 0.7877 \\
          2 & ${M_{Gas}}$ & 0.8243 \\
          3 & Z${_{Gas}}$ & 0.8401 \\
          4 & SFR${_{10}}$ & 0.8493 \\
          5 & Z${_{\star}}$ & 0.8507 \\
          6 & ${M_\star}$  & 0.8511 \\
          7 & Stellar Age & 0.8519 \\
          8 & SFR${_{100}}$ & 0.8517 \\
          9 & ${R_{Gal}}$ & 0.8516 \\
          \hline
          \end{tabular}
    \end{minipage}
    \begin{minipage}{.32\linewidth}
      \centering
        \caption*{}
        \begin{tabular}{|c|c|c|}
        \hline
        \multicolumn{3}{|c|}{{\flyc}} \\
        \hline
        Rank  & Variable &  Adjusted ${R^2}$  \\
        \hline
          1 & {\Llyaout} & 0.2983 \\
          2 & SFR${_{10}}$ & 0.5397 \\
          3 & ${M_{Gas}}$ & 0.6269 \\
          4 & Z${_{Gas}}$ & 0.6545 \\
          5 & Z${_\star}$ & 0.6563 \\
          6 & SFR${_{100}}$ & 0.6569 \\
          7 & Stellar Age & 0.6568 \\
          8 & M${_\star}$ & 0.6565 \\
          9 & ${R_{Gal}}$  & 0.6561 \\
          \hline
          \end{tabular}
    \end{minipage} 
    \label{table:ranking}
\end{table*}

\begin{table*}[!htb]
\caption{$R^2$ for predicting different variables with the minimal model (Model 3)}
   \vspace{-0.5 cm}
   \caption*{}
    \centering
    \begin{tabular}{|c|c|c|c|c|c|c|}
    \hline
      Model & 
      \Llyain & \Llyaout & \flya &
      \Llycin & \Llycout & \flyc \\
      \hline
      \hline
       \makecell[l]{3. $M_{\star} + SFR_{10}$ + \Llyaout }
       & 0.8827	&	NA		&  0.6242	&   0.8877	&   0.8030	&	0.5498\\
       \hline
    \end{tabular}
    \label{tab:r2table_model3}
\end{table*}

\begin{table}[!htb]
    \caption{Most important variables for predicting \Llycout with  Minimal Model (Model 3)}
    \vspace{-0.5 cm}
      \centering
        \caption*{}
        \begin{tabular}{|c|c|c|}
        \hline
        \multicolumn{3}{|c|}{\Llycout} \\
        \hline
        Rank  & Variable &  Adjusted $R^2$  \\
        \hline
          1 & \Llyaout & 0.7877 \\
          2 & SFR$_{10}$ & 0.8007 \\
          3 & $M_\star$ &  0.8030 \\
          \hline
          \end{tabular}
  \label{table:ranking_model3}
\end{table}
 

\subsubsection{Cross Validation of the models}
\label{sec:cross_valid}

We have built these models using all 940 eligible galaxies available in our simulation dataset. To check the model validity, we need to estimate the accuracy of these models when applied on other, new data, that is not part of the dataset used in building the models. The most straightforward way to do this is to apply this model to other new datasets where all of our desired input and output variables are available in order to readily test the difference between the prediction from models and the actual values. However, such full datasets can only be obtained from high resolution reionization simulations and currently we do not have other datasets. Instead, we can use repeated k-fold cross validation method (described in \ref{sec:cross_valid_method}) to gauge the performance of our models. 


In this work we have used k = 10, so we divide the dataset into 10 random subsets and calculate the average \radj for our response variables. We repeat this process 3 times and get an average of \radj from these runs.
We have calculated the k-fold \radj for each model and found that the \radj from the k-fold test is always very similar to the \radj we calculated when building the model with our whole dataset. For example, when we perform the cross validation for Model 2, for predicting \Llyain, \Llycin and \Llycout we get an average \radj of 0.8996, 0.8945 and 0.8471 respectively, compared to 0.9006, 0.8956 and 0.8466 from our full model, as shown in Table 2. These respective \radj values are very close to each other which shows that the our proposed models are indeed stable.


\section{Discussion}
\label{sec:discuss}

In this study we have explored the relationship between \lya and LyC emission from simulated EoR galaxies and we have shown that it is possible to predict LyC emission of galaxies using their physical and \lya properties. However, there are some important limitations of this study that we discuss below.


\textbf{Limitations of the simulation: }
Our simulation has a box size of 10 Mpc and the most luminous LAE in our sample of 1933 galaxies has a luminosity of $\Llyaout = 1.37\times10^{42}$ erg/s. As we have discussed in \S~\ref{sec:lyalyc_lum} and shown in figures \ref{fig:lya_lyc_prop_hist} and \ref{fig:lya_lyc_lum_inout}, recent observations of MUSE LAEs and low redshift LyC leakers ( table \ref{tab:observations}) are starting to overlap with the brightest end of our sample of simulated galaxies. However, our sample is at $z \geq 6$ and at these very high redshifts, the lower limit of observed \lya luminosity is around $\sim10^{43}$ erg/s, still more luminous than our brightest galaxies. 
These detections are probably not representative of the underlying LAEs populations. Although they may play a central role in reionizing the Universe, as demonstrated by the recent discovery of an extremely bright LCE at z$\sim 3$ \citep{Chaves2021}, the lack of very bright LAEs in our sample prevents us from making quantitative predictions for the contribution of very bright LAEs to reionization. As a consequence, our estimate of the fraction of the ionizing photons budget provided by galaxies with \Llyaout$ > 10^{41}$ erg.s$^{-1}$ in \S~\ref{sec:lyalyc_reion} could well be a lower limit.
In order to directly compare our predictions with observational data and to make better statistical predictions for bright galaxies, we need to analyze more luminous galaxies, for which we need to simulate a larger volume.  
The next generation of SPHINX will simulate a volume eight times larger than in the current study (i.e. 20 cMpc in width), which will include halos with stellar masses (virial masses) up to about $10^{10}\Msun (10^{11}\Msun)$ at z=6.


\textbf{IGM attenuation: }
In this work we have not considered the effects of the IGM absorption. The IGM is an important factor in determining the observability of \lya emission at these high redshifts, because in order to be observable LAEs, \lya must
be transmitted through a partially neutral IGM which can easily scatter \lya photons off the line-of-sight. This can considerably reduce the visibility of LAEs during the EoR, as hinted by the drop of the LAE fraction at z>6 \citep{Schenker2014, Kusakabe2020, Garel2021}.
Our results in this paper depict both \lya and LyC luminosities as they would be observed just outside of the halo virial radius. In practice some correction for IGM can be applied to the data before applying our model to estimate \Llycout of galaxies. 
Furthermore, the absence of IGM absorption has allowed us to compare our simulation results to low redshift observations of LCEs.
For more realistic modeling and direct comparison with high-redshift observations, we need to consider IGM absorption. \cite{Garel2021} predicts that the IGM transmission in Sphinx decreases from a factor of $\sim$ 2 at z=6 to $\sim$ 10 at z=9. 
Nevertheless, this study is the first necessary step to assess the link between \lya and LyC escape from galaxies. Since there are known LAEs at $z > 6$ (e.g. \cite{Meyer2021} and references therein), depending on the topology of the reionization, \lya emission may still go through large ionized bubbles at high redshift \citep{Dijkstra2014, Mason2020, Gronke2020}, and could serve as a tracer for LyC escape from galaxies at the cosmic dawn.

\textbf{Directional variation: }
In this study, we chose to consider global, theoretical, estimates of the \lya and LyC quantities, since they are the quantities which matter to determine ionizing photons budget, and study the process of reionization.

\begin{figure*}
    \centering
    \includegraphics[width=0.45\textwidth]{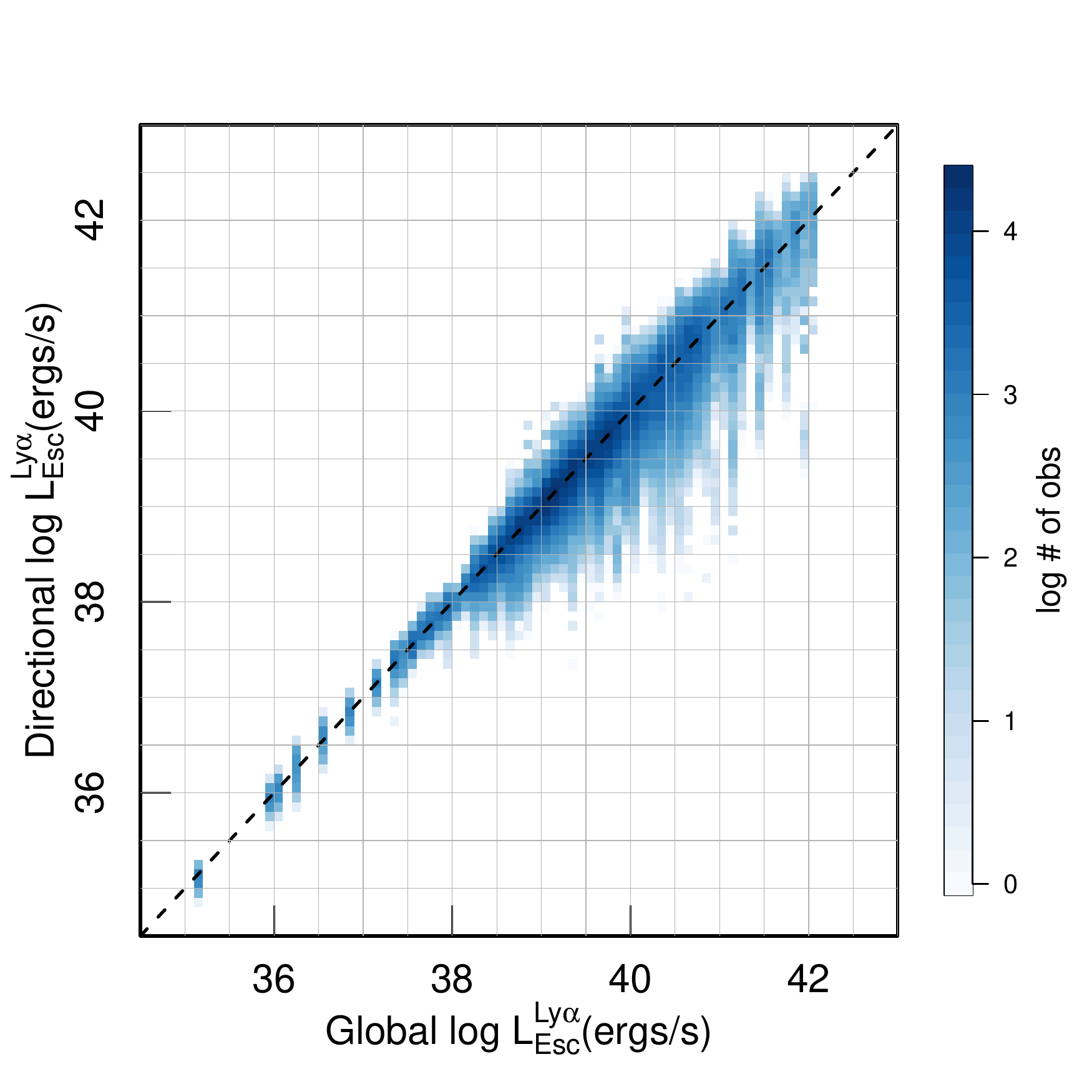}
    \includegraphics[width=0.45\textwidth]{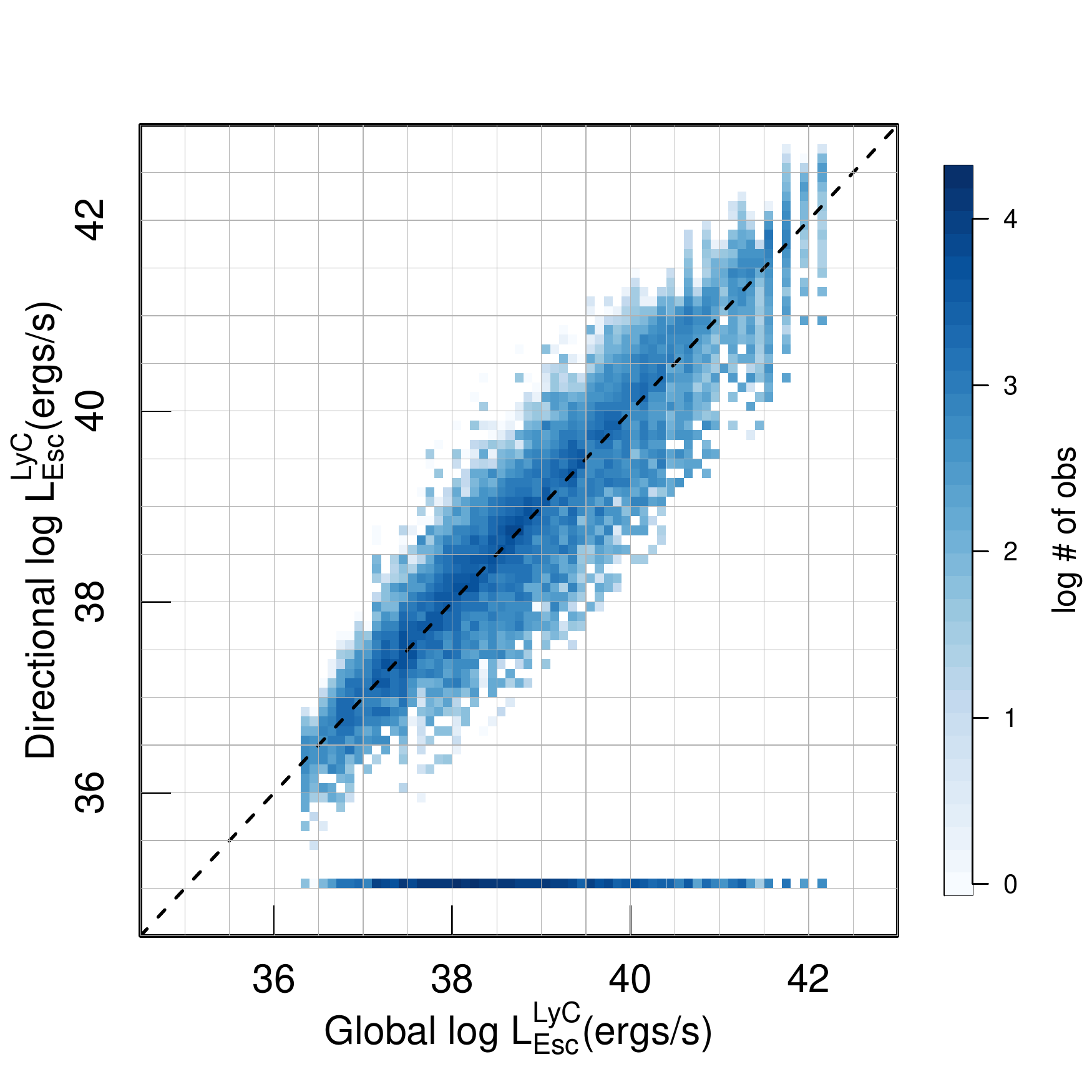}
    \caption{Left: Directional \Llyaout of 1933 sphinx galaxies as a function of their real global \Llyaout. We imagine a sphere around a halo at the halo virial radius and divide the surface area of the sphere into 1728 equal area pixels. We calculate the \Llyaout through each of them and for each pixel direction then we have the directional luminosity, $L_{\rm{esc}}^{\rm{\lya, directional}} = 1728\times L_{\rm{esc}}^{\rm{\lya, pixel}}$. Right: Same plot but for LyC where we show directional \Llycout vs the global \Llycout. The directions with no LyC escape are indicated with an artificial LyC luminosity of $10^{35}$ erg/s.
}
    \label{fig:healpix}
\end{figure*}

However, when we observe galaxies we will, of course, only be able to observe them from one direction (along our line-of-sight). Furthermore, the \lya and LyC luminosities and escape fraction of the same galaxy can differ significantly from direction to direction \citep[Chuniaud et al. in prep]{Cen2015, Mauerhofer2021}. To capture this added complexity, we will need to do directional analysis of our galaxies. As a first attempt to quantify the angular variations of \lya and LyC luminosities escaping from our simulated galaxies, we imagine a sphere around a halo at the halo virial radius and divide the surface area of the sphere into 1728 equal area pixels.  We then count the \lya and LyC photons that escape each of these pixels and calculate the \Llyaout and \Llycout through each of them. For each pixel direction then we have the directional luminosity ($L_{\rm{esc}}^{\rm{\lya \,(or\,LyC), directional}} = 1728\times L_{\rm{esc}}^{\rm{\lya\,(or\,LyC), pixel}}$). In Figure~\ref{fig:healpix} we show the distribution of the directional \lya and LyC luminosities (1728 directions for each galaxy) of the 1933 sphinx galaxies as a function of their actual global luminosities. Interestingly, we find that \lya-bright galaxies can vary up to a factor of $\sim$ 100 compared to their angle-averaged \Llyaout, whereas faint galaxies are more isotropic. On the other hand, the directional LyC luminosities vary quite a lot at all angle-averaged LyC luminosities. As we discussed in \S~3.5, \lya photons can scatter numerous times before escaping, hence they have a higher chance of finding channels in the ISM with low column density, hence their directional distribution is generally more isotropic. Conversely LyC photons generally escape close to the galaxy center where they are mainly produced, so they have lower probabilities of finding many channels, which can result in a more anisotropic distribution of directional luminosities.

The broad variety of Lyman alpha spectral shapes and strengths observed from galaxies is also one of the main probes of strong directional variations. Indeed, several recent observational studies \citep{Verhamme2017, Steidel2018, Izotov2021} have found that spectral features of \lya line profiles, such as high rest-frame equivalent width and a narrow separation between the blue and red peak of \lya spectra correlates positively with escape of LyC. Testing these directional spectral features are beyond the scope of this article. But these two approaches are complementary of each other and we would ideally need both to get a complete picture of the contribution of the galaxies along our line of sight, and globally, to the reionization process. To that end, in the next step, we will employ peeling off algorithms on our galaxies and observe them from several directions. Then we can build mock observations to compare directly with existing and future observations and comment on how to employ our predictive models based on observed directional properties.





\begin{figure*}[ht]
    \centering                 
    \includegraphics[width=0.4\textwidth]{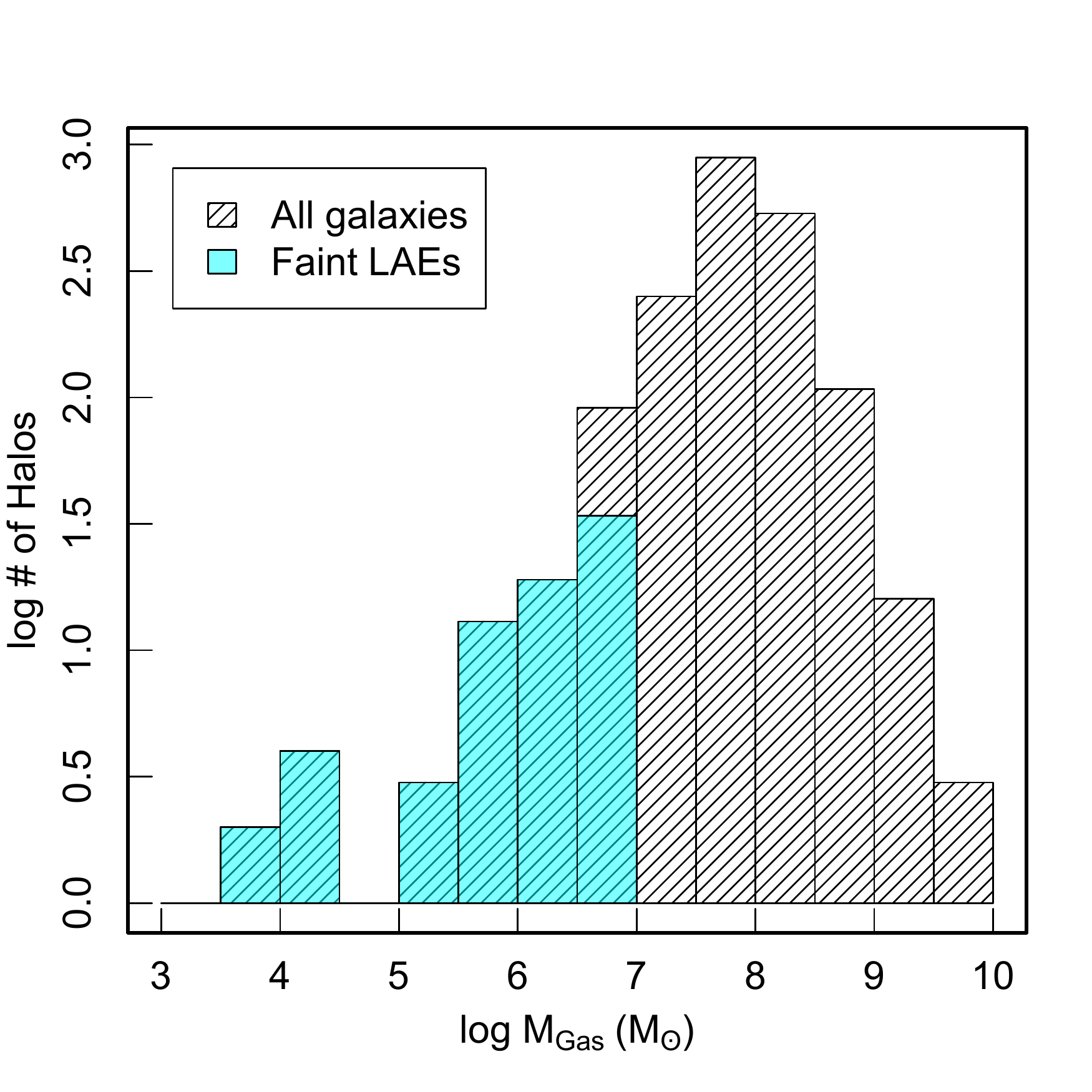}
    \includegraphics[width=0.4\textwidth]{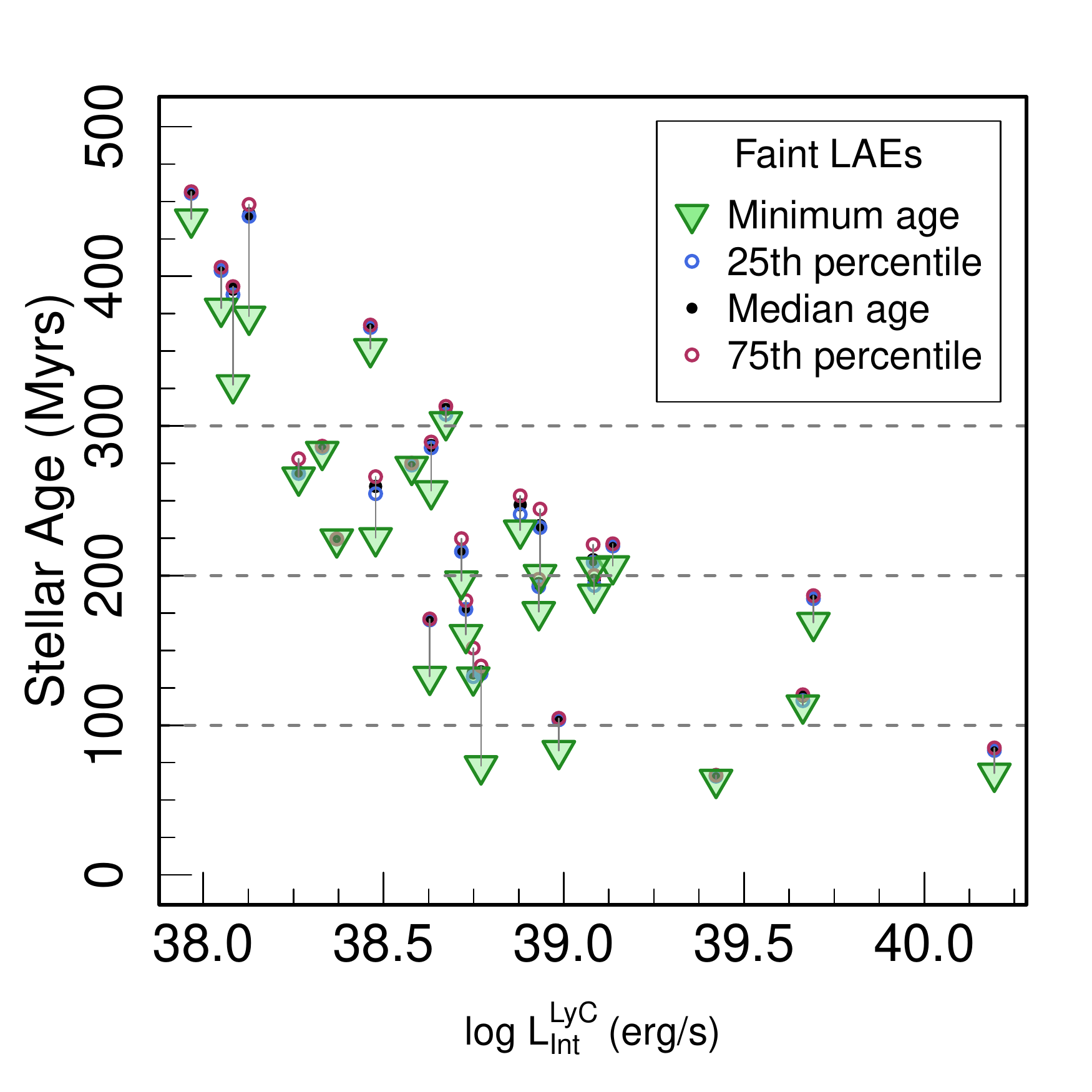}
    \caption{Left : Distribution of gas mass in the faint LAEs (\Llyain $<10^{38}$ erg/s, in sky blue) and the whole sample (shaded). Right : Stellar ages of these faint LAEs as a function of their intrinsic LyC luminosity. Figure shows the minimum stellar age, 25th percentile, 75th percentile and the median age of the stars for each of these faint galaxies.}
   \label{fig:faint_LAEs}
\end{figure*}

\textbf{Uncertainties in the intrinsic LyC spectral distributions:} 

The shape of the ionizing spectrum of galaxies is still poorly constrained.  
The LCEs detected so far have all been observed close to the Lyman limit \citep[e.g.]{Steidel2018, Izotov2021, Flury2022}. The only exception is the recent discovery of a z$\sim 1.4$ galaxy leaking ionizing radiation at 600\AA\ rest-frame with the Astrosat \citep{Saha2020}. The theoretical predictions from population synthesis models is also a debated topic so far. The SPHINX simulation uses BPASS models \citep{Stanway2016} to build the SEDs of galaxies, and in this version of SPHINX, all stars are binary systems. The binary star systems can emit more LyC photons for a longer time compared to single stellar populations, which impacts the full reionization history \citep{sphinx}. While binaries appear as a central ingredient in stellar radiation modelling at the EoR, the fraction of binary stars in the early Universe remains uncertain, as well as their exact spectral contribution.


While discussing the relationship of \lya and LyC intrinsic luminosities in \S~\ref{sec:lyalyc_lum} we have noted in Figure~\ref{fig:lya_lyc_lum_inout} that galaxies (77/1933 or $3.98\%$ of the population) at the very faint end of \lya ($\Llyain < 10^{38}$ erg/s) have LyC luminosity in the range of $10^{38} - 10^{40}$ erg/s. These faint LAEs are extremely gas deficient compared to the rest of the population, as shown in Figure~\ref{fig:faint_LAEs}. So we find that in these systems there is not enough gas in the ISM to produce \lya photons, resulting in very low \Llyain. In contrast, these galaxies do have some residual LyC production although there have been no star formation in them in the last 10 Myrs (i.e. SFR$_{10} = 0$). We show the stellar ages of these systems in Figure~\ref{fig:faint_LAEs} and find that their median ages range from 100 - 300 Myrs and even their minimum stellar ages are very high. Furthermore, in all of them the 25th, 50th, and 75th percentile of ages are very close in values. This indicates that these systems are very old and their star formation finished within a short amount of time. \cite{Stanway2016} (Figure 1) demonstrates that, for binary populations in BPASS models with an instantaneous star formation model, it is possible for stellar populations to emit $\sim$ $10^{49}$ LyC photons/s at an age of $~ 100$ Myr. So in these faint LAEs, it is feasible that even though the galaxies have very old stellar systems, the LyC production is non-negligible. If these simulated galaxies exist in the real Universe, their LyC contribution to the re-ionization photons budget cannot be captured by their \lya emission, and they will be missed by our prediction models.

\section{Summary}
\label{sec:summary}

We explore the connection between LyC and \lya emission from EoR galaxies using a sample of 1933 simulated galaxies in the SPHINX radiation hydrodynamical simulation. We post process these galaxies using the radiative transfer code RASCAS to obtain their \lya emission properties. 

We first investigate the link between \lya and LyC radiation from galaxies and our main results are as follows:

\begin{itemize}
    \item The intrinsic \lya and LyC luminosities are strongly correlated. The total LyC (0 - 912\AA) escaping luminosities, 
    are also correlated with escaping \lya luminosity, although the dispersion is higher, especially in faint LAEs.
    
    \item Given a threshold in observed LyC luminosity, as galaxies become brighter in \lya, the fraction of observable LCEs among LAE samples increases.
    
    \item In bright LAEs ($\Llyaout > 10^{41}$ erg/s) escape fractions of \lya and LyC are correlated, and in good agreement with the observed LCEs. However, when we consider all galaxies, including the fainter ones, there is no correlation, which suggests that the observed correlation is likely a selection effect.
    
    \item The median \flya of galaxies gradually decreases with their \lya luminosity and at the bright end with $\Llyaout \approx 10^{41.5} - 10^{42}$ erg/s, the median \flya $\approx 0.3$. Median value of \flyc is low 
    for all \lya luminosities with the bright LAEs ($\Llyaout > 10^{40.5}$ erg/s) having median $\flyc \sim 0.1$.
    
    \item Although very faint galaxies are more numerous, the relatively bright LAEs contribute more to reionization. In our SPHINX volume, LAEs with $\Llyaout \geq 10^{40}$ erg/s account for about $90\%$ of the total ionizing luminosity in the simulation box, even though they are only $6.8\%$ of the population.  
\end{itemize}
    
We explored models for predicting LyC emission from galaxies using their physical and \lya properties. We apply multivariate linear models on our sample of simulated galaxies and the main results are summarized below:

\begin{itemize}
        
    
    \item We build a set of models using different sets of galaxy properties as input parameters and predict LyC luminosities and escape fraction. In our fiducial model (Model 2) we give 8 galaxy physical properties (gas mass, stellar mass, galaxy $R_{vir}$, SFR$_{10}$, SFR$_{100}$, stellar age and stellar and gas metallicity) and \Llyaout as input parameters. The resulting model can predict \Llycin and \Llycout very well, with high (adjusted) $R^2$ values of 0.8969 and 0.8516 respectively. The \flyc is also predicted fairly well.
    
    \item We also determine the most important input variables for predicting LyC and find that the top four predictors of \Llycout are \Llyaout, gas mass, gas metallicity, and SFR$_{10}$. 
\end{itemize}


These results and the predictive models can be very useful in predicting the LyC emission from EoR galaxies and thus help us to determine the primary sources of reionization. We can apply these models on the upcoming EoR galaxy observations of JWST and other future surveys. They can also facilitate the selection and detection of LyC leakers. These models can be helpful to plan future direct LCE observation missions at lower redshifts. In a future work, we will investigate the effect of directional variation of \lya and LyC escape from galaxies, and IGM attenuation, on our predictions.


\section{Acknowledgement}
We thank the anonymous referee for valuable comments and suggestions that have substantially improved the paper.

MM, AV and TG are supported by the ERC Starting grant 757258 ‘TRIPLE’. AV acknowledges support from SNF Professorship PP00P2\_176808. TK was supported by the National Research Foundation of Korea (NRF-2019K2A9A1A0609137711 and NRF-2020R1C1C1007079).

We have performed the radiative transfer calculations in the LESTA and BAOBAB high-performance computing clusters of University of Geneva, and the RT post-processing for 1933 halos took approximately $\sim 37000$ CPU hours. The SPHINX simulation results of this research have been
achieved using the PRACE Research Infrastructure resource SuperMUC based in Garching, Germany, under PRACE grant 2016153539. We additionally acknowledge support and computational resources from the Common Computing Facility (CCF) of the LABEX Lyon Institute of Origins (ANR-10-LABX-66).


\bibliographystyle{aa} 
\bibliography{mybiblio} 

\begin{thebibliography}{107}
\expandafter\ifx\csname natexlab\endcsname\relax\def\natexlab#1{#1}\fi

\bibitem[{{Aubert} {et~al.}(2004){Aubert}, {Pichon}, \& {Colombi}}]{Aubert2004}
{Aubert}, D., {Pichon}, C., \& {Colombi}, S. 2004, \mnras, 352, 376

\bibitem[{{Bacon} {et~al.}(2015){Bacon}, {Brinchmann}, {Richard}, {Contini},
  {Drake}, {Franx}, {Tacchella}, {Vernet}, {Wisotzki}, {Blaizot}, {Bouch{\'e}},
  {Bouwens}, {Cantalupo}, {Carollo}, {Carton}, {Caruana}, {Cl{\'e}ment},
  {Dreizler}, {Epinat}, {Guiderdoni}, {Herenz}, {Husser}, {Kamann}, {Kerutt},
  {Kollatschny}, {Krajnovic}, {Lilly}, {Martinsson}, {Michel-Dansac},
  {Patricio}, {Schaye}, {Shirazi}, {Soto}, {Soucail}, {Steinmetz}, {Urrutia},
  {Weilbacher}, \& {de Zeeuw}}]{Bacon2015}
{Bacon}, R., {Brinchmann}, J., {Richard}, J., {et~al.} 2015, \aap, 575, A75

\bibitem[{{Bassett} {et~al.}(2019){Bassett}, {Ryan-Weber}, {Cooke}, {Diaz},
  {Nanayakkara}, {Yuan}, {Spitler}, {Me{\v{s}}tri{\'c}}, {Garel}, {Sawicki},
  {Gwyn}, \& {Golob}}]{Bassett2019}
{Bassett}, R., {Ryan-Weber}, E.~V., {Cooke}, J., {et~al.} 2019, \mnras, 483,
  5223

\bibitem[{{Behrens} \& {Braun}(2014)}]{Behrens2014}
{Behrens}, C. \& {Braun}, H. 2014, \aap, 572, A74

\bibitem[{{Borthakur} {et~al.}(2014){Borthakur}, {Heckman}, {Leitherer}, \&
  {Overzier}}]{Borthakur2014}
{Borthakur}, S., {Heckman}, T.~M., {Leitherer}, C., \& {Overzier}, R.~A. 2014,
  Science, 346, 216

\bibitem[{{Cantalupo} {et~al.}(2008){Cantalupo}, {Porciani}, \&
  {Lilly}}]{Cantalupo2008}
{Cantalupo}, S., {Porciani}, C., \& {Lilly}, S.~J. 2008, \apj, 672, 48

\bibitem[{{Cen} \& {Kimm}(2015)}]{Cen2015}
{Cen}, R. \& {Kimm}, T. 2015, \apjl, 801, L25

\bibitem[{{Chisholm} {et~al.}(2017){Chisholm}, {Orlitov{\'a}}, {Schaerer},
  {Verhamme}, {Worseck}, {Izotov}, {Thuan}, \& {Guseva}}]{Chisholm2017}
{Chisholm}, J., {Orlitov{\'a}}, I., {Schaerer}, D., {et~al.} 2017, \aap, 605,
  A67

\bibitem[{{Cowie} {et~al.}(2009){Cowie}, {Barger}, \& {Trouille}}]{Cowie2009}
{Cowie}, L.~L., {Barger}, A.~J., \& {Trouille}, L. 2009, \apj, 692, 1476

\bibitem[{{Dijkstra}(2014{\natexlab{a}})}]{Dijkstra2014}
{Dijkstra}, M. 2014{\natexlab{a}}, \pasa, 31, e040

\bibitem[{{Dijkstra}(2014{\natexlab{b}})}]{Dijkstra2014review}
{Dijkstra}, M. 2014{\natexlab{b}}, \pasa, 31, e040

\bibitem[{{Dijkstra} {et~al.}(2016){Dijkstra}, {Gronke}, \&
  {Venkatesan}}]{Dijkstra2016}
{Dijkstra}, M., {Gronke}, M., \& {Venkatesan}, A. 2016, \apj, 828, 71

\bibitem[{{Drake} {et~al.}(2017){Drake}, {Garel}, {Wisotzki}, {Leclercq},
  {Hashimoto}, {Richard}, {Bacon}, {Blaizot}, {Caruana}, {Conseil}, {Contini},
  {Guiderdoni}, {Herenz}, {Inami}, {Lewis}, {Mahler}, {Marino}, {Pello},
  {Schaye}, {Verhamme}, {Ventou}, \& {Weilbacher}}]{Drake2017}
{Drake}, A.~B., {Garel}, T., {Wisotzki}, L., {et~al.} 2017, \aap, 608, A6

\bibitem[{{Erb}(2015)}]{Erb2015}
{Erb}, D.~K. 2015, \nat, 523, 169

\bibitem[{{Erb} {et~al.}(2011){Erb}, {Bogosavljevi{\'c}}, \&
  {Steidel}}]{Erb2011}
{Erb}, D.~K., {Bogosavljevi{\'c}}, M., \& {Steidel}, C.~C. 2011, \apjl, 740,
  L31

\bibitem[{{Faucher-Gigu{\`e}re}(2020)}]{Faucher2020}
{Faucher-Gigu{\`e}re}, C.-A. 2020, \mnras, 493, 1614

\bibitem[{{Faucher-Gigu{\`e}re} {et~al.}(2010){Faucher-Gigu{\`e}re},
  {Kere{\v{s}}}, {Dijkstra}, {Hernquist}, \& {Zaldarriaga}}]{Faucher2010}
{Faucher-Gigu{\`e}re}, C.-A., {Kere{\v{s}}}, D., {Dijkstra}, M., {Hernquist},
  L., \& {Zaldarriaga}, M. 2010, \apj, 725, 633

\bibitem[{Feigelson \& Babu(2012)}]{feigelson_babu_2012}
Feigelson, E.~D. \& Babu, G.~J. 2012, Modern Statistical Methods for Astronomy:
  With R Applications (Cambridge University Press)

\bibitem[{{Finkelstein} {et~al.}(2013){Finkelstein}, {Papovich}, {Dickinson},
  {Song}, {Tilvi}, {Koekemoer}, {Finkelstein}, {Mobasher}, {Ferguson},
  {Giavalisco}, {Reddy}, {Ashby}, {Dekel}, {Fazio}, {Fontana}, {Grogin},
  {Huang}, {Kocevski}, {Rafelski}, {Weiner}, \& {Willner}}]{Finkelstein2013}
{Finkelstein}, S.~L., {Papovich}, C., {Dickinson}, M., {et~al.} 2013, \nat,
  502, 524

\bibitem[{{Flury} {et~al.}(2022){Flury}, {Jaskot}, {Ferguson}, {Worseck},
  {Makan}, {Chisholm}, {Saldana-Lopez}, {Schaerer}, {McCandless}, {Wang},
  {Ford}, {Heckman}, {Ji}, {Giavalisco}, {Amorin}, {Atek}, {Blaizot},
  {Borthakur}, {Carr}, {Castellano}, {Cristiani}, {de Barros}, {Dickinson},
  {Finkelstein}, {Fleming}, {Fontanot}, {Garel}, {Grazian}, {Hayes}, {Henry},
  {Mauerhofer}, {Micheva}, {Oey}, {Ostlin}, {Papovich}, {Pentericci},
  {Ravindranath}, {Rosdahl}, {Rutkowski}, {Santini}, {Scarlata}, {Teplitz},
  {Thuan}, {Trebitsch}, {Vanzella}, {Verhamme}, \& {Xu}}]{Flury2022}
{Flury}, S.~R., {Jaskot}, A.~E., {Ferguson}, H.~C., {et~al.} 2022, accepted to
  ApJS, arXiv:2201.11716

\bibitem[{{Fontanot} {et~al.}(2014){Fontanot}, {Cristiani}, {Pfrommer},
  {Cupani}, \& {Vanzella}}]{Fontanot2014}
{Fontanot}, F., {Cristiani}, S., {Pfrommer}, C., {Cupani}, G., \& {Vanzella},
  E. 2014, \mnras, 438, 2097

\bibitem[{{Fontanot} {et~al.}(2012){Fontanot}, {Cristiani}, \&
  {Vanzella}}]{Fontanot2012}
{Fontanot}, F., {Cristiani}, S., \& {Vanzella}, E. 2012, \mnras, 425, 1413

\bibitem[{{Garel} {et~al.}(2021){Garel}, {Blaizot}, {Rosdahl}, {Michel-Dansac},
  {Haehnelt}, {Katz}, {Kimm}, \& {Verhamme}}]{Garel2021}
{Garel}, T., {Blaizot}, J., {Rosdahl}, J., {et~al.} 2021, \mnras, 504, 1902

\bibitem[{{Gazagnes} {et~al.}(2020){Gazagnes}, {Chisholm}, {Schaerer},
  {Verhamme}, \& {Izotov}}]{Gazagnes2020}
{Gazagnes}, S., {Chisholm}, J., {Schaerer}, D., {Verhamme}, A., \& {Izotov}, Y.
  2020, \aap, 639, A85

\bibitem[{{Goerdt} {et~al.}(2010){Goerdt}, {Dekel}, {Sternberg}, {Ceverino},
  {Teyssier}, \& {Primack}}]{Goerdt2010}
{Goerdt}, T., {Dekel}, A., {Sternberg}, A., {et~al.} 2010, \mnras, 407, 613

\bibitem[{Gronke {et~al.}(2021)Gronke, Ocvirk, Mason, Matthee, Bosman, Sorce,
  Lewis, Ahn, Aubert, Dawoodbhoy, Iliev, Shapiro, \& Yepes}]{Gronke2020}
Gronke, M., Ocvirk, P., Mason, C., {et~al.} 2021, MNRAS, 508, 3697

\bibitem[{{Hayes} {et~al.}(2013){Hayes}, {{\"O}stlin}, {Schaerer}, {Verhamme},
  {Mas-Hesse}, {Adamo}, {Atek}, {Cannon}, {Duval}, {Guaita}, {Herenz}, {Kunth},
  {Laursen}, {Melinder}, {Orlitov{\'a}}, {Ot{\'\i}-Floranes}, \&
  {Sandberg}}]{Hayes2013}
{Hayes}, M., {{\"O}stlin}, G., {Schaerer}, D., {et~al.} 2013, \apjl, 765, L27

\bibitem[{{Heckman} {et~al.}(2015){Heckman}, {Alexandroff}, {Borthakur},
  {Overzier}, \& {Leitherer}}]{Heckman2015}
{Heckman}, T.~M., {Alexandroff}, R.~M., {Borthakur}, S., {Overzier}, R., \&
  {Leitherer}, C. 2015, \apj, 809, 147

\bibitem[{{Heckman} {et~al.}(2011){Heckman}, {Borthakur}, {Overzier},
  {Kauffmann}, {Basu-Zych}, {Leitherer}, {Sembach}, {Martin}, {Rich},
  {Schiminovich}, \& {Seibert}}]{Heckman2011}
{Heckman}, T.~M., {Borthakur}, S., {Overzier}, R., {et~al.} 2011, \apj, 730, 5

\bibitem[{{Henry} {et~al.}(2015){Henry}, {Scarlata}, {Martin}, \&
  {Erb}}]{Henry2015}
{Henry}, A., {Scarlata}, C., {Martin}, C.~L., \& {Erb}, D. 2015, \apj, 809, 19

\bibitem[{{Inoue} {et~al.}(2018){Inoue}, {Hasegawa}, {Ishiyama}, {Yajima},
  {Shimizu}, {Umemura}, {Konno}, {Harikane}, {Shibuya}, {Ouchi}, {Shimasaku},
  {Ono}, {Kusakabe}, {Higuchi}, \& {Lee}}]{Inoue2018}
{Inoue}, A.~K., {Hasegawa}, K., {Ishiyama}, T., {et~al.} 2018, \pasj, 70, 55

\bibitem[{{Inoue} {et~al.}(2014){Inoue}, {Shimizu}, {Iwata}, \&
  {Tanaka}}]{Inoue2014}
{Inoue}, A.~K., {Shimizu}, I., {Iwata}, I., \& {Tanaka}, M. 2014, \mnras, 442,
  1805

\bibitem[{{Itoh} {et~al.}(2018){Itoh}, {Ouchi}, {Zhang}, {Inoue}, {Mawatari},
  {Shibuya}, {Harikane}, {Ono}, {Kusakabe}, {Shimasaku}, {Fujimoto}, {Iwata},
  {Kajisawa}, {Kashikawa}, {Kawanomoto}, {Komiyama}, {Lee}, {Nagao}, \&
  {Taniguchi}}]{Itoh2018}
{Itoh}, R., {Ouchi}, M., {Zhang}, H., {et~al.} 2018, \apj, 867, 46

\bibitem[{{Izotov} {et~al.}(2016{\natexlab{a}}){Izotov}, {Orlitov{\'a}},
  {Schaerer}, {Thuan}, {Verhamme}, {Guseva}, \& {Worseck}}]{Izotov2016a}
{Izotov}, Y.~I., {Orlitov{\'a}}, I., {Schaerer}, D., {et~al.}
  2016{\natexlab{a}}, \nat, 529, 178

\bibitem[{{Izotov} {et~al.}(2016{\natexlab{b}}){Izotov}, {Schaerer}, {Thuan},
  {Worseck}, {Guseva}, {Orlitov{\'a}}, \& {Verhamme}}]{Izotov2016b}
{Izotov}, Y.~I., {Schaerer}, D., {Thuan}, T.~X., {et~al.} 2016{\natexlab{b}},
  \mnras, 461, 3683

\bibitem[{{Izotov} {et~al.}(2018{\natexlab{a}}){Izotov}, {Schaerer}, {Worseck},
  {Guseva}, {Thuan}, {Verhamme}, {Orlitov{\'a}}, \& {Fricke}}]{Izotov2018a}
{Izotov}, Y.~I., {Schaerer}, D., {Worseck}, G., {et~al.} 2018{\natexlab{a}},
  \mnras, 474, 4514

\bibitem[{Izotov {et~al.}(2021)Izotov, Worseck, Schaerer, Guseva, Chisholm,
  Thuan, Fricke, \& Verhamme}]{Izotov2021}
Izotov, Y.~I., Worseck, G., Schaerer, D., {et~al.} 2021, MNRAS, 503, 1734

\bibitem[{{Izotov} {et~al.}(2018{\natexlab{b}}){Izotov}, {Worseck}, {Schaerer},
  {Guseva}, {Thuan}, {Fricke}, \& {Orlitov{\'a}}}]{Izotov2018b}
{Izotov}, Y.~I., {Worseck}, G., {Schaerer}, D., {et~al.} 2018{\natexlab{b}},
  \mnras, 478, 4851

\bibitem[{{Jaskot} \& {Oey}(2013)}]{Jaskot2013}
{Jaskot}, A.~E. \& {Oey}, M.~S. 2013, \apj, 766, 91

\bibitem[{{Jung} {et~al.}(2019){Jung}, {Finkelstein}, {Dickinson}, {Hutchison},
  {Larson}, {Papovich}, {Pentericci}, {Song}, {Ferguson}, {Guo}, {Malhotra},
  {Mobasher}, {Rhoads}, {Tilvi}, \& {Wold}}]{Jung2019}
{Jung}, I., {Finkelstein}, S.~L., {Dickinson}, M., {et~al.} 2019, \apj, 877,
  146

\bibitem[{{Katz} {et~al.}(2019){Katz}, {Galligan}, {Kimm}, {Rosdahl},
  {Haehnelt}, {Blaizot}, {Devriendt}, {Slyz}, {Laporte}, \& {Ellis}}]{Katz2019}
{Katz}, H., {Galligan}, T.~P., {Kimm}, T., {et~al.} 2019, \mnras, 487, 5902

\bibitem[{{Katz} {et~al.}(2020){Katz}, {{\v{D}}urov{\v{c}}{\'\i}kov{\'a}},
  {Kimm}, {Rosdahl}, {Blaizot}, {Haehnelt}, {Devriendt}, {Slyz}, {Ellis}, \&
  {Laporte}}]{Katz2020}
{Katz}, H., {{\v{D}}urov{\v{c}}{\'\i}kov{\'a}}, D., {Kimm}, T., {et~al.} 2020,
  \mnras, 498, 164

\bibitem[{{Kimm} {et~al.}(2019){Kimm}, {Blaizot}, {Garel}, {Michel-Dansac},
  {Katz}, {Rosdahl}, {Verhamme}, \& {Haehnelt}}]{Kimm2019}
{Kimm}, T., {Blaizot}, J., {Garel}, T., {et~al.} 2019, \mnras, 486, 2215

\bibitem[{Konno {et~al.}(2014)Konno, Ouchi, Ono, Shimasaku, Shibuya, Furusawa,
  Nakajima, Naito, Momose, Yuma, \& Iye}]{Konno2014}
Konno, A., Ouchi, M., Ono, Y., {et~al.} 2014, 797, 16

\bibitem[{{Kulkarni} {et~al.}(2019){Kulkarni}, {Worseck}, \&
  {Hennawi}}]{Kulkarni2019}
{Kulkarni}, G., {Worseck}, G., \& {Hennawi}, J.~F. 2019, \mnras, 488, 1035

\bibitem[{{Kusakabe} {et~al.}(2020){Kusakabe}, {Blaizot}, {Garel}, {Verhamme},
  {Bacon}, {Richard}, {Hashimoto}, {Inami}, {Conseil}, {Guiderdoni}, {Drake},
  {Christian Herenz}, {Schaye}, {Oesch}, {Matthee}, {Anna Marino}, {Borello
  Schmidt}, {Pell{\'o}}, {Maseda}, {Leclercq}, {Kerutt}, \&
  {Mahler}}]{Kusakabe2020}
{Kusakabe}, H., {Blaizot}, J., {Garel}, T., {et~al.} 2020, \aap, 638, A12

\bibitem[{{Laursen} {et~al.}(2009){Laursen}, {Sommer-Larsen}, \&
  {Andersen}}]{Laursen2009}
{Laursen}, P., {Sommer-Larsen}, J., \& {Andersen}, A.~C. 2009, \apj, 704, 1640

\bibitem[{{Laursen} {et~al.}(2019){Laursen}, {Sommer-Larsen}, {Milvang-Jensen},
  {Fynbo}, \& {Razoumov}}]{Laursen2019}
{Laursen}, P., {Sommer-Larsen}, J., {Milvang-Jensen}, B., {Fynbo}, J. P.~U., \&
  {Razoumov}, A.~O. 2019, \aap, 627, A84

\bibitem[{{Leitet} {et~al.}(2013){Leitet}, {Bergvall}, {Hayes}, {Linn{\'e}}, \&
  {Zackrisson}}]{Leitet2013}
{Leitet}, E., {Bergvall}, N., {Hayes}, M., {Linn{\'e}}, S., \& {Zackrisson}, E.
  2013, \aap, 553, A106

\bibitem[{{Leitet} {et~al.}(2011){Leitet}, {Bergvall}, {Piskunov}, \&
  {Andersson}}]{Leitet2011}
{Leitet}, E., {Bergvall}, N., {Piskunov}, N., \& {Andersson}, B.~G. 2011, \aap,
  532, A107

\bibitem[{{Loeb} \& {Barkana}(2001)}]{Loeb2001}
{Loeb}, A. \& {Barkana}, R. 2001, \araa, 39, 19

\bibitem[{{Madau}(1995)}]{Madau1995}
{Madau}, P. 1995, \apj, 441, 18

\bibitem[{{Marques-Chaves} {et~al.}(2021){Marques-Chaves}, {Schaerer},
  {{\'A}lvarez-M{\'a}rquez}, {Colina}, {Dessauges-Zavadsky},
  {P{\'e}rez-Fournon}, {Saldana-Lopez}, \& {Verhamme}}]{Chaves2021}
{Marques-Chaves}, R., {Schaerer}, D., {{\'A}lvarez-M{\'a}rquez}, J., {et~al.}
  2021, \mnras, 507, 524

\bibitem[{{Mason} \& {Gronke}(2020)}]{Mason2020}
{Mason}, C.~A. \& {Gronke}, M. 2020, \mnras, 499, 1395

\bibitem[{{Matthee} {et~al.}(2020){Matthee}, {Pezzulli}, {Mackenzie},
  {Cantalupo}, {Kusakabe}, {Leclercq}, {Sobral}, {Richard}, {Wisotzki},
  {Lilly}, {Boogaard}, {Marino}, {Maseda}, \& {Nanayakkara}}]{Matthee2020}
{Matthee}, J., {Pezzulli}, G., {Mackenzie}, R., {et~al.} 2020, \mnras, 498,
  3043

\bibitem[{{Matthee} {et~al.}(2017){Matthee}, {Sobral}, {Darvish}, {Santos},
  {Mobasher}, {Paulino-Afonso}, {R{\"o}ttgering}, \& {Alegre}}]{Matthee2017}
{Matthee}, J., {Sobral}, D., {Darvish}, B., {et~al.} 2017, \mnras, 472, 772

\bibitem[{{Matthee} {et~al.}(2018){Matthee}, {Sobral}, {Gronke},
  {Paulino-Afonso}, {Stefanon}, \& {R{\"o}ttgering}}]{Matthee2018}
{Matthee}, J., {Sobral}, D., {Gronke}, M., {et~al.} 2018, \aap, 619, A136

\bibitem[{{Mauerhofer} {et~al.}(2021){Mauerhofer}, {Verhamme}, {Blaizot},
  {Garel}, {Kimm}, {Michel-Dansac}, \& {Rosdahl}}]{Mauerhofer2021}
{Mauerhofer}, V., {Verhamme}, A., {Blaizot}, J., {et~al.} 2021, \aap, 646, A80

\bibitem[{{Meyer} {et~al.}(2021){Meyer}, {Laporte}, {Ellis}, {Verhamme}, \&
  {Garel}}]{Meyer2021}
{Meyer}, R.~A., {Laporte}, N., {Ellis}, R.~S., {Verhamme}, A., \& {Garel}, T.
  2021, \mnras, 500, 558

\bibitem[{{Michel-Dansac} {et~al.}(2020){Michel-Dansac}, {Blaizot}, {Garel},
  {Verhamme}, {Kimm}, \& {Trebitsch}}]{rascas2020}
{Michel-Dansac}, L., {Blaizot}, J., {Garel}, T., {et~al.} 2020, \aap, 635, A154

\bibitem[{{Micheva} {et~al.}(2010){Micheva}, {Zackrisson}, {{\"O}stlin},
  {Bergvall}, \& {Pursimo}}]{Micheva2010}
{Micheva}, G., {Zackrisson}, E., {{\"O}stlin}, G., {Bergvall}, N., \&
  {Pursimo}, T. 2010, \mnras, 405, 1203

\bibitem[{{Nakajima} \& {Ouchi}(2014)}]{Nakajima2014}
{Nakajima}, K. \& {Ouchi}, M. 2014, \mnras, 442, 900

\bibitem[{{Ocvirk} {et~al.}(2016){Ocvirk}, {Gillet}, {Shapiro}, {Aubert},
  {Iliev}, {Teyssier}, {Yepes}, {Choi}, {Sullivan}, {Knebe}, {Gottl{\"o}ber},
  {D'Aloisio}, {Park}, {Hoffman}, \& {Stranex}}]{Ocvirk2016}
{Ocvirk}, P., {Gillet}, N., {Shapiro}, P.~R., {et~al.} 2016, \mnras, 463, 1462

\bibitem[{{Oesch} {et~al.}(2015){Oesch}, {van Dokkum}, {Illingworth},
  {Bouwens}, {Momcheva}, {Holden}, {Roberts-Borsani}, {Smit}, {Franx},
  {Labb{\'e}}, {Gonz{\'a}lez}, \& {Magee}}]{Oesch2015}
{Oesch}, P.~A., {van Dokkum}, P.~G., {Illingworth}, G.~D., {et~al.} 2015,
  \apjl, 804, L30

\bibitem[{{Ono} {et~al.}(2012){Ono}, {Ouchi}, {Mobasher}, {Dickinson},
  {Penner}, {Shimasaku}, {Weiner}, {Kartaltepe}, {Nakajima}, {Nayyeri},
  {Stern}, {Kashikawa}, \& {Spinrad}}]{Ono2012}
{Ono}, Y., {Ouchi}, M., {Mobasher}, B., {et~al.} 2012, \apj, 744, 83

\bibitem[{{{\"O}stlin} {et~al.}(2014){{\"O}stlin}, {Hayes}, {Duval}, {Sand
  berg}, {Rivera-Thorsen}, {Marquart}, {Orlitov{\'a}}, {Adamo}, {Melinder},
  {Guaita}, {Atek}, {Cannon}, {Gruyters}, {Herenz}, {Kunth}, {Laursen},
  {Mas-Hesse}, {Micheva}, {Ot{\'\i}-Floranes}, {Pardy}, {Roth}, {Schaerer}, \&
  {Verhamme}}]{Ostlin2014}
{{\"O}stlin}, G., {Hayes}, M., {Duval}, F., {et~al.} 2014, \apj, 797, 11

\bibitem[{{Ouchi} {et~al.}(2018){Ouchi}, {Harikane}, {Shibuya}, {Shimasaku},
  {Taniguchi}, {Konno}, {Kobayashi}, {Kajisawa}, {Nagao}, {Ono}, {Inoue},
  {Umemura}, {Mori}, {Hasegawa}, {Higuchi}, {Komiyama}, {Matsuda}, {Nakajima},
  {Saito}, \& {Wang}}]{Ouchi2018}
{Ouchi}, M., {Harikane}, Y., {Shibuya}, T., {et~al.} 2018, \pasj, 70, S13

\bibitem[{{Pardy} {et~al.}(2016){Pardy}, {Cannon}, {{\"O}stlin}, {Hayes}, \&
  {Bergvall}}]{Pardy2016}
{Pardy}, S.~A., {Cannon}, J.~M., {{\"O}stlin}, G., {Hayes}, M., \& {Bergvall},
  N. 2016, \aj, 152, 178

\bibitem[{{Partridge} \& {Peebles}(1967)}]{Partridge1967}
{Partridge}, R.~B. \& {Peebles}, P.~J.~E. 1967, \apj, 147, 868

\bibitem[{{Planck Collaboration} {et~al.}(2014){Planck Collaboration}, {Ade},
  {Aghanim}, {Armitage-Caplan}, {Arnaud}, {Ashdown}, {Atrio-Barandela},
  {Aumont}, {Baccigalupi}, {Banday}, {Barreiro}, {Bartlett}, {Battaner},
  {Benabed}, {Beno{\^\i}t}, {Benoit-L{\'e}vy}, {Bernard}, {Bersanelli},
  {Bielewicz}, {Bobin}, {Bock}, {Bonaldi}, {Bond}, {Borrill}, {Bouchet},
  {Bridges}, {Bucher}, {Burigana}, {Butler}, {Calabrese}, {Cappellini},
  {Cardoso}, {Catalano}, {Challinor}, {Chamballu}, {Chary}, {Chen}, {Chiang},
  {Chiang}, {Christensen}, {Church}, {Clements}, {Colombi}, {Colombo},
  {Couchot}, {Coulais}, {Crill}, {Curto}, {Cuttaia}, {Danese}, {Davies},
  {Davis}, {de Bernardis}, {de Rosa}, {de Zotti}, {Delabrouille}, {Delouis},
  {D{\'e}sert}, {Dickinson}, {Diego}, {Dolag}, {Dole}, {Donzelli}, {Dor{\'e}},
  {Douspis}, {Dunkley}, {Dupac}, {Efstathiou}, {Elsner}, {En{\ss}lin},
  {Eriksen}, {Finelli}, {Forni}, {Frailis}, {Fraisse}, {Franceschi}, {Gaier},
  {Galeotta}, {Galli}, {Ganga}, {Giard}, {Giardino}, {Giraud-H{\'e}raud},
  {Gjerl{\o}w}, {Gonz{\'a}lez-Nuevo}, {G{\'o}rski}, {Gratton}, {Gregorio},
  {Gruppuso}, {Gudmundsson}, {Haissinski}, {Hamann}, {Hansen}, {Hanson},
  {Harrison}, {Henrot-Versill{\'e}}, {Hern{\'a}ndez-Monteagudo}, {Herranz},
  {Hildebrandt}, {Hivon}, {Hobson}, {Holmes}, {Hornstrup}, {Hou}, {Hovest},
  {Huffenberger}, {Jaffe}, {Jaffe}, {Jewell}, {Jones}, {Juvela},
  {Keih{\"a}nen}, {Keskitalo}, {Kisner}, {Kneissl}, {Knoche}, {Knox}, {Kunz},
  {Kurki-Suonio}, {Lagache}, {L{\"a}hteenm{\"a}ki}, {Lamarre}, {Lasenby},
  {Lattanzi}, {Laureijs}, {Lawrence}, {Leach}, {Leahy}, {Leonardi},
  {Le{\'o}n-Tavares}, {Lesgourgues}, {Lewis}, {Liguori}, {Lilje},
  {Linden-V{\o}rnle}, {L{\'o}pez-Caniego}, {Lubin}, {Mac{\'\i}as-P{\'e}rez},
  {Maffei}, {Maino}, {Mandolesi}, {Maris}, {Marshall}, {Martin},
  {Mart{\'\i}nez-Gonz{\'a}lez}, {Masi}, {Massardi}, {Matarrese}, {Matthai},
  {Mazzotta}, {Meinhold}, {Melchiorri}, {Melin}, {Mendes}, {Menegoni},
  {Mennella}, {Migliaccio}, {Millea}, {Mitra}, {Miville-Desch{\^e}nes},
  {Moneti}, {Montier}, {Morgante}, {Mortlock}, {Moss}, {Munshi}, {Murphy},
  {Naselsky}, {Nati}, {Natoli}, {Netterfield}, {N{\o}rgaard-Nielsen},
  {Noviello}, {Novikov}, {Novikov}, {O'Dwyer}, {Osborne}, {Oxborrow}, {Paci},
  {Pagano}, {Pajot}, {Paladini}, {Paoletti}, {Partridge}, {Pasian},
  {Patanchon}, {Pearson}, {Pearson}, {Peiris}, {Perdereau}, {Perotto},
  {Perrotta}, {Pettorino}, {Piacentini}, {Piat}, {Pierpaoli}, {Pietrobon},
  {Plaszczynski}, {Platania}, {Pointecouteau}, {Polenta}, {Ponthieu}, {Popa},
  {Poutanen}, {Pratt}, {Pr{\'e}zeau}, {Prunet}, {Puget}, {Rachen}, {Reach},
  {Rebolo}, {Reinecke}, {Remazeilles}, {Renault}, {Ricciardi}, {Riller},
  {Ristorcelli}, {Rocha}, {Rosset}, {Roudier}, {Rowan-Robinson},
  {Rubi{\~n}o-Mart{\'\i}n}, {Rusholme}, {Sandri}, {Santos}, {Savelainen},
  {Savini}, {Scott}, {Seiffert}, {Shellard}, {Spencer}, {Starck}, {Stolyarov},
  {Stompor}, {Sudiwala}, {Sunyaev}, {Sureau}, {Sutton}, {Suur-Uski}, {Sygnet},
  {Tauber}, {Tavagnacco}, {Terenzi}, {Toffolatti}, {Tomasi}, {Tristram},
  {Tucci}, {Tuovinen}, {T{\"u}rler}, {Umana}, {Valenziano}, {Valiviita}, {Van
  Tent}, {Vielva}, {Villa}, {Vittorio}, {Wade}, {Wandelt}, {Wehus}, {White},
  {White}, {Wilkinson}, {Yvon}, {Zacchei}, \& {Zonca}}]{Planck2014}
{Planck Collaboration}, {Ade}, P.~A.~R., {Aghanim}, N., {et~al.} 2014, \aap,
  571, A16

\bibitem[{{Puschnig} {et~al.}(2017){Puschnig}, {Hayes}, {{\"O}stlin},
  {Rivera-Thorsen}, {Melinder}, {Cannon}, {Menacho}, {Zackrisson}, {Bergvall},
  \& {Leitet}}]{Puschnig2017}
{Puschnig}, J., {Hayes}, M., {{\"O}stlin}, G., {et~al.} 2017, \mnras, 469, 3252

\bibitem[{{Raiter} {et~al.}(2010){Raiter}, {Fosbury}, \&
  {Teimoorinia}}]{Raiter2010}
{Raiter}, A., {Fosbury}, R.~A.~E., \& {Teimoorinia}, H. 2010, \aap, 510, A109

\bibitem[{{Roberts-Borsani} {et~al.}(2016){Roberts-Borsani}, {Bouwens},
  {Oesch}, {Labbe}, {Smit}, {Illingworth}, {van Dokkum}, {Holden}, {Gonzalez},
  {Stefanon}, {Holwerda}, \& {Wilkins}}]{RobertsBorsani2016}
{Roberts-Borsani}, G.~W., {Bouwens}, R.~J., {Oesch}, P.~A., {et~al.} 2016,
  \apj, 823, 143

\bibitem[{{Rosdahl} \& {Blaizot}(2012)}]{Rosdahl2012}
{Rosdahl}, J. \& {Blaizot}, J. 2012, \mnras, 423, 344

\bibitem[{{Rosdahl} {et~al.}(2013){Rosdahl}, {Blaizot}, {Aubert}, {Stranex}, \&
  {Teyssier}}]{Rosdahl2013}
{Rosdahl}, J., {Blaizot}, J., {Aubert}, D., {Stranex}, T., \& {Teyssier}, R.
  2013, \mnras, 436, 2188

\bibitem[{{Rosdahl} {et~al.}(2018){Rosdahl}, {Katz}, {Blaizot}, {Kimm},
  {Michel-Dansac}, {Garel}, {Haehnelt}, {Ocvirk}, \& {Teyssier}}]{sphinx}
{Rosdahl}, J., {Katz}, H., {Blaizot}, J., {et~al.} 2018, \mnras, 479, 994

\bibitem[{{Runnholm} {et~al.}(2020){Runnholm}, {Hayes}, {Melinder},
  {Rivera-Thorsen}, {{\"O}stlin}, {Cannon}, \& {Kunth}}]{Runnholm2020}
{Runnholm}, A., {Hayes}, M., {Melinder}, J., {et~al.} 2020, \apj, 892, 48

\bibitem[{{Saha} {et~al.}(2020){Saha}, {Tandon}, {Simmonds}, {Verhamme},
  {Paswan}, {Schaerer}, {Rutkowski}, {Borgohain}, {Elmegreen}, {Inoue},
  {Combes}, {Elmegreen}, \& {Paalvast}}]{Saha2020}
{Saha}, K., {Tandon}, S.~N., {Simmonds}, C., {et~al.} 2020, Nature Astronomy,
  4, 1185

\bibitem[{{Schaerer}(2003)}]{Schaerer2003}
{Schaerer}, D. 2003, \aap, 397, 527

\bibitem[{{Schaerer} {et~al.}(2016){Schaerer}, {Izotov}, {Verhamme},
  {Orlitov{\'a}}, {Thuan}, {Worseck}, \& {Guseva}}]{Schaerer2016}
{Schaerer}, D., {Izotov}, Y.~I., {Verhamme}, A., {et~al.} 2016, \aap, 591, L8

\bibitem[{{Schenker} {et~al.}(2014){Schenker}, {Ellis}, {Konidaris}, \&
  {Stark}}]{Schenker2014}
{Schenker}, M.~A., {Ellis}, R.~S., {Konidaris}, N.~P., \& {Stark}, D.~P. 2014,
  \apj, 795, 20

\bibitem[{{Schenker} {et~al.}(2012){Schenker}, {Stark}, {Ellis}, {Robertson},
  {Dunlop}, {McLure}, {Kneib}, \& {Richard}}]{Schenker2012}
{Schenker}, M.~A., {Stark}, D.~P., {Ellis}, R.~S., {et~al.} 2012, \apj, 744,
  179

\bibitem[{{Shibuya} {et~al.}(2012){Shibuya}, {Kashikawa}, {Ota}, {Iye},
  {Ouchi}, {Furusawa}, {Shimasaku}, \& {Hattori}}]{Shibuya2012}
{Shibuya}, T., {Kashikawa}, N., {Ota}, K., {et~al.} 2012, \apj, 752, 114

\bibitem[{{Shibuya} {et~al.}(2018){Shibuya}, {Ouchi}, {Konno}, {Higuchi},
  {Harikane}, {Ono}, {Shimasaku}, {Taniguchi}, {Kobayashi}, {Kajisawa},
  {Nagao}, {Furusawa}, {Goto}, {Kashikawa}, {Komiyama}, {Kusakabe}, {Lee},
  {Momose}, {Nakajima}, {Tanaka}, {Wang}, \& {Yuma}}]{Shibuya2018}
{Shibuya}, T., {Ouchi}, M., {Konno}, A., {et~al.} 2018, \pasj, 70, S14

\bibitem[{{Smith} {et~al.}(2019){Smith}, {Ma}, {Bromm}, {Finkelstein},
  {Hopkins}, {Faucher-Gigu{\`e}re}, \& {Kere{\v{s}}}}]{Smith2019}
{Smith}, A., {Ma}, X., {Bromm}, V., {et~al.} 2019, \mnras, 484, 39

\bibitem[{{Song} {et~al.}(2016){Song}, {Finkelstein}, {Livermore}, {Capak},
  {Dickinson}, \& {Fontana}}]{Song2016}
{Song}, M., {Finkelstein}, S.~L., {Livermore}, R.~C., {et~al.} 2016, \apj, 826,
  113

\bibitem[{{Songaila} {et~al.}(2018){Songaila}, {Hu}, {Barger}, {Cowie},
  {Hasinger}, {Rosenwasser}, \& {Waters}}]{Songaila2018}
{Songaila}, A., {Hu}, E.~M., {Barger}, A.~J., {et~al.} 2018, \apj, 859, 91

\bibitem[{{Spitzer}(1978)}]{Spitzer1978}
{Spitzer}, L. 1978, {Physical processes in the interstellar medium}

\bibitem[{{Stanway} {et~al.}(2016){Stanway}, {Eldridge}, \&
  {Becker}}]{Stanway2016}
{Stanway}, E.~R., {Eldridge}, J.~J., \& {Becker}, G.~D. 2016, \mnras, 456, 485

\bibitem[{{Stark}(2016)}]{Stark2016}
{Stark}, D.~P. 2016, \araa, 54, 761

\bibitem[{{Stark} {et~al.}(2017){Stark}, {Ellis}, {Charlot}, {Chevallard},
  {Tang}, {Belli}, {Zitrin}, {Mainali}, {Gutkin}, {Vidal-Garc{\'\i}a},
  {Bouwens}, \& {Oesch}}]{Stark2017}
{Stark}, D.~P., {Ellis}, R.~S., {Charlot}, S., {et~al.} 2017, \mnras, 464, 469

\bibitem[{{Steidel} {et~al.}(2018){Steidel}, {Bogosavljevi{\'c}}, {Shapley},
  {Reddy}, {Rudie}, {Pettini}, {Trainor}, \& {Strom}}]{Steidel2018}
{Steidel}, C.~C., {Bogosavljevi{\'c}}, M., {Shapley}, A.~E., {et~al.} 2018,
  \apj, 869, 123

\bibitem[{{Teyssier}(2002)}]{Teyssier2002}
{Teyssier}, R. 2002, \aap, 385, 337

\bibitem[{{Trainor} {et~al.}(2015){Trainor}, {Steidel}, {Strom}, \&
  {Rudie}}]{Trainor2015}
{Trainor}, R.~F., {Steidel}, C.~C., {Strom}, A.~L., \& {Rudie}, G.~C. 2015,
  \apj, 809, 89

\bibitem[{{Trebitsch} {et~al.}(2016){Trebitsch}, {Verhamme}, {Blaizot}, \&
  {Rosdahl}}]{Trebitsch2016}
{Trebitsch}, M., {Verhamme}, A., {Blaizot}, J., \& {Rosdahl}, J. 2016, \aap,
  593, A122

\bibitem[{{Trebitsch, Maxime} {et~al.}(2021){Trebitsch, Maxime}, {Dubois,
  Yohan}, {Volonteri, Marta}, {Pfister, Hugo}, {Cadiou, Corentin}, {Katz,
  Harley}, {Rosdahl, Joakim}, {Kimm, Taysun}, {Pichon, Christophe}, {Beckmann,
  Ricarda S.}, {Devriendt, Julien}, \& {Slyz, Adrianne}}]{Trebitsch2020}
{Trebitsch, Maxime}, {Dubois, Yohan}, {Volonteri, Marta}, {et~al.} 2021, A\&A,
  653, A154

\bibitem[{{Tweed} {et~al.}(2009){Tweed}, {Devriendt}, {Blaizot}, {Colombi}, \&
  {Slyz}}]{Tweed2009}
{Tweed}, D., {Devriendt}, J., {Blaizot}, J., {Colombi}, S., \& {Slyz}, A. 2009,
  \aap, 506, 647

\bibitem[{{Urrutia} {et~al.}(2019){Urrutia}, {Wisotzki}, {Kerutt}, {Schmidt},
  {Herenz}, {Klar}, {Saust}, {Werhahn}, {Diener}, {Caruana}, {Krajnovi{\'c}},
  {Bacon}, {Boogaard}, {Brinchmann}, {Enke}, {Maseda}, {Nanayakkara},
  {Richard}, {Steinmetz}, \& {Weilbacher}}]{Urrutia2019}
{Urrutia}, T., {Wisotzki}, L., {Kerutt}, J., {et~al.} 2019, \aap, 624, A141

\bibitem[{{Vanzella} {et~al.}(2011){Vanzella}, {Pentericci}, {Fontana},
  {Grazian}, {Castellano}, {Boutsia}, {Cristiani}, {Dickinson}, {Gallozzi},
  {Giallongo}, {Giavalisco}, {Maiolino}, {Moorwood}, {Paris}, \&
  {Santini}}]{Vanzella2011}
{Vanzella}, E., {Pentericci}, L., {Fontana}, A., {et~al.} 2011, \apjl, 730, L35

\bibitem[{{Verhamme} {et~al.}(2012){Verhamme}, {Dubois}, {Blaizot}, {Garel},
  {Bacon}, {Devriendt}, {Guiderdoni}, \& {Slyz}}]{Verhamme2012}
{Verhamme}, A., {Dubois}, Y., {Blaizot}, J., {et~al.} 2012, \aap, 546, A111

\bibitem[{{Verhamme} {et~al.}(2015){Verhamme}, {Orlitov{\'a}}, {Schaerer}, \&
  {Hayes}}]{Verhamme2015}
{Verhamme}, A., {Orlitov{\'a}}, I., {Schaerer}, D., \& {Hayes}, M. 2015, \aap,
  578, A7

\bibitem[{{Verhamme} {et~al.}(2017){Verhamme}, {Orlitov{\'a}}, {Schaerer},
  {Izotov}, {Worseck}, {Thuan}, \& {Guseva}}]{Verhamme2017}
{Verhamme}, A., {Orlitov{\'a}}, I., {Schaerer}, D., {et~al.} 2017, \aap, 597,
  A13

\bibitem[{Wise(2019)}]{Wise2019}
Wise, J.~H. 2019, Contemporary Physics, 60, 145–163

\bibitem[{{Yajima} {et~al.}(2013){Yajima}, {Li}, \& {Zhu}}]{Yajima2013}
{Yajima}, H., {Li}, Y., \& {Zhu}, Q. 2013, \apj, 773, 151

\bibitem[{{Yajima} {et~al.}(2014){Yajima}, {Li}, {Zhu}, {Abel}, {Gronwall}, \&
  {Ciardullo}}]{Yajima2014}
{Yajima}, H., {Li}, Y., {Zhu}, Q., {et~al.} 2014, \mnras, 440, 776

\bibitem[{{Yang} {et~al.}(2017){Yang}, {Malhotra}, {Gronke}, {Rhoads},
  {Leitherer}, {Wofford}, {Jiang}, {Dijkstra}, {Tilvi}, \& {Wang}}]{Yang2017}
{Yang}, H., {Malhotra}, S., {Gronke}, M., {et~al.} 2017, \apj, 844, 171

\bibitem[{{Zitrin} {et~al.}(2015){Zitrin}, {Labb{\'e}}, {Belli}, {Bouwens},
  {Ellis}, {Roberts-Borsani}, {Stark}, {Oesch}, \& {Smit}}]{Zitrin2015}
{Zitrin}, A., {Labb{\'e}}, I., {Belli}, S., {et~al.} 2015, \apjl, 810, L12

\end{thebibliography}

\begin{appendix}
\section{Supplementary figures}
\subsection{Comparing z=6 sample to the stacked sample}
\label{appendix:z6z10}
In the main text, we combine our galaxy samples at different redshifts and explored the connection between LyC and \lya emission from galaxies. Herein we inspect if the selected populations of galaxies at different redshifts have significantly different properties. 
We compare two samples specifically, 674 galaxies at $z=6$, and the stacked sample of 1933 galaxies that combines all galaxies in all of the 5 redshifts (z = 6, 7, 8, 9, 10). 

\begin{figure*}[h]
    \centering
    \includegraphics[width=0.3\textwidth]{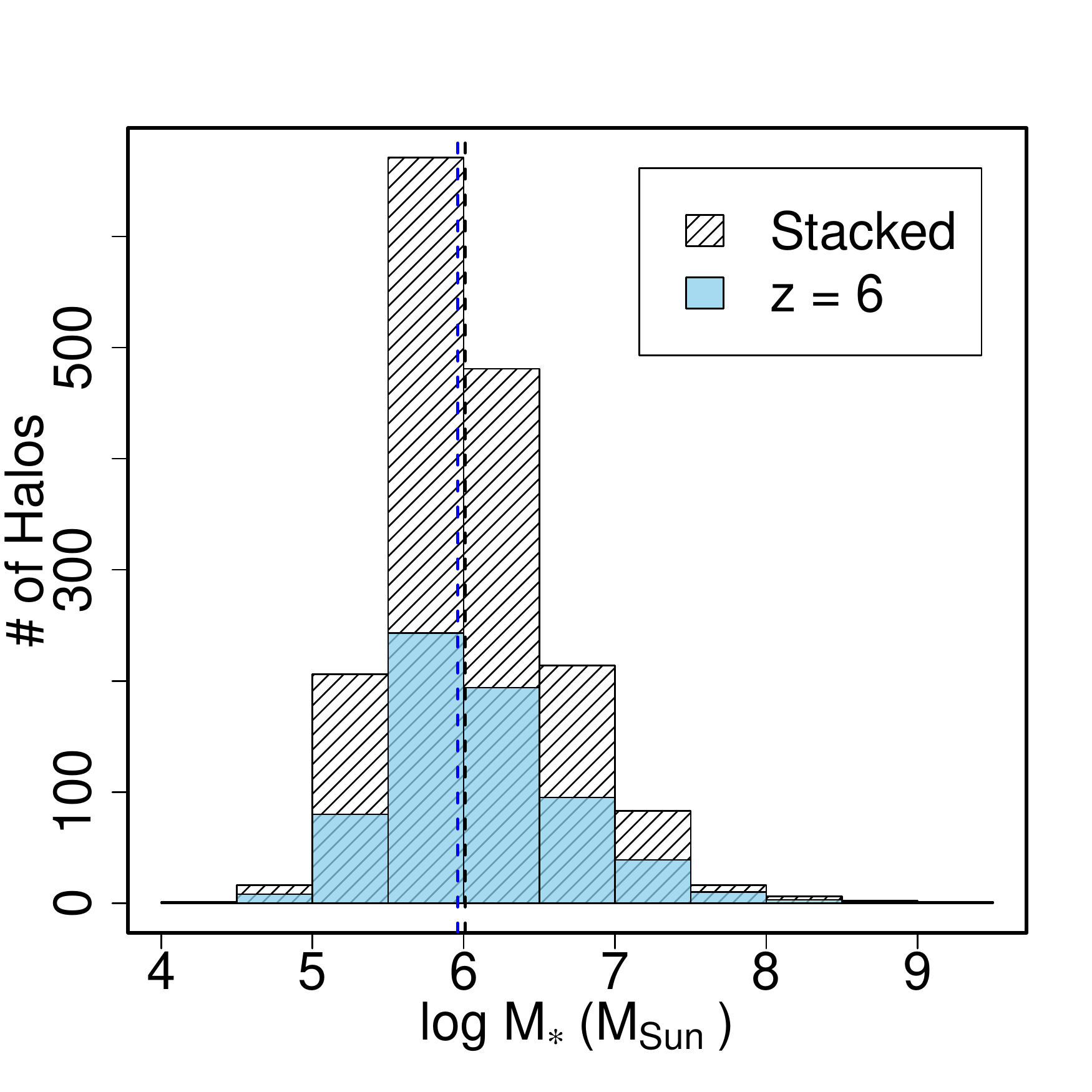}
    \includegraphics[width=0.3\textwidth]{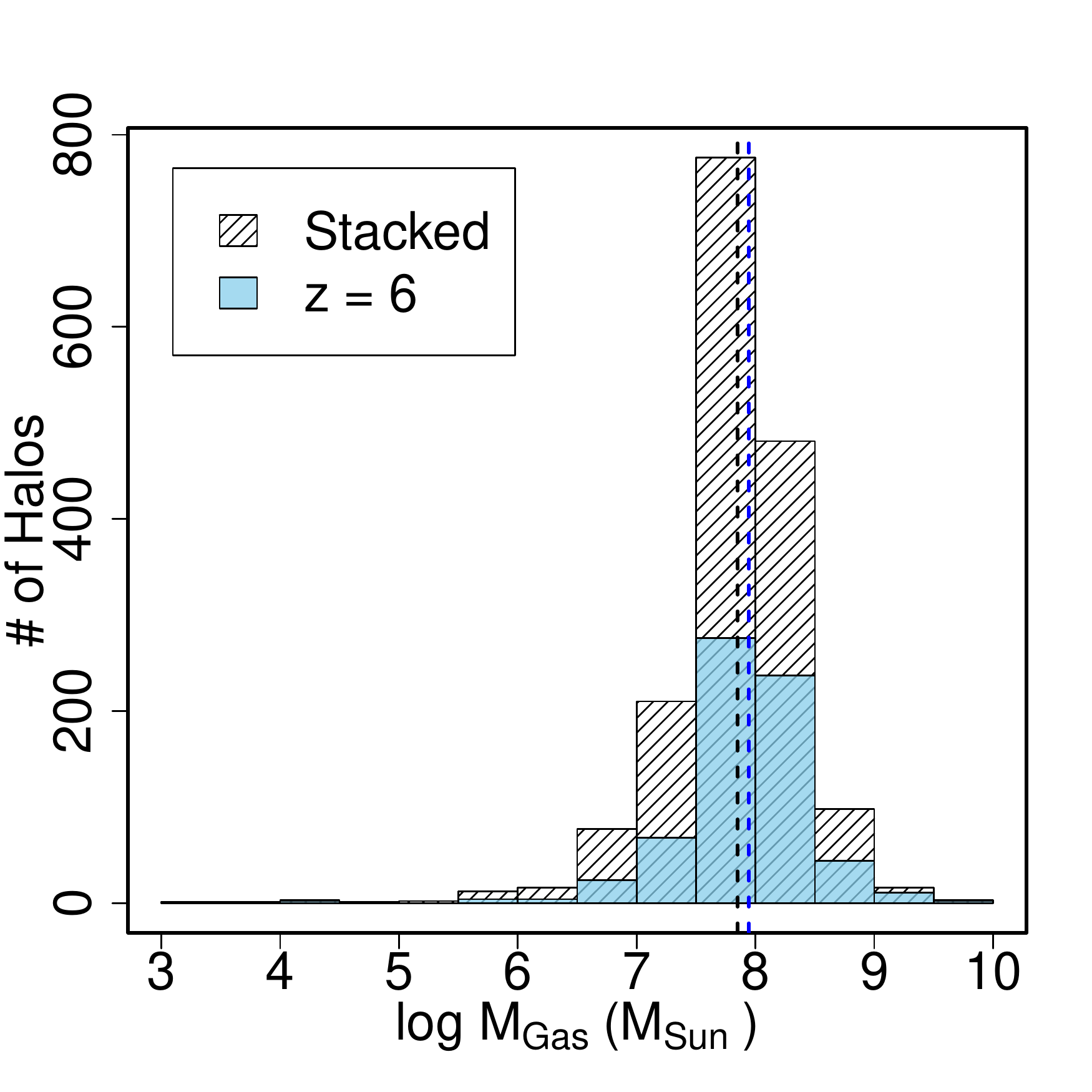}
    \includegraphics[width=0.3\textwidth]{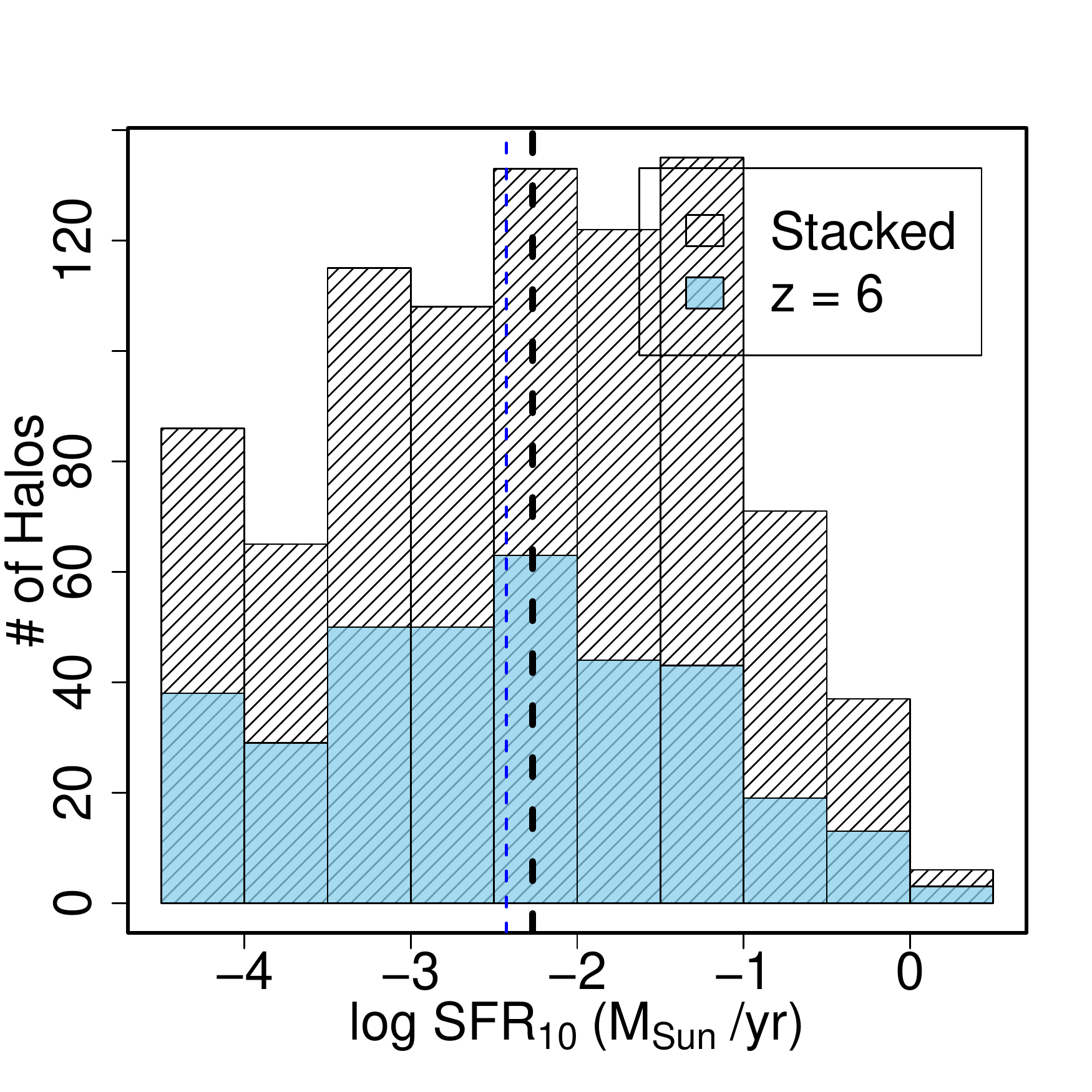}  \\
    \includegraphics[width=0.3\textwidth]{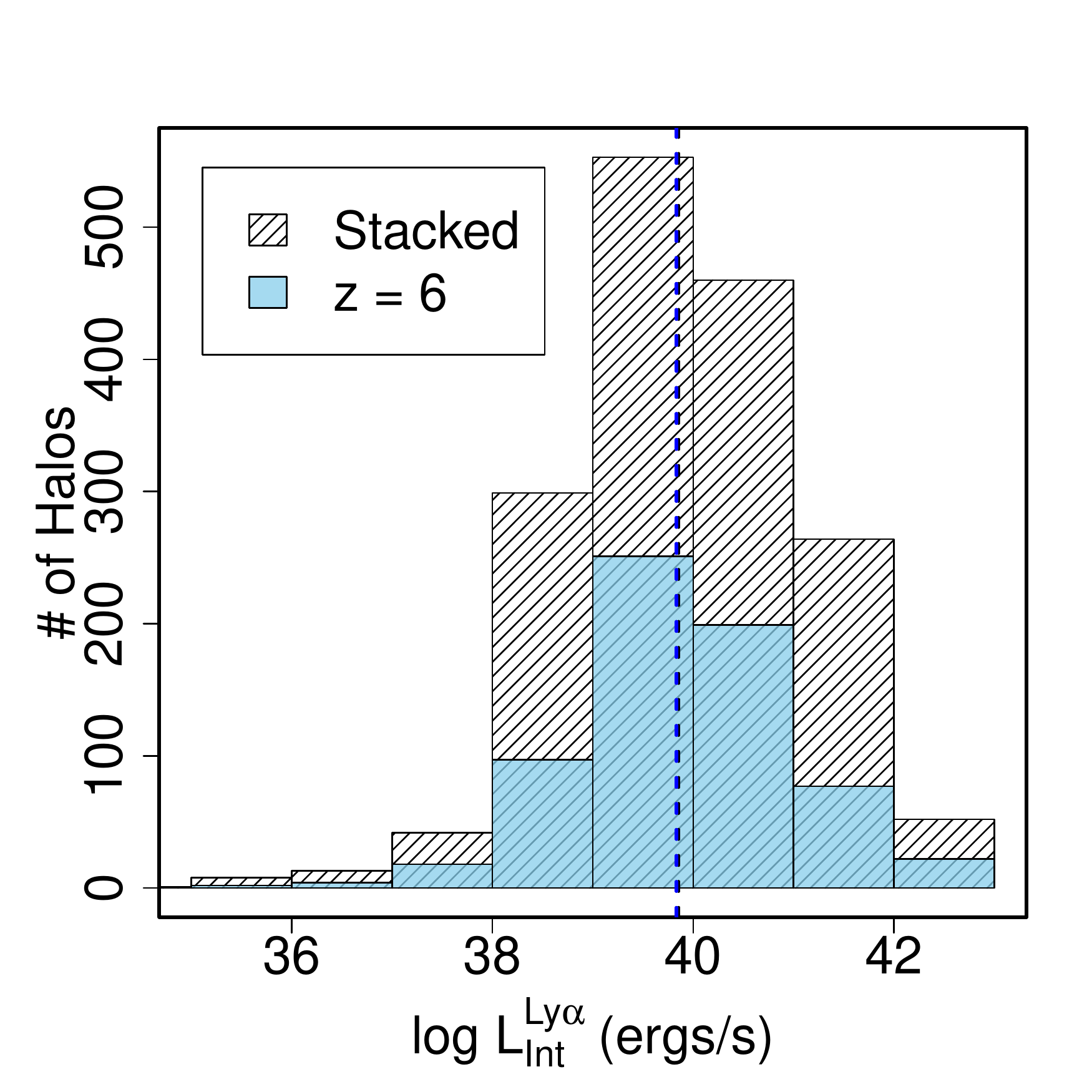}
    \includegraphics[width=0.3\textwidth]{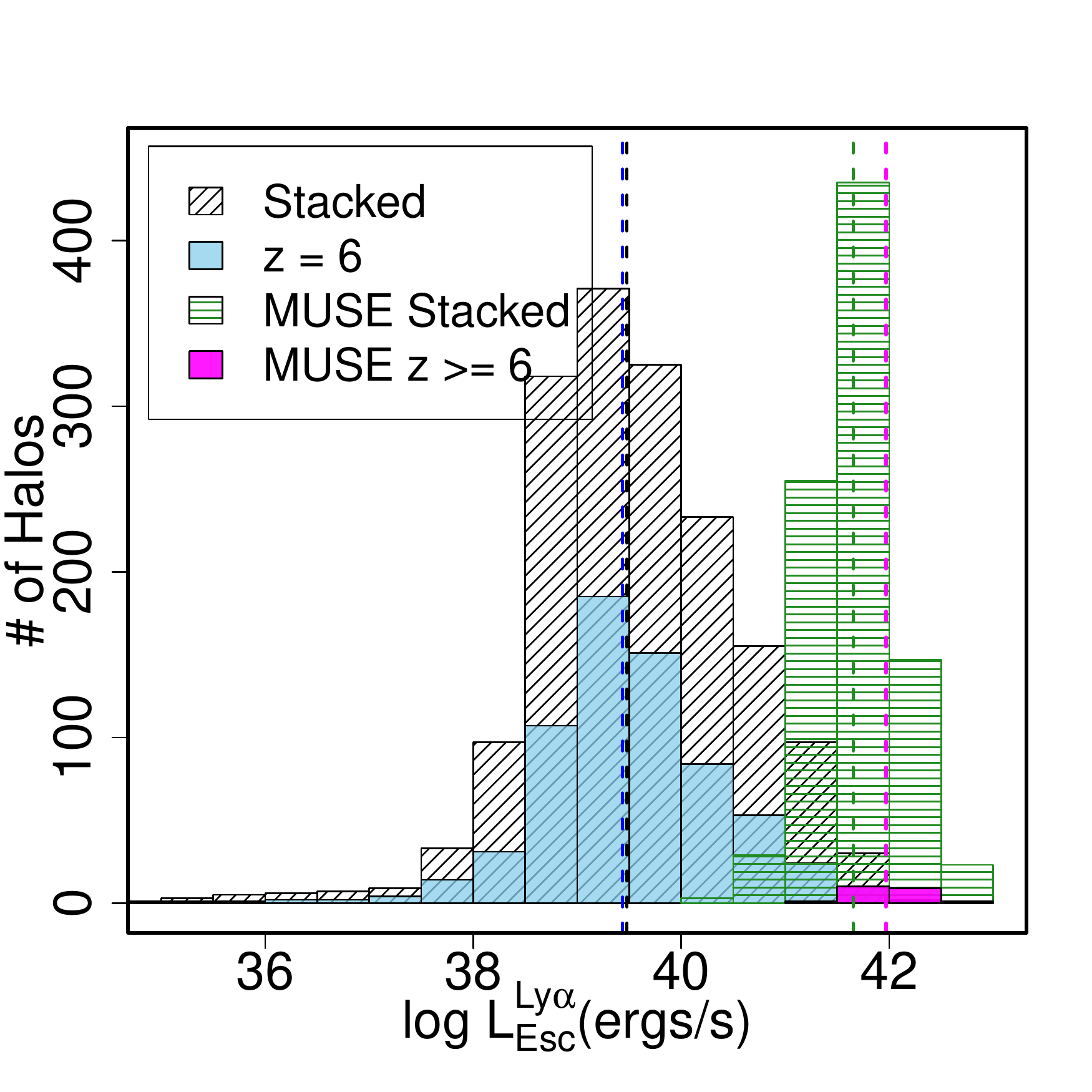}
    \includegraphics[width=0.3\textwidth]{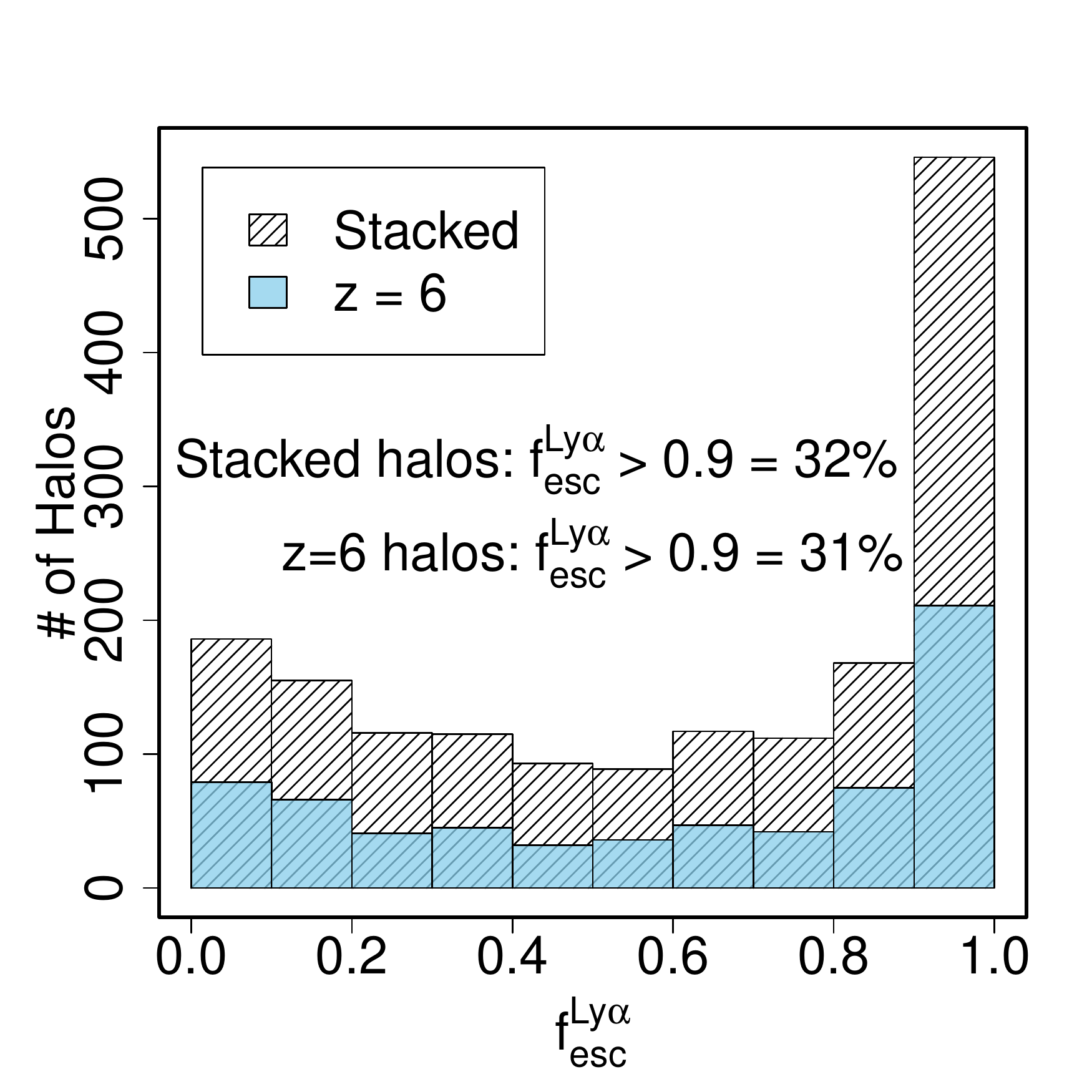}\\
    \includegraphics[width=0.3\textwidth]{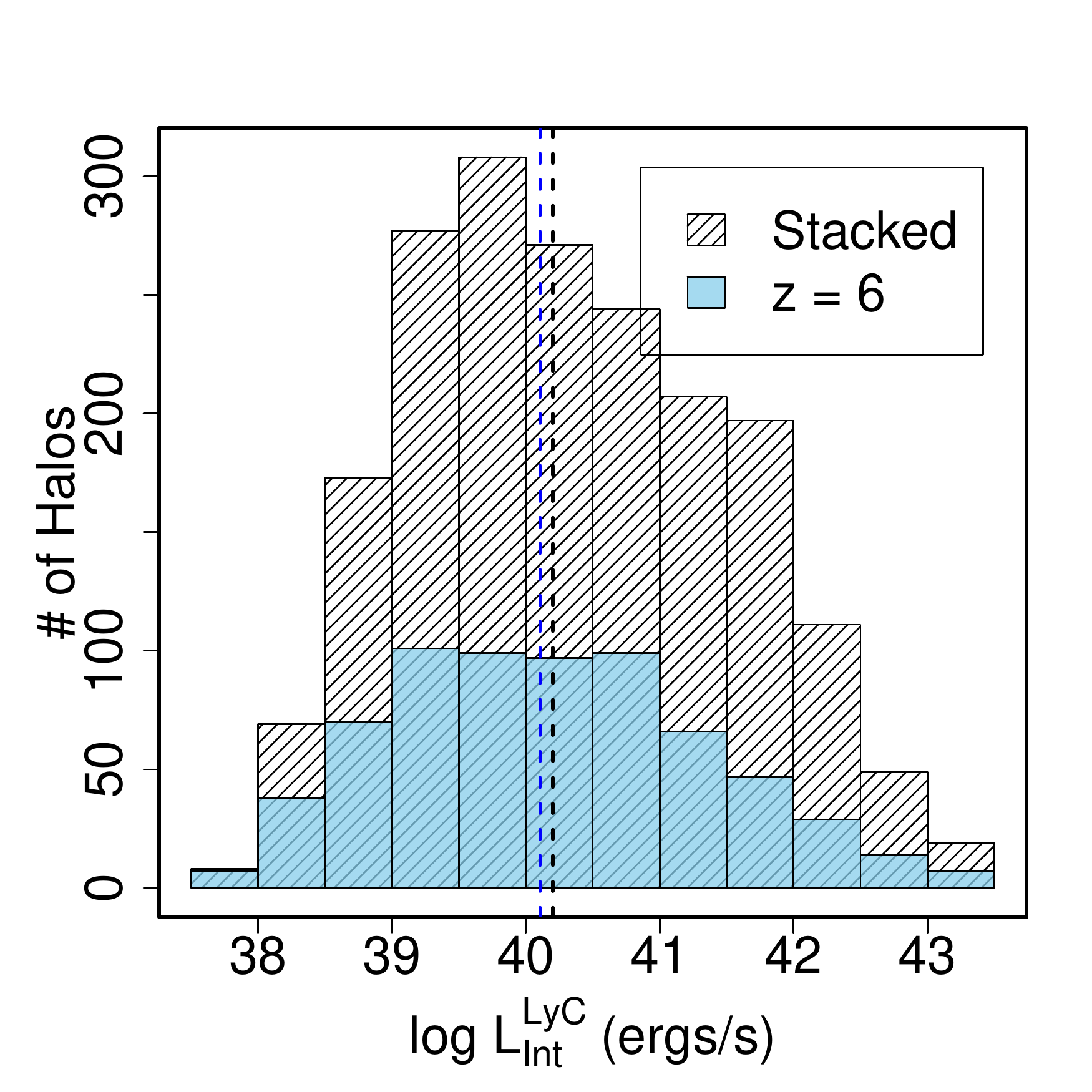}
    \includegraphics[width=0.3\textwidth]{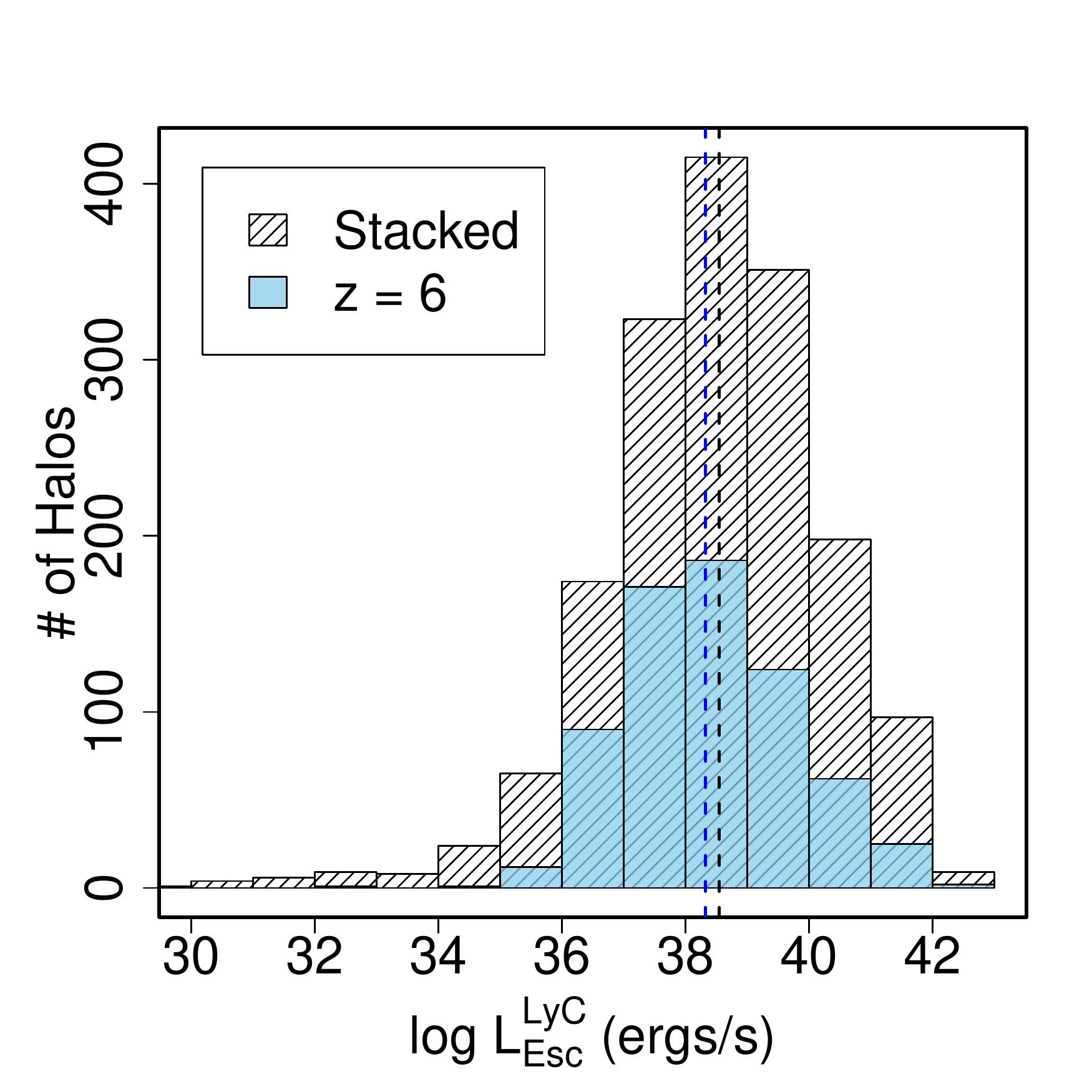}
    \includegraphics[width=0.3\textwidth]{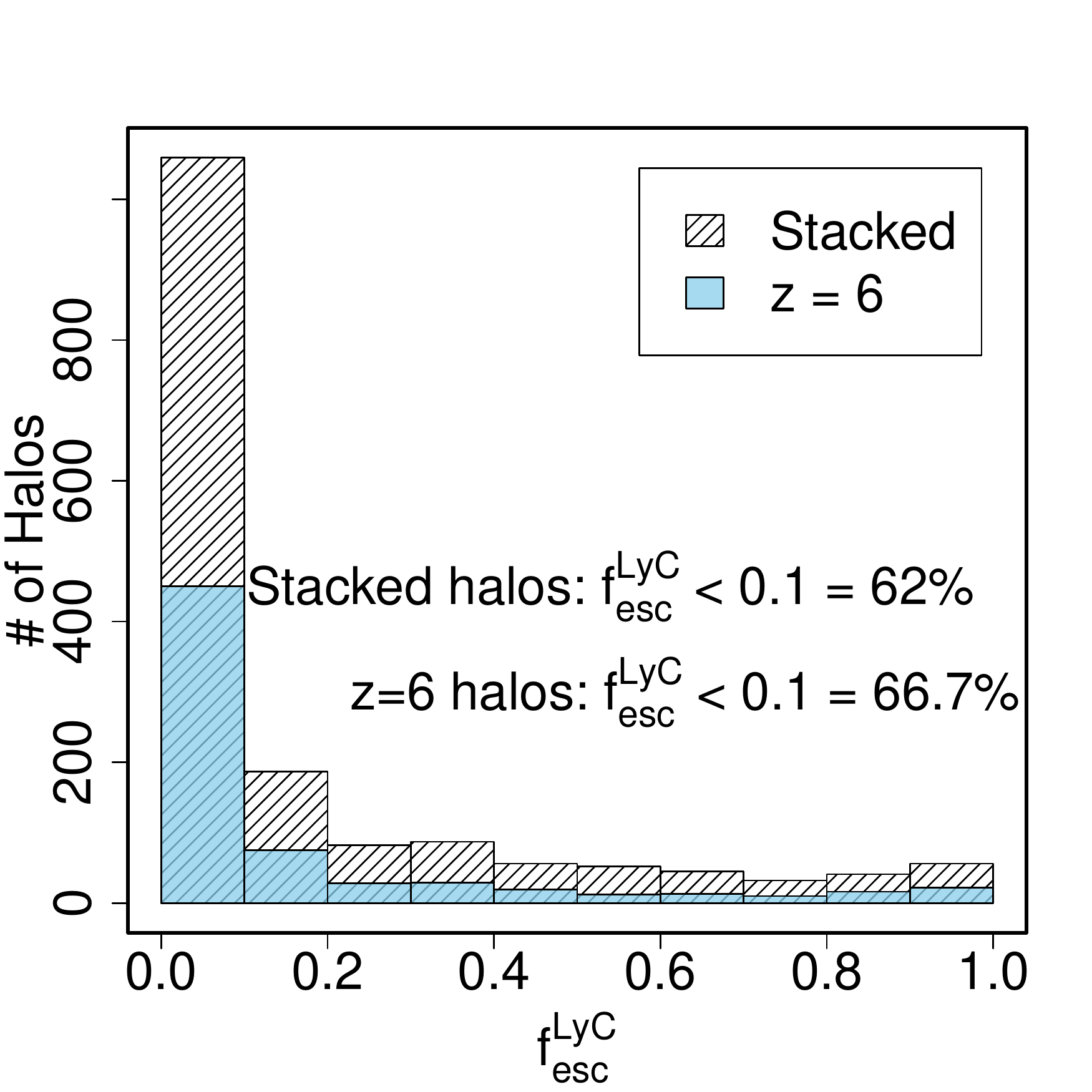}
  \caption{Comparing the physical, \lya and LyC properties of galaxies of the stacked sample (gray) with z=6 sample (blue). The top row shows compares stellar mass (left), gas mass (middle) and SFR$_{10} (right)$. The middle row compares \lya properties of the two samples with intrinsic luminosity (left), escaping luminosity (middle) and escape fraction (right). The bottom row shows the same properties but for LyC radiation. The dashed lines show the median value of the properties for both  stacked (black) and $z = 6$ sample (blue).}
    \label{fig:compare_stacked_z6}
\end{figure*}

We compare the physical properties, \lya properties and LyC properties of these two samples and present the results in Figure~\ref{fig:compare_stacked_z6}. In the top row, it shows comparisons of three physical galaxy properties, stellar mass, gas mass and SFR calculated over the last 10 Myrs (SFR$_{10}$). We find that in each case, the distributions are very similar and the median value of the mass and SFR$_{10}$ is also almost the same. We have also compared the halo mass, size of the galaxy (R$_{\rm vir, gal}$) and halo (R$_{\rm vir, halo}$) and SFR calculated over the last 100 Myr (SFR$_{100}$) and found that for each of these properties, the two samples have very similar values. Here we choose to show only the three properties mentioned as representative plots for brevity's sake. 

In the second and third row of Figure~\ref{fig:compare_stacked_z6}, we have compared the \lya and LyC properties of the two samples, showing for each intrinsic luminosity, escaping luminosity and the escape fraction. The plots clearly show that for both intrinsic and escaping luminosity, the distributions are again very similar with almost the same median values. 

For \flya and \flyc comparisons, we find that the distribution for either samples is not single peaked or gaussian like the other properties. The \flya distribution is close to a binomial with values biased towards close to either 0 or 1. \flyc distribution is also biased towards values close to 0.

Since a median of these two distributions would not be very meaningful, we calculate the percentage of the population that have very high \flya, defined as $\flya > 0.9$ and find that in $z=6$ sample $31\%$ fall in this category, whereas in the stacked sample the population is $32\%$. For \flyc, the distribution peaks towards extremely low values, so we calculate the percentage of population with $\flyc < 0.1$ and find it to be $66.7\%$ and $61\%$ for $z=6$ and stacked sample respectively. We find that the stacked sample is very similar to the $z=6$ sample of galaxies and there are no large systematic differences between them in terms of their physical or radiative properties. We note that the age of the Universe at z = 6 is 927 Myr and at z = 10 it is 470 Myr, 
so between the redshift range of 6 - 10, only 457 Myr pass. So it is not surprising that we find the statistical properties of the galaxies within this time frame do not change significantly in our simulation. Our results suggest that we can use our stacked sample of 1933 galaxies for our \lya and LyC analysis to study reionization era galaxies.

\subsection{Contribution of Recombination and Collision to \lya production}
\label{appendix:rec_col}

\begin{figure*}[h]
    \centering
    \includegraphics[width=0.5\textwidth]{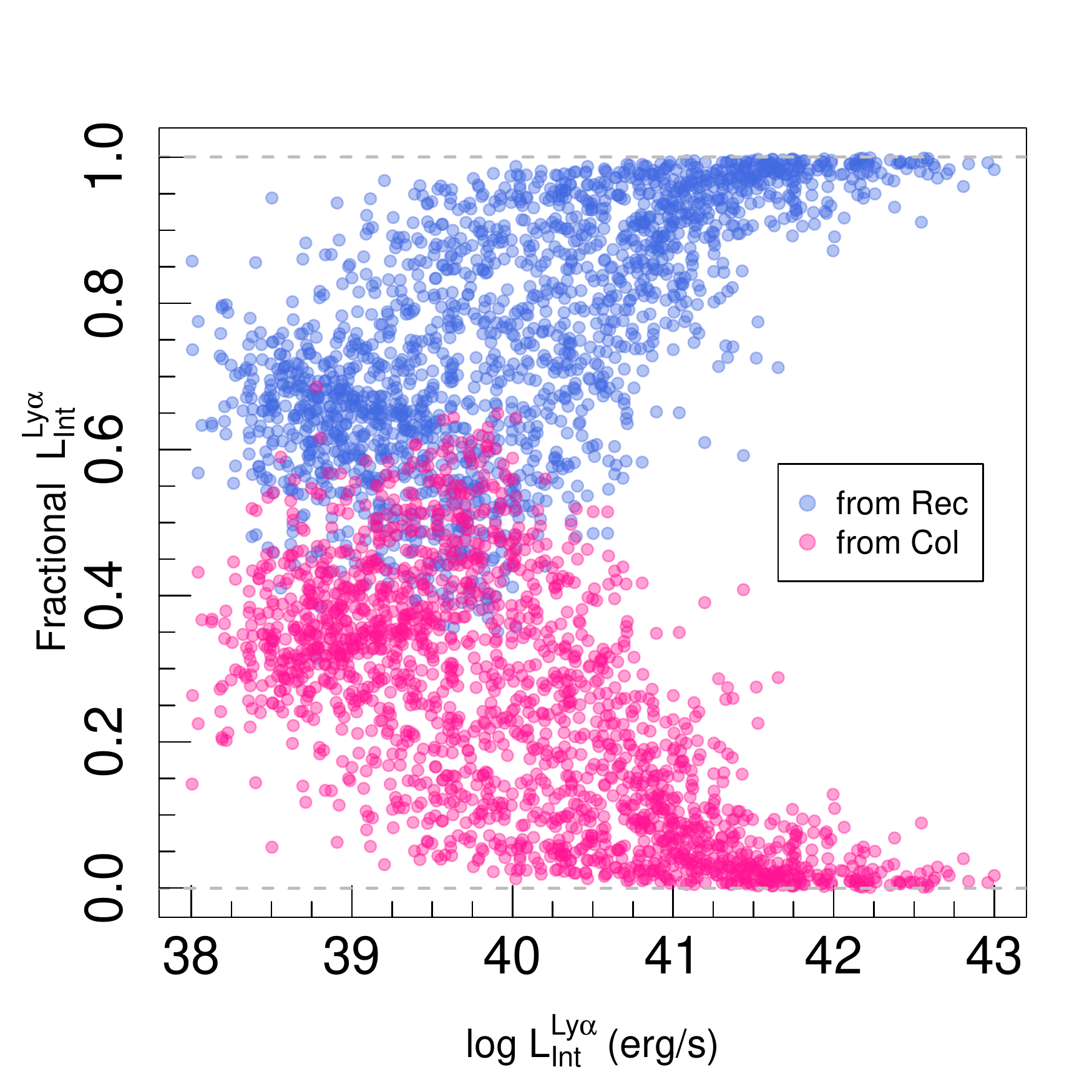}
    \caption{Fraction of intrinsic \lya luminosity generated by recombination (blue) and collision (pink) as a function of \Llyain.}
    \label{fig:frac_rec_col}
\end{figure*}

Figure~\ref{fig:frac_rec_col} shows the fraction of intrinsic \lya that comes from recombination and collision respectively. We find that in bright LAEs almost all of the \Llyain is generated from recombination. However, the contribution of collision becomes higher as galaxies becomes fainter. For example, in galaxies where $\Llyain > 10^{42}$ erg/s collisions contribute $\sim$ a few percent, but it can rise to $\sim 50\%$ in galaxies $10^{38} > \Llyain > 10^{40}$ erg/s.

\subsection{Variation of escape fractions with escaping luminosities}
\label{appendix:colorbyfesc}
\begin{figure*}[h]
    \centering
    \includegraphics[width=0.45\textwidth]{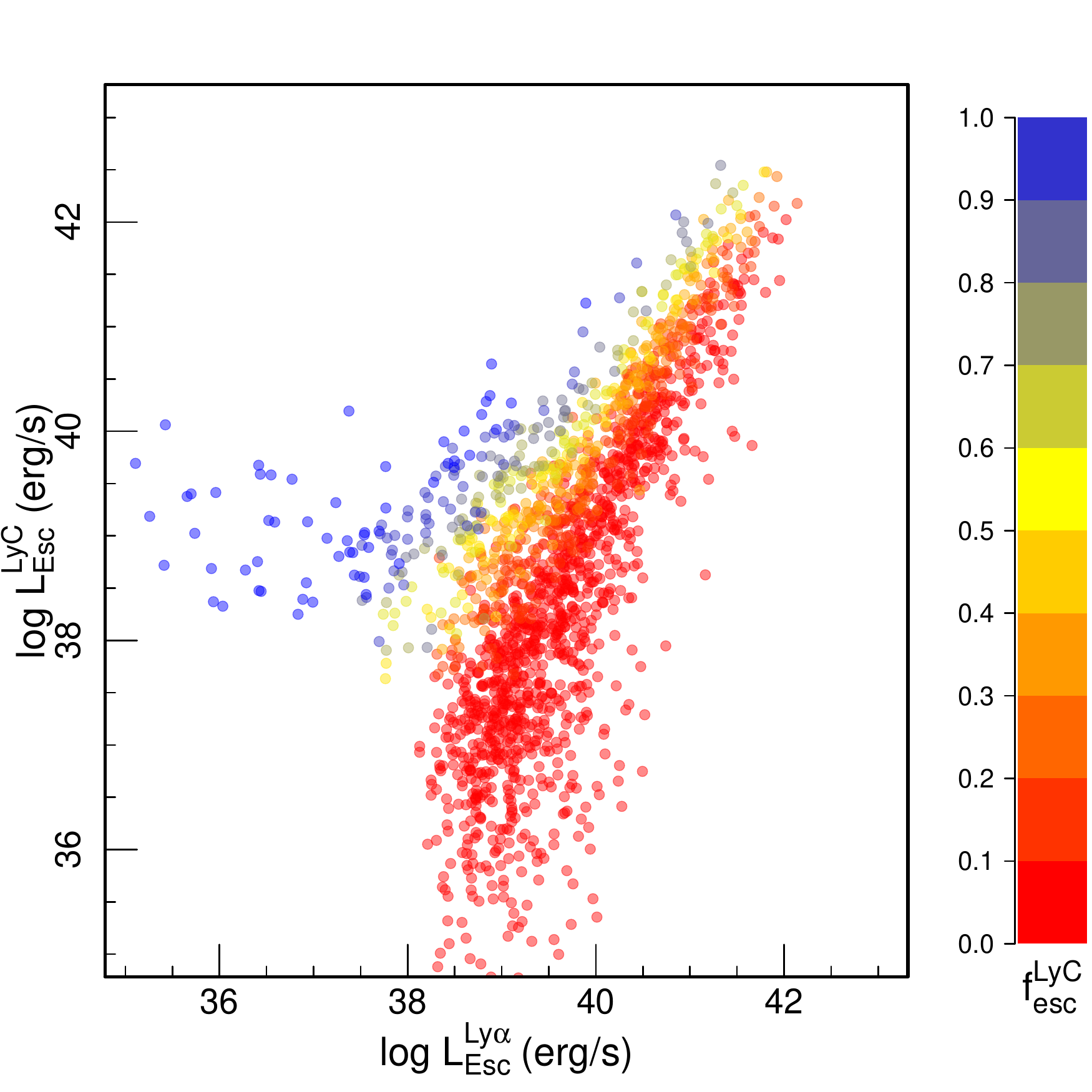}
    \includegraphics[width=0.45\textwidth]{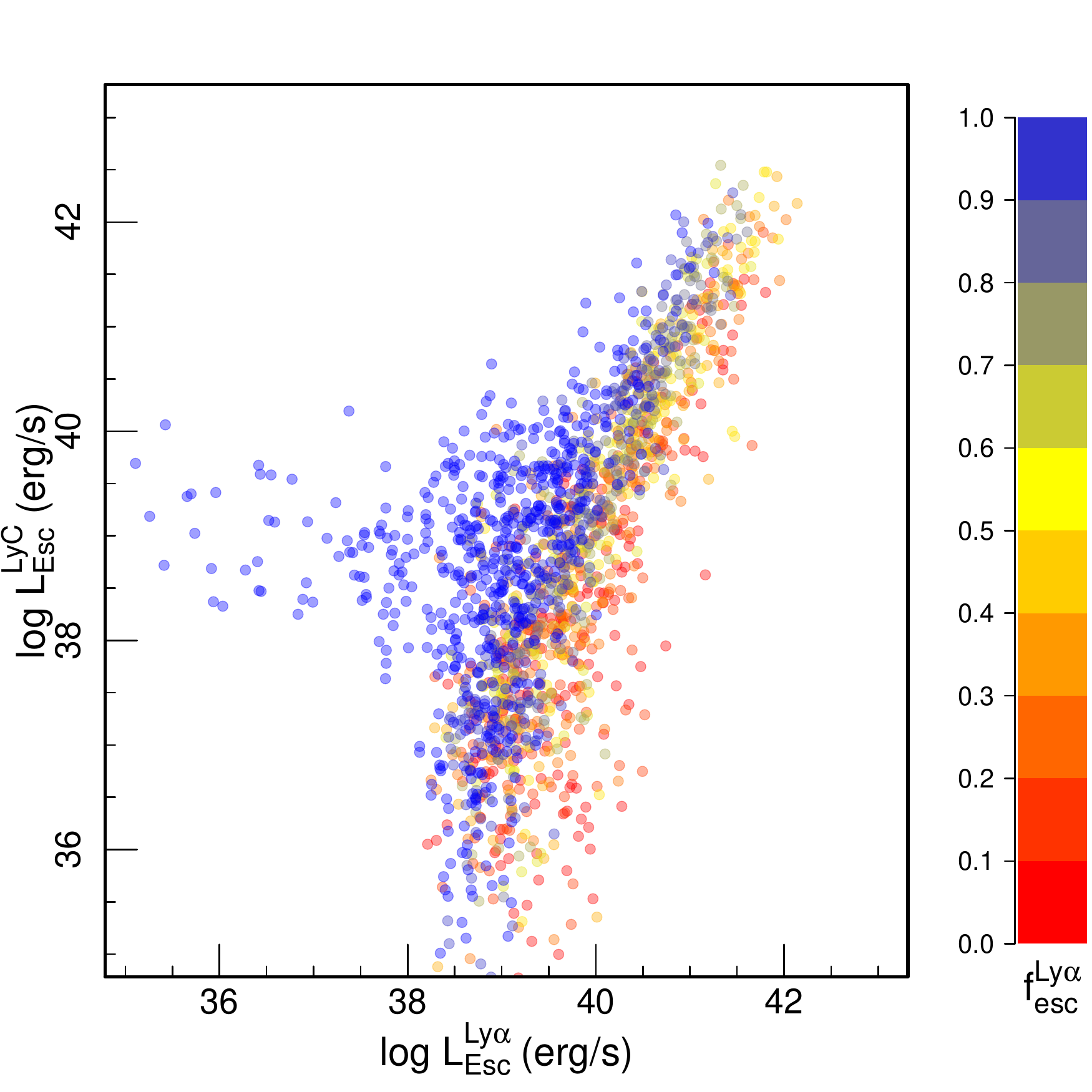}
    \caption{Escaping LyC luminosity of galaxies as a function of their escaping \lya luminosity. This is same as Fig~\ref{fig:lya_lyc_lum_inout}, but the points here are colored by their \flyc (left) and \flya (right).}
    \label{fig:lumout_color}
\end{figure*}

We have discussed the relationship between \lya and LyC luminosities and escape fractions in \S~\ref{sec:lyalyc_lum} and \S~\ref{sec:lyalyc_fesc} respectively. Here we revisit them and discuss how galaxy escape fractions vary with their luminosities. In fig~\ref{fig:lumout_color} we show \Llycout as a function of their \Llyaout, similar to Fig~\ref{fig:lya_lyc_lum_inout}, but here colored by their \flyc and \flya. We find that most of the galaxies have low \flyc and there is a clear trend that for a given \Llycout, brighter LAEs have lower \flyc. When \flyc is high, most of the LyC is escaping, so there are few LyC photons available to produce \lya, hence \lya luminosity is low. As \flyc decreases, more and more LyC photons are reprocessed into \lya, and \lya luminosity increases. On the other hand, most of the galaxies have high \flya. In general, faint LAEs have high \lya escape fraction, but there is significant scatter at each luminosities.

\subsection{Reionization accounting with lower mass limit}
\label{appedix:reion}
In \S~\ref{sec:lyalyc_reion} we have discussed the contribution of LAEs towards reionization and found that LAEs brighter than $10^{40}$ erg/s can account for ~$95\%$ of the total ionizing luminosity in the simulation, suggesting that bright LAEs may be the most important sources of reionization. However, in this analysis while counting the LyC contribution of LAEs, following our galaxy selection criterion in \S~\ref{sec:sample} we have considered all galaxies with $M_\star > 10^6$\Msun. It will be instructive to explore how the results will change if we impose a lower mass limit, e.g. $10^5$\Msun. In order to investigate this, we need to first run the \lya radiative transfer on all galaxies with $M_\star > 10^5$\Msun. Since the number of galaxies within $10^5 - 10^6$\Msun range is very high, post-processing all of them in the full stacked sample will be very expensive. Hence, we limit our investigation to galaxies in z = 6 snapshot only. At z=6, there are 674 and 1495 galaxies with $M_\star > 10^6$\Msun and $M_\star > 10^5$\Msun, respectively.

\begin{figure*}[h]
    \centering
    \includegraphics[width=0.75\textwidth]{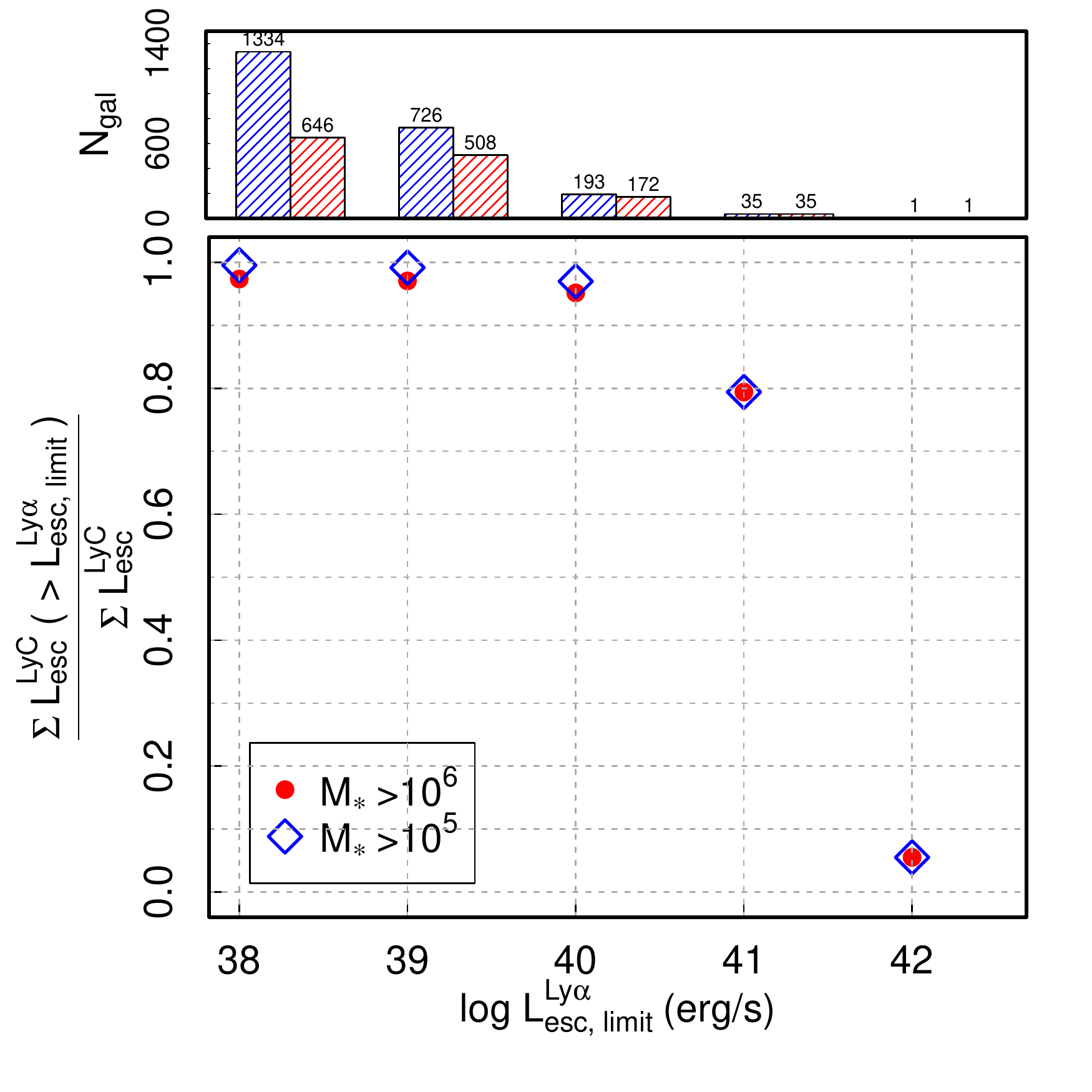}
    \caption{Fraction of the total escaping LyC luminosity emitted by galaxies brighter than a given \lya luminosity limit as a function of the \lya luminosity limit. Here we compare this fraction for two sets of galaxy sample: all galaxies at level 1 with $M_\star > 10^6 \Msun$ (red points), and all galaxies at level 1 with $M_\star > 10^5 \Msun$ (blue points). These galaxies are all taken from z=6 snapshot. So the denominator of the fraction is same in both cases, the total LyC emission by all galaxies (at level 1) at z = 6. The numerator calculates the total LyC luminosity of the galaxies brighter than a given \lya luminosity limit with the two samples, e.g. the total LyC emitted by all galaxies (at level 1) with $M_\star > 10^6$ (or $10^5$) \Msun and \Llyaout > $10^{40}$ erg/s. The histograms above show the number of galaxies brighter than the corresponding \lya luminosity limit, e.g. the number of galaxies with $\Llyaout > 10^{40}$ for the two mass limits. This is also the number of galaxies used to calculate the corresponding fractions shown in the main plot. We find that when we take all galaxies with $M_\star > 10^6$ ($10^5$)\Msun, LAEs brighter than $10^{40}$ erg/s can account for 95$\%$ (97$\%$) of the total ionizing luminosity.}
    \label{fig:reion_105_106}
\end{figure*}

Similar to our analysis in \S~\ref{sec:lyalyc_reion}, we first calculate the total LyC luminosity emitted by all (level 1) galaxies at z = 6. Then we calculate how much of this total LyC is emitted by galaxies with \Llyaout $> 10^{38}, 10^{39}, 10^{40}, 10^{41}$ and $10^{42}$ erg/s using samples with both stellar mass limits of $10^6$ and $10^5$ \Msun. Figure~\ref{fig:reion_105_106} show this cumulative fraction against the limiting \lya luminosity of the galaxies. We find that LAEs brighter than $10^{40}$ erg/s can account for $95\%$ of total LyC when counting only $M_\star > 10^6$\Msun galaxies, and if we lower the mass limit to $10^5$\Msun, this fraction increases to $97\%$. At the low luminosity limit, LAEs brighter than $10^{38}$ erg/s contribute $97\%$ (99\%) of the re-ionizing radiation. This results show that although lowering the mass limit slightly increase these fractions, the differences are very small. This indicates that the reionization accounting we have done in \S~\ref{sec:lyalyc_reion} with $10^6$\Msun mass limit is reasonably accurate. 

\subsection{Multivariate model: more exploratory analysis}

\begin{figure*}
    \centering
    \includegraphics[width=\textwidth]{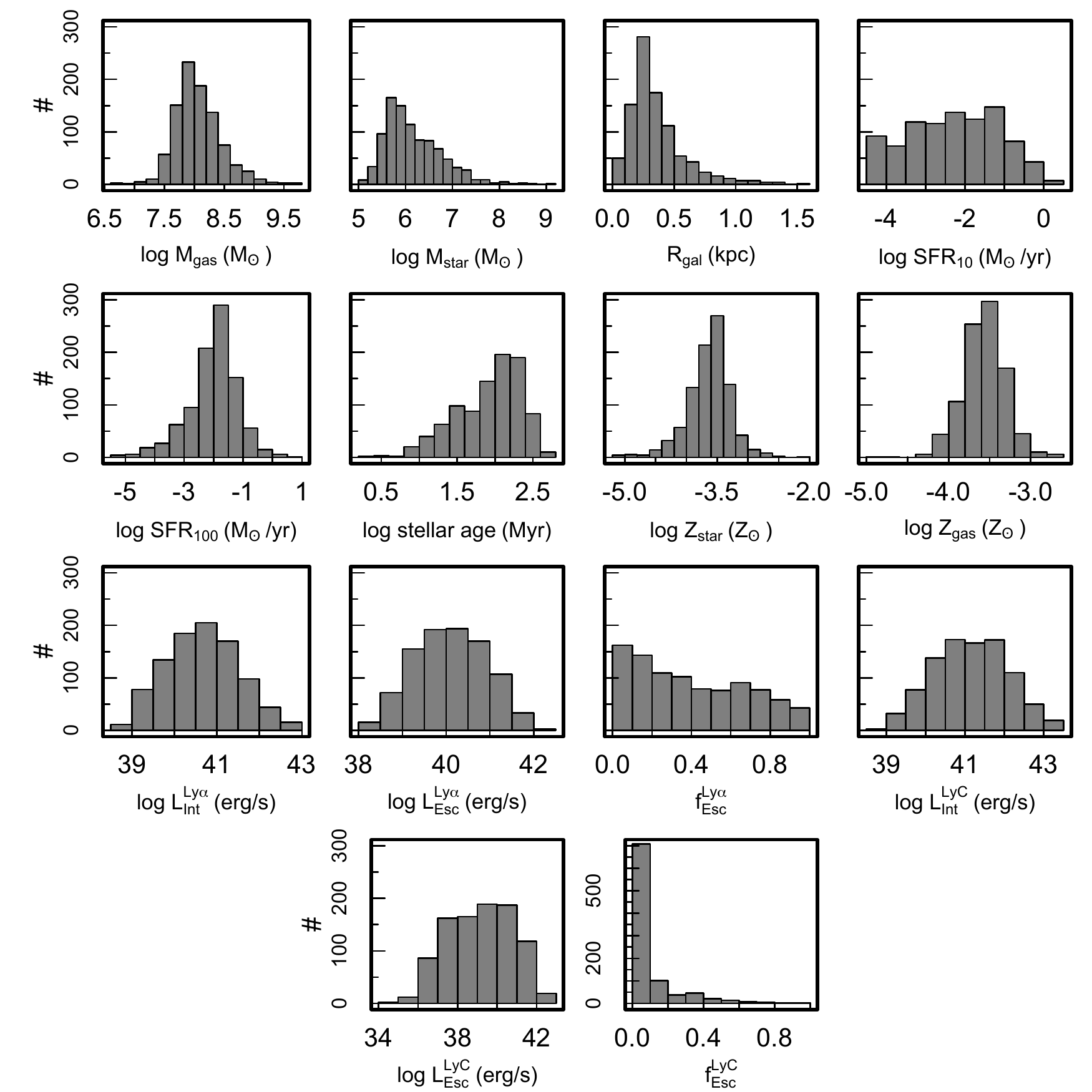}
    \caption{Histogram of the 14 galaxy properties (gas mass, stellar mass, galaxy radius, SFR$_{10}$, SFR$_{100}$, stellar age, stellar and gas metallicity, intrinsic and escaping luminosities and escape fractions of \lya and LyC, as described in \S~4.1.1) for our sample of 940 galaxies (\S~4.2.1) that were used to build the predictive models (\S~4.2.2).}
    \label{fig:hist_pred_prop}
\end{figure*}

\begin{figure*}[h]
    \centering
    \includegraphics[width=0.85\textwidth]{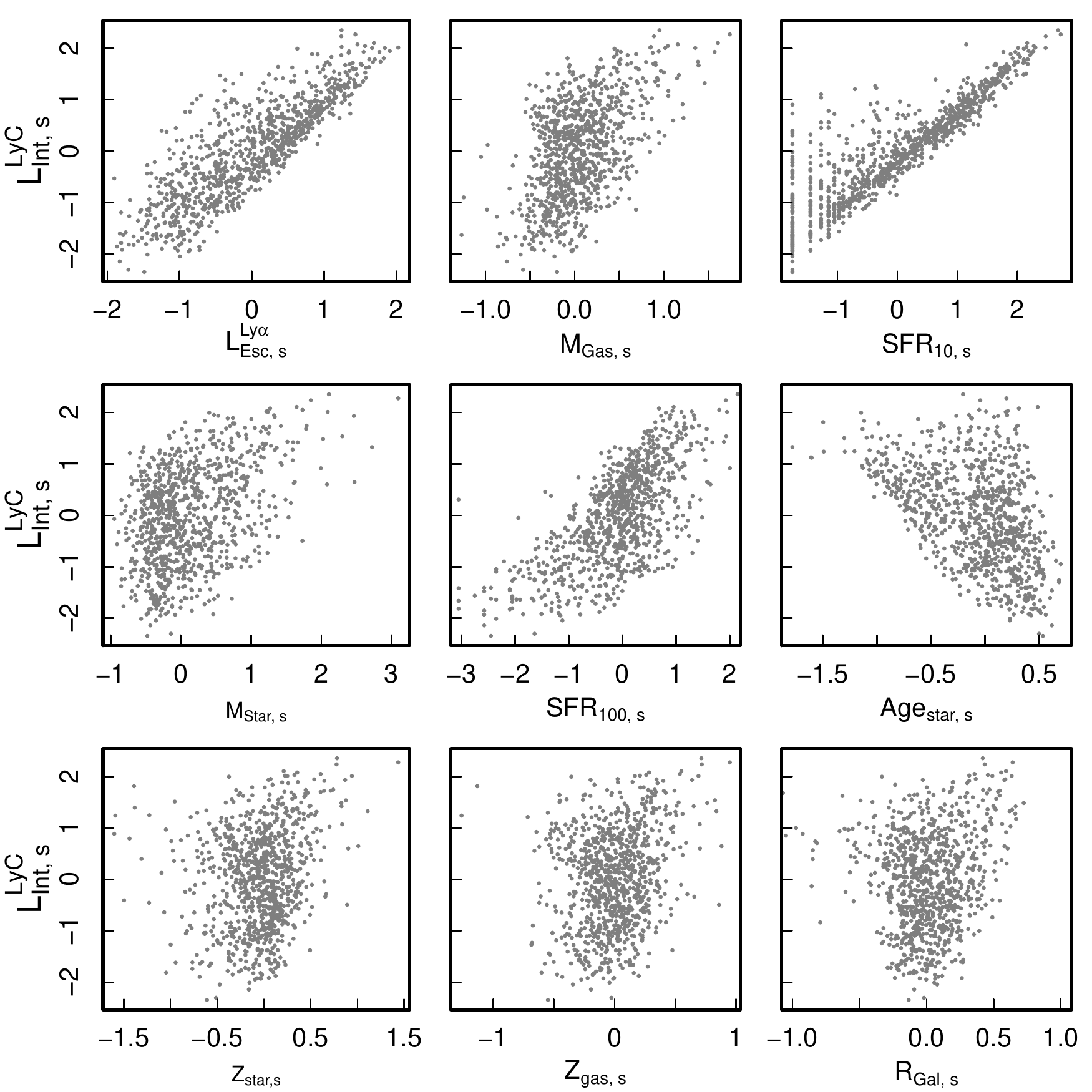}
    \caption{\Llycin vs. all \textit{x-}variables in our model. All quantities here are scaled as prescribed in $\S$ \ref{sec:var_selection}. We find that several properties, especially, SFR$_{10}$ and \Llyaout correlate very well with \Llycin, which suggests that the multivariate linear regression model will be a choice for predicting \Llycin.}
    \label{fig:Lin_vs_everything}
\end{figure*}

\begin{figure*}
    \centering
    \includegraphics[width=0.85\textwidth]{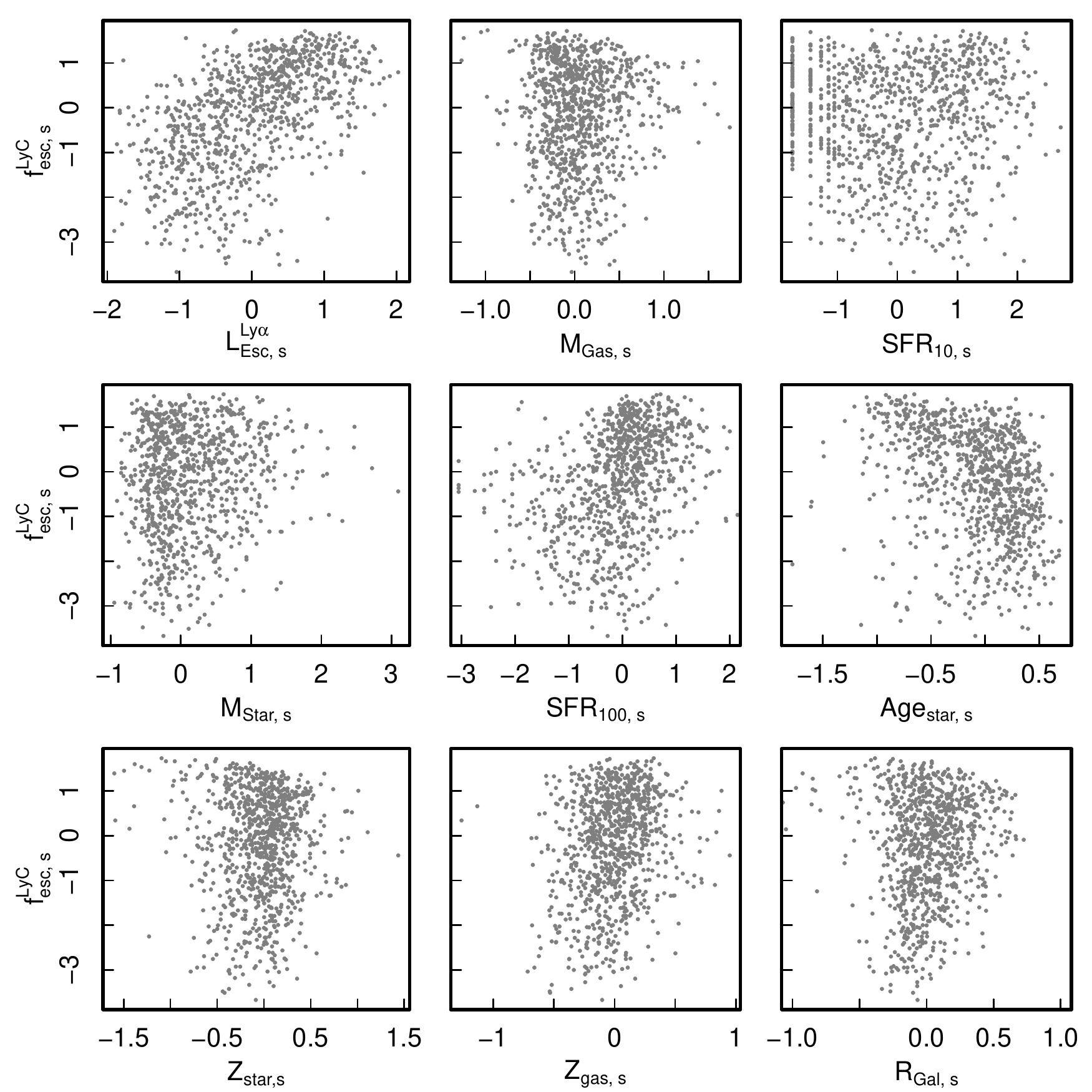}
    \caption{Same as \ref{fig:Lin_vs_everything} but for the response variable \flyc.}
    \label{fig:flyc_allx}
\end{figure*}

We show the histograms for the galaxy properties used in building our models (as listed in \S~\ref{sec:var_selection}) for our sample of 940 galaxies (\S~\ref{sec:gal_selection}) in Figure~\ref{fig:hist_pred_prop}. We have discussed in \S~\ref{sec:multivariate_model} that the before building a multivariate linear model to predict LyC properties, it is important to check if any of the proposed \textit{x-}variables or input variables have any correlation with the y-variable or response variable. Figure~\ref{fig:Lin_vs_everything} and Figure~\ref{fig:flyc_allx} show such exploratory plot of the response variable \Llycin and \flyc vs various galaxy properties, respectively.  We find that several properties, especially, SFR$_{10}$ and \Llyaout correlates very well with \Llycin. There are also weak correlations with gas mass, SFR$_{100}$ and stellar age. \flyc is also correlated with \Llyaout. These suggests that the multivariate linear regression model can be a good choice for predicting LyC emission from galaxies using these properties.

\end{appendix}

\end{document}